\documentclass[reprint,showpacs,superscriptaddress,twocolumn,english,notitlepage,longbibliography,nofootinbib]{revtex4-1}
\usepackage{mathrsfs}
\usepackage{graphicx}
\usepackage{dcolumn}
\usepackage{bm}
\usepackage{amsmath}
\usepackage{amsfonts}
\usepackage{color}
\usepackage{hyperref}
\usepackage{cleveref}

\def\beq#1\eeq{\begin{equation}#1\end{equation}}
\def\beqs#1\eeqs{\begin{align}#1\end{align}}

\def\dg{\dagger}

\def\pr{\prime}

\def\kk{\mathbf{k}}
\def\qq{\mathbf{q}}

\def\GG{\mathbf{G}}
\def\QQ{\mathbf{Q}}
\def\RR{\mathbf{R}}

\def\CC{\mathcal{P}}

\makeatother

\begin{document}

\title{Twisted Bilayer Graphene IV. Exact Insulator Ground States and Phase Diagram}

\author{Biao Lian}
\affiliation{Department of Physics, Princeton University, Princeton, New Jersey 08544, USA}
\author{Zhi-Da Song}
\affiliation{Department of Physics, Princeton University, Princeton, New Jersey 08544, USA}
\author{Nicolas Regnault}
\affiliation{Department of Physics, Princeton University, Princeton, New Jersey 08544, USA}
\affiliation{Laboratoire de Physique de l'Ecole normale superieure, ENS, Universit\'e PSL, CNRS, Sorbonne Universit\'e, Universit\'e Paris-Diderot, Sorbonne Paris Cit\'e, Paris, France}
\author{Dmitri K. Efetov}
\affiliation{ICFO – Institut de Ciencies Fotoniques, The Barcelona Institute of Science and Technology, Castelldefels, Barcelona, Spain}
\author{Ali Yazdani}
\affiliation{Department of Physics, Princeton University, Princeton, New Jersey 08544, USA}
\affiliation{Joseph Henry Laboratories, Princeton University, Princeton, New Jersey 08544, USA}
\author{B. Andrei Bernevig}
\affiliation{Department of Physics, Princeton University, Princeton, New Jersey 08544, USA}

\begin{abstract}
We derive the exact insulator ground states of the projected Hamiltonian of magic-angle twisted bilayer graphene (TBG) flat bands with Coulomb interactions in various limits, and study the perturbations away from these limits. We define the (first) chiral limit where the AA stacking hopping is zero, and a flat limit with exactly flat bands. In the chiral-flat limit, the TBG Hamiltonian has a U(4)$\times$U(4) symmetry, and we find that the exact ground states at integer filling $-4\le \nu\le 4$ relative to charge neutrality are Chern insulators of Chern numbers $\nu_C=4-|\nu|,2-|\nu|,\cdots,|\nu|-4$, all of which are degenerate. This confirms recent experiments where Chern insulators are found to be competitive low-energy states of TBG.  When the chiral-flat limit is reduced to the nonchiral-flat limit which has a U(4) symmetry,  we find $\nu=0,\pm2$ has exact ground states of Chern number $0$, while $\nu=\pm1,\pm3$ has perturbative ground states of Chern number $\nu_C=\pm1$, which are U(4) ferromagnetic. In the chiral-nonflat limit with a different U(4) symmetry, different Chern number states are degenerate up to second order perturbations. In the realistic nonchiral-nonflat case, we find that the perturbative insulator states with Chern number $\nu_C=0$ ($0<|\nu_C|<4-|\nu|$) at integer fillings $\nu$ are fully (partially) intervalley coherent, while the insulator states with Chern number $|\nu_C|=4-|\nu|$ are valley polarized. However, for $0<|\nu_C|\le4-|\nu|$, the fully intervalley coherent states are highly competitive (0.005meV/electron higher). At nonzero magnetic field $|B|>0$, a first-order phase transition for $\nu=\pm1,\pm2$ from Chern number $\nu_C=\text{sgn}(\nu B)(2-|\nu|)$ to $\nu_C=\text{sgn}(\nu B)(4-|\nu|)$ is expected, which agrees with recent experimental observations. Lastly, the TBG Hamiltonian reduces into an extended Hubbard model in the stabilizer code limit.
\end{abstract}

\date{\today}

\maketitle

\section{Introduction}\label{sec:introduction}

Recently, remarkable interacting phases have been observed in twisted bilayer graphene (TBG) near the magic angle $\theta\approx 1.1^\circ$, including correlated insulators, Chern insulators and superconductors \cite{bistritzer_moire_2011,cao_correlated_2018,cao_unconventional_2018, lu2019superconductors, yankowitz2019tuning, sharpe_emergent_2019, saito_independent_2020, stepanov_interplay_2020, liu2020tuning, arora_2020, serlin_QAH_2019, cao_strange_2020, polshyn_linear_2019,  xie2019spectroscopic, choi_imaging_2019, kerelsky_2019_stm, jiang_charge_2019,  wong_cascade_2020, zondiner_cascade_2020,  nuckolls_chern_2020, choi2020tracing, saito2020,das2020symmetry, wu_chern_2020,park2020flavour, saito2020isospin,rozen2020entropic, lu2020fingerprints, burg_correlated_2019,shen_correlated_2020, cao_tunable_2020, liu_spin-polarized_2019, chen_evidence_2019, chen_signatures_2019, chen_tunable_2020, burg2020evidence, tarnopolsky_origin_2019, zou2018, fu2018magicangle, liu2019pseudo, Efimkin2018TBG, kang_symmetry_2018, song_all_2019,po_faithful_2019,ahn_failure_2019,Slager2019WL, hejazi_multiple_2019, lian2020, hejazi_landau_2019, padhi2020transport, xu2018topological,  koshino_maximally_2018, ochi_possible_2018, xux2018, guinea2018, venderbos2018, you2019,  wu_collective_2020, Lian2019TBG,Wu2018TBG-BCS, isobe2018unconventional,liu2018chiral, bultinck2020, zhang2019nearly, liu2019quantum,  wux2018b, thomson2018triangular,  dodaro2018phases, gonzalez2019kohn, yuan2018model,kang_strong_2019,bultinck_ground_2020,seo_ferro_2019, hejazi2020hybrid, khalaf_charged_2020,po_origin_2018,xie_superfluid_2020,julku_superfluid_2020, hu2019_superfluid, kang_nonabelian_2020, soejima2020efficient, pixley2019, knig2020spin, christos2020superconductivity,lewandowski2020pairing, xie_HF_2020,liu2020theories, cea_band_2020,zhang_HF_2020,liu2020nematic, daliao_VBO_2019,daliao2020correlation, classen2019competing, kennes2018strong, eugenio2020dmrg, huang2020deconstructing, huang2019antiferromagnetically,guo2018pairing, ledwith2020, repellin_EDDMRG_2020,abouelkomsan2020,repellin_FCI_2020, vafek2020hidden, fernandes_nematic_2020,  Wilson2020TBG,wang2020chiral, ourpaper1, ourpaper2,ourpaper3,ourpaper5,ourpaper6}. At integer fillings of electrons per moir\'e unit cell (quarter fillings of the "active flat bands" around charge neutrality due to spin-valley degeneracy), a slew of interacting insulating phases has been observed. Since the system hosts 8 flat electron bands, strong many-body interactions are expected to be responsible for these unconventional phases, as suggested by experiments \cite{xie2019spectroscopic,wong_cascade_2020,zondiner_cascade_2020}.  Scanning tunneling spectroscopy experiments reveal a Coulomb repulsion strengh ($\sim 25$meV) \cite{xie2019spectroscopic, wong_cascade_2020} much larger than the electron bandwidths, and show that TBG (without hBN substrate alignment) develops strong correlation gaps in magnetic fields $B$ at integer fillings $\nu$ with respect to charge neutrality, which are topological with Chern numbers $\pm(4-|\nu|)$ \cite{nuckolls_chern_2020,choi2020tracing}. Chern insulators in magnetic fields have also been observed by transport experiments in TBG with \cite{serlin_QAH_2019, sharpe_emergent_2019} and without \cite{saito2020,das2020symmetry, wu_chern_2020,park2020flavour} hBN substrate alignment.
In this paper we explain these experimental findings  (which have so far only been explained by phenomenological theories) by deriving exact ground states of the projected interacting TBG Hamiltonian within the flat bands.

Among the  theoretical studies on TBG interacting phases \cite{xu2018topological,  koshino_maximally_2018, ochi_possible_2018, xux2018, guinea2018, venderbos2018, you2019,  wu_collective_2020, Lian2019TBG,Wu2018TBG-BCS, isobe2018unconventional,liu2018chiral, bultinck2020, zhang2019nearly, liu2019quantum,  wux2018b, thomson2018triangular,  dodaro2018phases, gonzalez2019kohn, yuan2018model,kang_strong_2019,bultinck_ground_2020,seo_ferro_2019, hejazi2020hybrid, khalaf_charged_2020,po_origin_2018,xie_superfluid_2020,julku_superfluid_2020, hu2019_superfluid, kang_nonabelian_2020, soejima2020efficient, pixley2019, knig2020spin, christos2020superconductivity,lewandowski2020pairing, xie_HF_2020,liu2020theories, cea_band_2020,zhang_HF_2020,liu2020nematic, daliao_VBO_2019,daliao2020correlation, classen2019competing, kennes2018strong, eugenio2020dmrg, huang2020deconstructing, huang2019antiferromagnetically,guo2018pairing, ledwith2020, repellin_EDDMRG_2020,abouelkomsan2020,repellin_FCI_2020, vafek2020hidden, fernandes_nematic_2020}, Kang and Vafek \cite{kang_strong_2019} first proposed an approximate U(4) symmetric  interacting positive semidefinite  Hamiltonian (PSDH) in a non-maximally-symmetric Wannier basis \cite{kang_symmetry_2018}, which allowed them to obtain an exact insulator ground states at filling $\nu=\pm 2$ electrons per unit cell. Bultinck et al. \cite{bultinck_ground_2020} further discussed the TBG ground state at even fillings by identifying a U(4)$\times$U(4) symmetry of TBG in the chiral limit (named the first chiral limit in Refs.~\cite{ourpaper2,ourpaper3} in contrast to the second chiral limit defined therein), and showed that an intervalley-coherent state (denoted as K-IVC state in Ref.~\cite{bultinck_ground_2020}) is favored at charge neutrality ($\nu=0$), and could also be favored at $\nu=\pm2$. However, the analytical calculation of other integer filling (per moir\'e unit cell) ground states in the strong interaction limit has not yet been done. In paper \cite{ourpaper3} we have showed that all projected Coulomb Hamiltonians in \emph{any} number of bands have the Kang-Vafek PSDH form - with U(4) $\times$ U(4) symmetry in two chiral limits, while U(4) subgroups of this symmetry group remain valid upon moving away from either of the two chiral limits, or upon introducing kinetic terms. In paper \cite{ourpaper1} we showed that a large number of the TBG matrix elements of the Coulomb interaction can be neglected. In paper \cite{ourpaper2, ourpaper3} we have also defined and gauge-fixed a Chern basis in the lowest 8 bands in both the chiral and nonchiral limits, which is also discussed by Refs.~\cite{bultinck_ground_2020,hejazi2020hybrid}.

In this paper, employing the momentum-space projected TBG Hamiltonian derived in Ref.~\cite{ourpaper3} which is of the PSDH  Kang-Vafek type \cite{kang_strong_2019}, we demonstrate that exact TBG insulator ground states (which are Fock states) and their perturbations can be derived in and away from various limits at integer fillings per moir\'e unit cell. We define the \emph{first chiral limit} (hereafter denoted as the ``chiral limit" when no ambiguity) \cite{tarnopolsky_origin_2019,ourpaper2,ourpaper3} as the limit where the AA and AB/BA stacking centers have hoppings $w_0=0$ and $w_1>0$ respectively, and the flat limit as the limit of exactly flat kinetic bands. We then study different combinations of these two limits. (We note that a \emph{second chiral limit} is also defined in Refs.~\cite{ourpaper2,ourpaper3} where $w_0>0$ and $w_1=0$, which is however far away from realistic TBG parameters. Throughout this paper, when we talk about the ``chiral limit", we refer to the first chiral limit.)

\subsection{Summary of results}

Our results can be conveniently presented in the Chern (band) basis we defined in Ref.~\cite{ourpaper3} (see also \cite{bultinck_ground_2020,liu2019pseudo}), which are defined by linearly recombining 2 flat bands of each spin-valley into 2 Chern bands of Chern number $\pm1$, respectively (thus in total 4 Chern number $+1$ bands and 4 Chern number $-1$ bands given by 2 valleys and 2 spins, see Eq.~(\ref{eq-Chernbasis})). In the (first) chiral-flat limit, the Hamiltonian has a valley-spin U(4)$\times$U(4) symmetry \cite{bultinck_ground_2020, ourpaper3}, and we find that provided a weak condition (\ref{eq-Mq=0}) called the \emph{flat metric condition} is not largely violated, the exact ground states at each integer filling $\nu$ ($|\nu|\le 4$) relative to the charge neutral point (CNP) are given by the fully occupying any $\nu+4$ Chern bands (of either Chern number $\pm1$), leading to exactly degenerate Chern insulator ground states with total Chern number $\nu_C=4-|\nu|,2-|\nu|,\cdots,|\nu|-4$. This degeneracy between different Chern number states is lifted when going away from the chiral-flat limit. When reduced to the nonchiral-flat limit, the Hamiltonian still has a valley-spin U(4) rotational symmetry \cite{bultinck_ground_2020, ourpaper3}, and we find that the lowest possible Chern number is favored: all the even fillings $\nu=0,\pm2$ have Chern number $0$ insulator ground states which are exactly solvable, while all the odd fillings $\nu=\pm1,\pm3$ have Chern number $\pm1$ insulator ground states by perturbation analysis. All of these ground states in the nonchiral-flat limit are U(4) ferromagnetic (FM). If the kinetic energy (nonflatness) is further turned on, the symmetry of the system will be broken into U(2)$\times$U(2) \cite{bultinck_ground_2020, ourpaper3}. In this case, we find the U(4) FM insulator states with Chern number $\nu_C=0$ at even fillings $\nu$ (e.g., the $\nu=0,\pm2$ ground states which have $\nu_C=0$) are fully intervalley coherent (with a maximal in-plane polarization in the valley Bloch sphere, Fig. \ref{fig:valley}(e)), the states with Chern number $0<|\nu_C|<4-|\nu|$ at integer fillings $\nu$ (e.g., the $\nu=\pm1$ ground states with $\nu_C=\pm1$) are partially intervalley coherent, while the insulator states with Chern number $|\nu_C|=4-|\nu|$ (e.g., the $\nu=\pm3$ ground states with $\nu_C=\pm1$) are valley polarized (with maximal $z$ direction polarization in the valley Bloch sphere). At even fillings $\nu=0,\pm2$, our results agree with the energy argument and the K-IVC states proposed at $\nu=0,\pm2$ in Refs.~\cite{bultinck_ground_2020}. The ground state valley coherence/polarization we found at all integer fillings also agrees with that found by the Hartree-Fock calculation in Ref.~\cite{zhang_HF_2020}. However, the ground state Chern numbers are not discussed in Ref.~\cite{zhang_HF_2020}.  At $\nu=\pm1,\pm3$, the energy difference between the Chern number $\pm1$ ground states (valley polarized at $\nu=\pm3$, partially intervalley coherent at $\nu=\pm1$) and the corresponding Chern number $\pm1$ fully intervalley coherent states is very small (of order $0.005$meV per electron), making the latter still a competitive state. 

The other perturbation away from the (first) chiral-flat limit is the (first) chiral-nonflat limit with a nonzero kinetic energy, which also has a valley-spin U(4) rotational symmetry (different from the nonchiral-flat U(4)) \cite{bultinck_ground_2020, ourpaper3}. In this case, we find all the different Chern number states at a fixed integer filling $\nu$ are degenerate up to second order perturbations. Without symmetry protections, their degeneracy will be lifted by higher order perturbations, and we show numerically in a different paper \cite{ourpaper6} that the lowest Chern number (absolute value) is favored. Besides, we find the ground state in the chiral-nonflat limit favors filling only one of the two Chern bands in each valley and spin, in agreement with the analysis in Ref.~\cite{bultinck_ground_2020}. As a result, the occupied Chern $+1$ basis and the occupied Chern $-1$ basis tend to have distinct spin-valley polarizations, thus are antiferrmomagnetic (AFM) between each other from the perspective of the chiral-nonflat U(4) group. When the nonchiral perturbation is further turned on and the symmetry is reduced to U(2)$\times$U(2), we find the same valley coherence/polarization as that from perturbing the nonchiral-flat insulating states, namely, fully intervalley coherent states are favored when the Chern number $\nu_C=0$, partially intervalley coherent states are favored when $0<|\nu_C|<4-|\nu|$, and valley polarized states are favored when $|\nu_C|=4-|\nu|$. 

In particular, in the nonchiral-nonflat case, the perturbative ground states we obtained by adding nonchiral perturbation to the chiral-nonflat limit are the same as those we obtained by adding nonflat perturbation to the nonchiral-flat limit, which indicate the uniqueness of the ground state when both nonchiral interaction and kinetic energy are small. Note that the spin-valley U(4) polarizations (magnetizations) of all the insulating states we found in this paper have an orbital magnetism origin, due to the absence of spin-orbital coupling in TBG.

Crucially, all of our statements about the ground states at nonzero integer fillings are checked by exact diagonalization techniques in Ref.~\cite{ourpaper6} (fully verified for $\nu=-3$, and showing agreement within limited Hilbert spaces/parameters for $\nu=-2$ and $-1$).

Furthermore, by a free energy estimation, we predict that for $\nu=\pm1,\pm2$, an interaction-driven first-order phase transition from the lowest Chern number $\nu_C=\text{sgn}(\nu B)(2-|\nu|)$ intervalley coherent state to the highest Chern number $\nu_C=\text{sgn}(\nu B)(4-|\nu|)$ valley polarized state happens at a finite out-of-plane magnetic field $|B|=B_\nu^*$ (with $B_\nu^*$ of order $0.5$T, see Fig. \ref{fig:PD}), where $\text{sgn}(x)$ is the sign function. At filling $\nu=\pm 3$, the only possible Chern number is $\nu_C=\text{sgn}(\nu B)(4-|\nu|)=\text{sgn}(\nu B)$ (valley polarized). At filling $\nu=0$, such a transition is absent. 
Remarkably, this is supported by the experimental findings by scanning tunneling spectroscopy \cite{nuckolls_chern_2020,choi2020tracing} as well as observed in transport experiments \cite{saito2020,das2020symmetry, wu_chern_2020}, where correlated gaps of Chern number $\nu_C=\text{sgn}(\nu B)(4-|\nu|)$ emerge above a certain magnetic field for all integer fillings $\nu\neq 0$. Besides, hysteresis loop has been observed by transport experiment in magnetic field $B>0$ near $\nu=\pm1$ \cite{lu2019superconductors,das2020symmetry}, after which the system enters a Chern number $+3$ phase. Moreover, Pomeranchuk effect in magnetic fields is also observed near $\nu=\pm1$ in transport experiments \cite{saito2020isospin,rozen2020entropic}. These evidences strongly support our prediction of the in-field first-order phase transitions at $\nu=\pm1,\pm2$.

Lastly, we study the stabilizer code limit we identified in Ref.~\cite{ourpaper3}, where the TBG Hamiltonian becomes the sum of mutually commuting terms. We solve exactly the entire spectrum in this limit by showing it is equivalent to an extended Hubbard model. This limit, although not satisfied by realistic parameters, gives a heuristic understanding of the TBG spectra as Hubbard subbands, as revealed by the cascade spectral features observed by scanning tunneling spectroscopy \cite{xie2019spectroscopic,wong_cascade_2020,nuckolls_chern_2020,choi2020tracing}.

\subsection{Paper organization}

The paper is organized as follows. In Sec.~\ref{sec:H}, we briefly review the projected TBG Hamiltonian in the lowest 8 flat bands derived in Ref.~\cite{ourpaper3}. We first study the exact insulating ground states at integer fillings $\nu$ in the (first) chiral-flat limit in Sec.~\ref{sec:chernchiralflatband}. We then derive either the exact or perturbative insulating ground states/low energy states away from the (first) chiral-flat limit (in the nonchiral-flat limit, chiral-nonflat limit and nonchiral-nonflat case) in Secs. \ref{sec:gsnonchiralflat}-\ref{sec:nonchiralnonflat}. Sec.~\ref{sec:phasetransitionbfield} is then devoted to examine the Chern insulator phase transitions in magnetic fields near the integer fillings. In Sec.~\ref{sec:stabilizercode}, we  exactly solve all the eigenstates in the stabilizer code limit of TBG. The discussion and conclusion are then given in Sec.~\ref{sec:discussion}.

\section{The Positive Semi-definite Projected TBG Hamiltonian}\label{sec:H}

In Ref.~\cite{ourpaper3}, we derived the projected Hamiltonian in the 8 (2 per spin-valley) moir\'e flat bands of the magic angle TBG under Coulomb interactions. In this paper, we study the ground states of such a projected TBG Hamiltonian, which can be written into kinetic and interaction parts as $H=H_0+H_I$. The kinetic term is (App.~\ref{app:notationsonebody})
\begin{equation}\label{eq-pH0}
H_0=\sum_{n=\pm 1}\sum_{\mathbf{k} \eta s} \epsilon_{n,\eta}(\mathbf{k}) c^\dag_{\mathbf{k},n,\eta, s} c_{\mathbf{k},n,\eta, s}\ ,
\end{equation}
where $\eta=\pm$ denote graphene valleys $K$ and $K'$, $s=\uparrow,\downarrow$ denote the electron spin, and $n=\pm1$ denote the conduction/valence flat bands in each spin-valley flavor. $c^\dag_{\mathbf{k},n,\eta, s}$ is the electron creation operator of energy band $n$, and the origin of $\mathbf{k}$ is chosen at $\Gamma$ point of the moir\'e Brillouin zone (MBZ). The single-particle energy $\epsilon_{n,\eta}(\mathbf{k})$ depends on the twist angle $\theta$ and two interlayer hopping parameters \cite{bistritzer_moire_2011,tarnopolsky_origin_2019,ourpaper1,ourpaper2} (definition given in Eq. \ref{seq-Tj}):
\begin{equation}\label{eq-Tj}
\begin{split}
&w_0\ge0:\ \  \text{AA hopping},\\ &w_1\ge0:\ \ \text{AB/BA hopping},
\end{split}
\end{equation}
The lowest 8 moir\'e bands (2 per spin-valley) become extremely flat near the magic angle manifold $w_0\le w_1\approx v_Fk_\theta/\sqrt{3}$ \cite{bistritzer_moire_2011,tarnopolsky_origin_2019,ourpaper1}, where $v_F$ is the monolayer graphene Fermi velocity, and $k_\theta=8\pi\sin(\theta/2) /3a_0$ with $a_0=0.246$nm being the graphene lattice constant. The realistic TBG generically have $w_0< w_1$ due to lattice corrugations and relaxations \cite{Uchida_corrugation,Wijk_corrugation,dai_corrugation, jain_corrugation}, while the isotropic case $w_0=w_1$ correspond to TBG without relaxation or corrugation \cite{bistritzer_moire_2011}. In this paper, we shall assume $w_1=110$meV is fixed, and $w_0$ is tunable.

The projected Coulomb interaction term $H_I$ within the lowest 8 moir\'e bands then takes the form (see Ref.~\cite{ourpaper3} for details, see App.~\ref{app:interactinghamiltonian} for a brief review)
\begin{equation}\label{eq-pHI}
H_I=\frac{1}{2\Omega_{\text{tot}}}\sum_{\mathbf{q}\in\text{MBZ}}\sum_{\mathbf{G}\in\mathcal{Q}_0} O_{\mathbf{-q,-G}} O_{\mathbf{q,G}}\ ,
\end{equation}
where $\Omega_{\text{tot}}$ is the total area of TBG, $\GG$ belongs to the triangular moir\'e reciprocal lattice $\mathcal{Q}_0$ of TBG, and
\begin{equation}\label{eq-OqG}
\begin{split}
O_{\mathbf{q,G}} =&\sum_{\mathbf{k}\eta s}\sum_{m,n=\pm1} \sqrt{V(\mathbf{q+G})} M_{m,n}^{\left(\eta\right)} \left(\mathbf{k},\mathbf{q}+\mathbf{G}\right) \\
&\quad \times \left(\rho_{\mathbf{k,q},m,n,s}^\eta-\frac{1}{2}\delta_{\mathbf{q,0}}\delta_{m,n}\right)\ .
\end{split}
\end{equation}
Here we have defined the Coulomb potential $V(\mathbf{q})=2\pi e^2 \tanh (\xi |\mathbf{q}|/2)/ \epsilon |\mathbf{q}|$ for an effective dielectric constant $\epsilon\ (\sim 6)$, and screening length $\xi$ ($\sim 10$nm) from the top and bottom gates. The coefficients $M_{m,n}^{\left(\eta\right)} \left(\mathbf{k},\mathbf{q}+\mathbf{G}\right)=\sum_{\alpha,\mathbf{Q}\in\mathcal{Q}_\pm}u^*_{\mathbf{Q-G},\alpha m\eta}(\mathbf{k+q})u_{\mathbf{Q},\alpha n\eta}(\mathbf{k})$ are called the \emph{form factors (overlaps)}, where $u_{\mathbf{Q},\alpha n\eta}(\mathbf{k})$ is the wavefunction of band $n$ at valley $\eta$. $\rho_{\mathbf{k,q},m,n,s}^\eta=c^\dag_{\mathbf{k+q},m,\eta, s}c_{\mathbf{k},n,\eta, s}$ is the density operator. In particular, since $O_{-\qq,-\GG}=O_{\qq,\GG}^\dag$, the interaction $H_I$ in Eq.~(\ref{eq-pHI}) is a Kang-Vafek type \cite{kang_strong_2019} positive semidefinite Hamiltonian (PSDH).

The TBG Hamiltonian has a rotational symmetry $C_{2z}$ and a time-reversal symmetry $T$, and a U(2)$\times$U(2) symmetry given by spin-charge rotations of each valley. Besides, there is a particle-hole (PH) symmetry $P$ satisfying $\{H_0,P\}=[H_I,P]=0$. The combined symmetry $C_{2z}P$ ensures $\epsilon_{n,\eta}(\kk)=-\epsilon_{-n,-\eta}(\kk)$. The full Hamiltonian $H$ also has a many-body charge conjugation symmetry $\mathcal{P}_c$, which ensures that all phenomena are PH symmetric about the CNP (see definition in App.~\ref{app:interactinghamiltonian} and proof in Ref.~\cite{ourpaper3}). 

Furthermore, in the first chiral limit with AA stacking hopping $w_0=0$, there is an additional chiral symmetry $C$ satisfying $\{H_0,C\}=[H_I,C]=0$ (see definition in App.~\ref{app:interactinghamiltonian}, and proof in Ref.~\cite{ourpaper2,ourpaper3} where $C$ is denoted as the first chiral symmetry, in contrast to a second chiral symmetry defined by $w_1=0$ therein). Since throughout this paper we will only be considering the first chiral limit, hereafter we will simply denote it as the ``chiral limit" unless there is an ambiguity.

Hereafter we will use $\zeta^a$, $\tau^a$, $s^a$ to denote the identity matrix ($a=0$) and Pauli matrices ($a=x,y,z$) in the flat band $n=\pm1$, graphene valley $\eta=\pm$ and spin $s=\uparrow,\downarrow$ spaces, respectively. Throughout this paper, we adopt the gauge fixing of the band basis that $C_{2z}c^\dag_{\mathbf{k},n,\eta, s}C_{2z}^{-1}=Tc^\dag_{\mathbf{k},n,\eta, s}T^{-1}=c^\dag_{-\mathbf{k},n,-\eta, s}$, and $(C_{2z}P)c^\dag_{\mathbf{k},n,\eta, s}(C_{2z}P)^{-1}=-n\eta c^\dag_{\mathbf{k},-n,-\eta, s}$. This fixes the form factors (overlaps) $M_{mn}^{(\eta)}\left(\mathbf{k},\mathbf{q}+\mathbf{G}\right)$ into
\begin{equation}\label{eq-Mmn}
M_{mn}^{(\eta)}\left(\mathbf{k},\mathbf{q}+\mathbf{G}\right)=\sum_{j=0}^3 (M_j)_{m,\eta;n,\eta} \alpha_j(\mathbf{k,q+G})\ ,
\end{equation}
where $\alpha_j(\mathbf{k,q+G})$ are real scalar functions, and we have defined $M_0=\zeta^0\tau^0$, $M_1=\zeta^x\tau^z$, $M_2=i\zeta^y\tau^0$, and $M_3=\zeta^z\tau^z$. In particular, for $\mathbf{q=0}$, one can prove that $\alpha_0(\mathbf{k,G})=\alpha_0(\mathbf{-k,G})$, and $\alpha_j(\mathbf{k,G})=-\alpha_j(\mathbf{-k,G})$ for $j=1,2,3$ (see proof in Ref.~\cite{ourpaper3} and brief review in App.~\ref{app:interactinghamiltonian}). Besides, we assume the energy band basis is further fixed by the continuous condition Eq.~(\ref{seq:c-continuous}) (see also \cite{ourpaper3}).

To study the ground states of TBG, it is useful to note that $H_I$ in Eq.~(\ref{eq-pHI}) can be rewritten as (App.~\ref{app:chempotentialshift})
\begin{equation}\label{eq-HIrewritten}
\begin{split}
&H_I=\frac{1}{2\Omega_{\text{tot}}}\sum_{\mathbf{G}} \Big[ 2N_MA_{\mathbf{-G}}O_{\mathbf{0,G}} -N_M^2 A_{\mathbf{-G}}A_{\mathbf{G}} + \\
& \sum_{\mathbf{q}}(O_{\mathbf{-q,-G}}-N_MA_{-\mathbf{G}}\delta_{\mathbf{q,0}}) (O_{\mathbf{q,G}}-N_MA_{\mathbf{G}}\delta_{\mathbf{q,0}})\Big]
\end{split}
\end{equation}
where $N_M$ is the total number of moir\'e unit cells, and $A_{\mathbf{G}}$ can be any $\mathbf{G}$ dependent coefficient. Since $O_{\mathbf{-q,-G}}=O_{\mathbf{q,G}}^\dag$, the last term on the right-hand-side of Eq.~(\ref{eq-HIrewritten}) is always nonnegative. In particular, if for $\mathbf{q=0}$ one has the \cite{ourpaper1}
\begin{equation}\label{eq-Mq=0}
\text{Flat Metric Condition: }\; M_{mn}^{(\eta)}\left(\mathbf{k},\mathbf{G}\right)=\xi(\mathbf{G})\delta_{m,n}
\end{equation}
being independent of $\mathbf{k}$, where $\xi(\mathbf{G})$ is some function of $\mathbf{G}$, one would have $O_{\mathbf{0,G}}=\sqrt{V(\mathbf{G})}\xi(\mathbf{G})\nu N_M$, where $-4\le \nu\le 4$ is the number of electrons per moir\'e unit cell relative to the CNP. Therefore, for a fixed filling $\nu$, if either the flat metric condition (\ref{eq-Mq=0}) holds \emph{or} if $A_{\mathbf{G}}=0$, the first two terms in Eq.~(\ref{eq-HIrewritten}) will be constant, and thus a state annihilated by $O_{\mathbf{q,G}}-N_MA_{\mathbf{G}}\delta_{\mathbf{q,0}}$ for all $\mathbf{q}$ and $\mathbf{G}$ will necessarily be a ground state of $H_I$. Based on this idea, we will identify the ground states of strongly interacting TBG at integer fillings $\nu$. 

We note that as shown in Ref.~\cite{ourpaper1}, the flat metric condition in Eq.~(\ref{eq-Mq=0}) is always satisfied for $\GG=\mathbf{0}$, and is approximately satisfied by the TBG single-particle Hamiltonian for $|\GG|>\sqrt{3}k_\theta$ due to the fast exponential decay of the form factors with respect to $|\GG|$. Therefore, to a good approximation, the only $\GG$ which violate the flat metric condition Eq.~(\ref{eq-Mq=0}) are the six smallest nonzero moir\'e reciprocal lattice sites with $|\GG|=\sqrt{3}k_\theta$.

\section{Chern insulators in the (first) chiral-flat limit}\label{sec:chernchiralflatband}

We first study the (first) chiral-flat limit, for which the projected kinetic term is $H_0=0$, and $w_0=0<w_1$. The Hamiltonian thus has the (first) chiral symmetry $C$ (which ensures $\epsilon_{n,\eta}(\mathbf{k})=-\epsilon_{-n,\eta}(\mathbf{k})$). Here we choose the gauge fixing $Cc^\dag_{\mathbf{k},n,\eta, s}C^{-1}=in\eta c^\dag_{\mathbf{k},-n,\eta, s}$ for $n=\pm1$ (see Ref.~\cite{ourpaper3}, see also App.~\ref{app:ihamiltoniangaugefixing}).  
In total, due to the $C$ and $C_{2z}P$ symmetries, the projected Hamiltonian $H=H_I$ in this limit has a U(4)$\times$U(4) symmetry in the band-valley-spin space, which has 32  generators $S^{ab}_{\pm}=\sum_{\mathbf{k}} (s_\pm^{ab})_{m,\eta,s;n,\eta',s'}c_{\mathbf{k},m,\eta,s}^\dag c_{\mathbf{k},n,\eta',s'}$ ($a,b=0,x,y,z$), where $s_\pm^{ab}=\left(\zeta^0\pm\zeta^y\right)\tau^a s^b/2$ (see Ref.~\cite{ourpaper3}, see also the brief review in App.~\ref{app:U(4)U(4)-cf}). 

It is convenient to transform into another basis which we call the Chern (band) basis defined in Refs.~\cite{ourpaper2,ourpaper3} (see also App.~\ref{app:chernbasis}):
\begin{equation}\label{eq-Chernbasis}
d^\dag_{\mathbf{k},e_Y,\eta, s}=\frac{c^\dag_{\mathbf{k},1,\eta, s}+ie_Y c^\dag_{\mathbf{k},-1,\eta, s}}{\sqrt{2}}\ , \quad (e_Y=\pm1).
\end{equation}
As proved in Refs.~\cite{ourpaper2,ourpaper3}, for fixed $\eta$ and $s$, $d^\dag_{\mathbf{k},e_Y,\eta, s}$ form the basis of a Chern number $e_Y$ band. We note that the Chern basis (\ref{eq-Chernbasis}) is adiabatically equivalent to the Chern basis defined in the first chiral limit in Ref.~\cite{bultinck_ground_2020}, while we show in Ref.~\cite{ourpaper2} that this Chern basis can still be defined by Eq.~(\ref{eq-Chernbasis}) away from the first chiral limit, which is also discussed in Ref.~\cite{hejazi2020hybrid}. 

With the chiral symmetry $C$, one can show that $\alpha_1(\mathbf{k,q+G})=\alpha_3(\mathbf{k,q+G})=0$ in Eq.~(\ref{eq-Mmn}). Therefore, under basis (\ref{eq-Chernbasis}), the operator $O_{\mathbf{q,G}}$ in Eq.~(\ref{eq-OqG}) reduces to
\begin{equation}\label{eq-OqG0}
\begin{split}
&O_{\mathbf{q,G}} =O_{\mathbf{q,G}}^0=\sum_{\mathbf{k}\eta s}\sum_{e_Y=\pm1} \sqrt{V(\mathbf{q+G})}  \\
& \times M_{e_Y}(\kk,\qq+\GG) \left(d^\dag_{\mathbf{k+q},e_Y,\eta,s}d_{\mathbf{k},e_Y,\eta,s}-\frac{1}{2}\delta_{\mathbf{q,0}}\right),
\end{split}
\end{equation}
where we have defined $M_{e_Y}(\kk,\qq+\GG)=\alpha_0 \left(\mathbf{k},\mathbf{q}+\mathbf{G}\right)+ie_Y\alpha_2 \left(\mathbf{k},\mathbf{q}+\mathbf{G}\right)$. 

At integer filling $\nu$ relative to the CNP ($-4\le\nu\le 4$), we define a spin-valley polarized Fock state
\begin{equation}\label{eq-Psi-nu-nup-num}
|\Psi_{\nu}^{\nu_+,\nu_-}\rangle =\prod_{\mathbf{k}} \prod_{j_1=1}^{\nu_+}d^\dag_{\mathbf{k},+1,\eta_{j_1},s_{j_1}} \prod_{j_2=1}^{\nu_-}d^\dag_{\mathbf{k},-1,\eta_{j_2}',s_{j_2}'}|0\rangle,
\end{equation}
where $\nu_+,\nu_-\in[0,4]$ are two integers satisfying $\nu_++\nu_-=\nu+4$, $\mathbf{k}$ runs over the MBZ, $|0\rangle$ is the zero electron state of flat bands, and $\{\eta_{j_1},s_{j_1}\}$ and $\{\eta_{j_2}',s_{j_2}'\}$ can be chosen arbitrarily. This is a state with $N_\nu=(\nu_++\nu_-)N_M$ electrons fully occupying $\nu_+$ Chern number $+1$ bands of valley and spin indices $\{\eta_{j_1},s_{j_1}\}$ and $\nu_-$ Chern number $-1$ bands of valley and spin indices $\{\eta_{j_2}',s_{j_2}'\}$, while all the other Chern bands are empty. It is straightforward to verify that if we define 
\begin{equation}\label{eq-AG}
A_\mathbf{G}=\frac{\nu }{N_M} \sqrt{V(\mathbf{G})} \sum_\mathbf{k} \alpha_0(\mathbf{k,G}), 
\end{equation}
we have for any $\mathbf{q}$ and $\mathbf{G}$:
\begin{equation}
(O_{\mathbf{q,G}}-N_MA_{\mathbf{G}}\delta_{\mathbf{q,0}})|\Psi_{\nu}^{\nu_+,\nu_-}\rangle=0\ .
\end{equation}
Assume either the flat metric condition (\ref{eq-Mq=0}) is satisfied, or that $\nu=0$ (where the flat metric condition (\ref{eq-Mq=0}) is not needed). By rewriting the Hamiltonian as Eq.~(\ref{eq-HIrewritten}), we see that the first two terms in Eq.~(\ref{eq-HIrewritten}) are constants only depending on $\nu$, and thus the state $|\Psi_{\nu}^{\nu_+,\nu_-}\rangle$ at integer filling $\nu$ with any $\nu_+,\nu_-$ is an exact ground state in the (first) chiral-flat limit. Furthermore, such a ground state with any $\nu,\nu_+,\nu_-$ has gapped charge excitations, as we will show in Refs.~\cite{ourpaper5,ourpaper6}. This is because there is no remaining symmetry protecting a gapless electron spectrum: the electrons in valley $\eta$ will be gapless if valley $\eta$ (for a fixed spin $s$) is half-filled and the spinless $C_{2z}T$ symmetry is preserved, due to the $C_{2z}T$ protected fragile topology \cite{po_origin_2018,song_all_2019,po_faithful_2019,ahn_failure_2019,lian2020,po_fragile_2018,cano_fragile_2018,Slager2019WL}. This is not satisfied by state $|\Psi_{\nu}^{\nu_+,\nu_-}\rangle$, since a half-filled valley $\eta$ (of a given spin $s$) always fully occupies one Chern band, which breaks the $C_{2z}T$ symmetry.

If the flat metric condition (\ref{eq-Mq=0}) is not satisfied, $|\Psi_{\nu}^{\nu_+,\nu_-}\rangle$ in Eq.~(\ref{eq-Psi-nu-nup-num}) is still an eigenstate of $H=H_I$, since $O_{\mathbf{0,G}}$ in the first term of Eq.~(\ref{eq-HIrewritten}) satisfies 
\begin{equation}
O_{\mathbf{0,G}}|\Psi_{\nu}^{\nu_+,\nu_-}\rangle=\nu\sqrt{V(\mathbf{G})}\sum_{\mathbf{k}}\alpha_0(\mathbf{k,G})|\Psi_{\nu}^{\nu_+,\nu_-}\rangle\ . 
\end{equation}
Therefore, $|\Psi_{\nu}^{\nu_+,\nu_-}\rangle$ will remain the ground state at filling $\nu$ unless the flat metric condition (\ref{eq-Mq=0}) is sufficiently violated to bring down the energy of another eigenstate into the lowest. (Notice that in Ref.~\cite{ourpaper5}, the spectrum is derived and gapless due to the goldstone mode; this mode has always energy above the $|\Psi_{\nu}^{\nu_+,\nu_-}\rangle$, even when it is not a ground state.) In particular, for $\nu=0$, $|\Psi_{0}^{\nu_+,\nu_-}\rangle$ is always the ground state without the flat metric condition (\ref{eq-Mq=0}), since $O_{\mathbf{0,G}}|\Psi_{\nu}^{\nu_+,\nu_-}\rangle=0$. We will show in a different paper \cite{ourpaper5,ourpaper6} that $|\Psi_{\nu}^{\nu_+,\nu_-}\rangle$ at all integer fillings $\nu\in[-4,4]$ remains the ground state for realistic parameters in the chiral-flat limit, although the flat metric condition (\ref{eq-Mq=0}) is violated.

By definition (\ref{eq-Psi-nu-nup-num}), the ground state $|\Psi_{\nu}^{\nu_+,\nu_-}\rangle$ carries a Chern number 
\begin{equation}
\nu_C=\nu_+-\nu_-\ , 
\end{equation}
which can take values $\nu_C=4-|\nu|,2-|\nu|,\cdots,|\nu|-4$. These ground states with different Chern numbers $\nu_C$ at filling $\nu$ are exactly degenerate in the chiral-flat limit.

Due to the U(4)$\times$U(4) symmetry in the chiral-flat limit, the ground states fall into irreducible representations (irreps) of U(4)$\times$U(4). 
For instance, different choices of $\{\eta_{j_1},s_{j_1}\}$ and $\{\eta_{j_2}',s_{j_2}'\}$ in Eq.~(\ref{eq-Psi-nu-nup-num}) give different states in the same irrep multiplet. We  label 
the irreps of the U(4) group by their Young tableau as $[\lambda_1,\lambda_2,\cdots]_4$ (abbreviated as $[\{\lambda_i\}]_4$), where $\lambda_i$ ($i\le4$, $\lambda_i \ge \lambda_{i+1}$) is the number of boxes in row $i$ of the Young tableau ($\lambda_i$ will be omitted if $\lambda_i=0$) (see App.\ref{app:U4irrep}). Note that the usual Young tableau notation for U(4) only requires three rows, here for convenience $[\lambda_1,\lambda_2,\lambda_3,\lambda_4]_4$ should be understood as $[\lambda_1-\lambda_4,\lambda_2-\lambda_4,\lambda_3-\lambda_4]_4$. The irreps of U(4)$\times$U(4) are then given by the tensor products of irreps $[\{\lambda_{1,i}\}]_4$ of the first U(4) and irreps $[\{\lambda_{2,i}\}]_4$ of the second U(4), which we denote as $([\{\lambda_{1,i}\}]_4,[\{\lambda_{2,i}\}]_4)$ (see App.\ref{app:u4u4irreps}). In Ref.~\cite{ourpaper3}, we showed that for each $\mathbf{k}$, the creation operators $d^\dag_{\mathbf{k},+1,\eta, s}$ and $d^\dag_{\mathbf{k},-1,\eta, s}$ occupy U(4)$\times$U(4) irreps $([1]_4,[0]_4)$ and $([0]_4,[1]_4)$, respectively, where $[1]_4$ and $[0]_4$ are the fundamental irrep and identity irrep of U(4). The U(4)$\times$U(4) irrep of ground state $|\Psi_{\nu}^{\nu_+,\nu_-}\rangle$ can then be shown to be (see App.~\ref{app:c-f-eigenstates})
\begin{equation}
([N_M^{\nu_+}]_4,[N_M^{\nu_-}]_4)\ , 
\end{equation}
where $[\lambda^p]_4$ ($p\le 4$) is short for the U(4) irrep $[\lambda,\lambda,\cdots]_4$ with $p$ number of $\lambda$. Within each U(4) of Chern bands $e_Y$, the U(4) irrep $[N_M^{\nu_{e_Y}}]_4$ of $|\Psi_{\nu}^{\nu_+,\nu_-}\rangle$ is maximally symmetric,
thus is a U(4) FM with the maximal possible spin-valley polarization at filling $\nu_{e_Y}$ within the Chern basis $e_Y$. The U(4) FM polarizations of the $e_Y=\pm1$ Chern basis are unrelated in the chiral-flat limit (Fig.~\ref{fig:valley}(b)). Thus, the ground state $|\Psi_{\nu}^{\nu_+,\nu_-}\rangle$ is a U(4)$\times$U(4) FM state.

We further derive the Hartree-Fock Hamiltonian of state $|\Psi_{\nu}^{\nu_+,\nu_-}\rangle$ in Eq. (\ref{eq-Psi-nu-nup-num}) under the Chern band basis, which is of the form
\begin{equation}
H_{HF}=\sum_{\kk}\sum_{\eta,s,e_Y} h_{HF,\nu,\nu_C}^{e_Y,\eta,s} (\kk) d^\dagger_{\kk,e_Y,\eta,s} d_{\kk,e_Y,\eta,s}\ .
\end{equation}
The expression of $h_{HF,\nu,\nu_C}^{e_Y,\eta,s} (\kk)$ is given in Eq. (\ref{seq:HF-chiral-flat}) of App.~\ref{app:HF-Hamiltonian}. In particular, the Hartree-Fock band energies are equal to that of a branch of exact charge $\pm1$ excitations we derived in \cite{ourpaper5}.

\section{Ground states in the nonchiral-flat limit}\label{sec:gsnonchiralflat}

We now turn to the nonchiral-flat limit, where the projected kinetic term $H_0=0$, and $w_1\ge w_0>0$. Without chiral symmetry $C$, the Hamiltonian $H=H_I$ only has a global U(4) symmetry  \cite{ourpaper3}, with generators $S^{ab}=\sum_{\mathbf{k}} (s^{ab})_{m,\eta,s;n,\eta',s'}c_{\mathbf{k},m,\eta,s}^\dag c_{\mathbf{k},n,\eta',s'}$, where $s^{ab}= \{\zeta^0\tau^{0,z} s^b,\zeta^y\tau^{x,y} s^b\}$ ($b=0,x,y,z$) (see derivation in Ref.~\cite{ourpaper3}, and App.~\ref{app:U4-nc-f} for brief review). One can still define the Chern band basis (\ref{eq-Chernbasis}), under which the operator $O_{\mathbf{q,G}}$ in Eq.~(\ref{eq-OqG}) decomposes into 
\begin{equation}\label{eq-OqG-nf}
O_{\mathbf{q,G}}=O_{\mathbf{q,G}}^0+O_{\mathbf{q,G}}^1, 
\end{equation}
where $O_{\mathbf{q,G}}^0$ is given in Eq.~(\ref{eq-OqG0}), and
\begin{equation}\label{eq-OqG1}
\begin{split}
O_{\mathbf{q,G}}^1=&\sum_{\mathbf{k}\eta s}\sum_{e_Y=\pm1} \eta \sqrt{V(\qq+\GG)} \\ 
&\times F_{e_Y}(\kk,\qq+\GG)  d^\dag_{\mathbf{k+q},-e_Y,\eta,s}d_{\mathbf{k},e_Y,\eta,s},
\end{split}
\end{equation}
with the coefficient $F_{e_Y}(\kk,\qq+\GG)=\alpha_1 \left(\mathbf{k},\mathbf{q}+\mathbf{G}\right)+ie_Y\alpha_3 \left(\mathbf{k},\mathbf{q}+\mathbf{G}\right)$.

We first show that exact ground states can be obtained for even integer fillings $\nu=0,\pm2$ (the trivial band insulators at $\nu=\pm4$ are not discussed). We define a state with Chern number zero at even filling $\nu$ as:
\begin{equation}\label{eq-Psi-nu}
|\Psi_\nu\rangle=\prod_{\mathbf{k}} \prod_{j=1}^{(\nu+4)/2}d^\dag_{\mathbf{k},+1,\eta_{j},s_{j}} d^\dag_{\mathbf{k},-1,\eta_{j},s_{j}}|0\rangle\ ,
\end{equation}
where $\{\eta_j,s_j\}$ can be chosen arbitrarily. This state has $(\nu+4)/2$ valley-spin flavors $\{\eta_j,s_j\}$ fully occupied, and the rest $(4-\nu)/2$ valley-spin flavors fully empty. By choosing coefficients $A_\mathbf{G}$ as given in Eq.~(\ref{eq-AG}), one can verify that
\begin{equation}
(O_{\mathbf{q,G}}-N_MA_{\mathbf{G}}\delta_{\mathbf{q,0}})|\Psi_{\nu}\rangle=0\ ,
\end{equation}
with $O_{\mathbf{q,G}}$ given in Eq.~(\ref{eq-OqG-nf}). Therefore, if the flat metric condition (\ref{eq-Mq=0}) is satisfied or $\nu=0$, the state $|\Psi_\nu\rangle$ gives an exact ground state in the nonchiral-flat limit.  Since no valley-spin flavor is half-occupied (although $C_{2z}T$ symmetry may persist), we expect state $|\Psi_\nu\rangle$ to be a gapped insulator.

If the flat metric condition (\ref{eq-Mq=0}) is unsatisfied, $|\Psi_\nu\rangle$ is still an eigenstate of $H=H_I$. For $\nu=0$, $|\Psi_0\rangle$ is a ground state with or without the flat metric condition (\ref{eq-Mq=0}). For $\nu=\pm2$, we expect the states $|\Psi_{\pm2}\rangle$ to remain ground states unless the flat metric condition (\ref{eq-Mq=0}) is violated beyond a threshold.

In Ref.~\cite{ourpaper3}, we showed that the Chern band basis $d^\dag_{\mathbf{k},e_Y,\eta,s}$ for each $\mathbf{k}$ and $e_Y$ occupies a fundamental U(4) irrep $[1]_4$ of the nonchiral-flat U(4) symmetry, with the generators represented by
\begin{equation}\label{eq-SabeY}
s^{ab}(e_Y)=\{ \tau^0s^b,\ e_Y\tau^x s^b,\ e_Y\tau^y s^b,\ \tau^zs^b \}.
\end{equation}
Accordingly, the ground state $|\Psi_\nu\rangle$ in Eq.~(\ref{eq-Psi-nu}) occupies a maximally symmetric U(4) irrep (see App.~\ref{app:exactgsnonchiralflat})
\begin{equation}
[(2N_M)^{(\nu+4)/2}]_4\ , 
\end{equation}
which is thus a U(4) FM state with maximal possible U(4) polarization. However, the physical valley polarizations of the $e_Y=\pm1$ Chern band basis differ by a $\pi$ valley rotation about $z$ axis (Fig.~\ref{fig:valley}(c)), as can be seen from their representation matrices in Eq.~(\ref{eq-SabeY}) (which are relatively twisted between $e_Y=\pm1$ by a unitary transformation $\tau_z$), although they occupy the same fundamental irrep of the nonchiral-flat U(4).

Such exact analytical many-body eigenstates, however, do not occur at odd integer fillings or states carrying a nonzero Chern number at even integer fillings. Therefore, we consider all integer fillings $\nu$ at small $w_0>0$, where we can treat $O_{\mathbf{q,G}}^1$ in Eq.~(\ref{eq-OqG-nf}) as perturbation to the (first) chiral-flat limit. We note that such a perturbation analysis becomes exact for zero Chern number states at even fillings, leading to the exact ground states in Eq.~(\ref{eq-Psi-nu}). 

This perturbation $O_{\mathbf{q,G}}^1$ favors as many fully occupied or fully empty valley-spin flavors $\{\eta,s\}$ as possible (see App.~\ref{app:perturbation-energy-chern} for details), as also showed by Ref.~\cite{bultinck_ground_2020}, since it gives zero when acting on a fully occupied (empty) valley-spin flavor $\{\eta,s\}$ and lowers the total interaction energy $H_I$. 
Thus, it selects the following subset of states in the chiral-flat limit multiplet $|\Psi_{\nu}^{\nu_+,\nu_-}\rangle$ ($\nu_++\nu_-=\nu+4$) as the lowest states of Chern number $\nu_C=\nu_+-\nu_-$:
\begin{equation}\label{eq-Psi-nu-nuC}
|\Psi_{\nu,\nu_C}\rangle =\prod_{\mathbf{k}} \prod_{j=1}^{\nu_+}d^\dag_{\mathbf{k},+1,\eta_{j},s_{j}} \prod_{j=1}^{\nu_-}d^\dag_{\mathbf{k},-1,\eta_{j},s_{j}}|0\rangle\ ,
\end{equation}
which fully occupies valley-spin flavors $\{\eta_j,s_j\}$ with $1\le j\le \text{min}(\nu_+,\nu_-)$. The nonchiral-flat U(4) irrep of state $|\Psi_{\nu,\nu_C}\rangle$ is (see App.~\ref{app:perturbation-energy-chern})
\begin{equation}
[(2N_M)^{(\nu-|\nu_C|+4)/2},N_M^{|\nu_C|}]_4\ , 
\end{equation}
which can be understood as a nonchiral-flat U(4) FM with the maximal possible U(4) polarization for fixed $\nu$ and $\nu_C$. The $e_Y=\pm1$ subspaces of Chern basis thus have a nonchiral-flat U(4) FM coupling between them. However, since the representation matrices of the Chern basis $e_Y=\pm1$ differ by a unitary transformation $\tau_z$, the physical valley polarizations of Chern basis $e_Y=\pm1$ differ by a $\pi$ valley rotation about the $z$-axis (Fig.~\ref{fig:valley}(c)). 
The perturbation energy of the state $|\Psi_{\nu,\nu_C}\rangle$ up to order $w_0^2$ is (see App.~\ref{app:perturbation-energy-chern})
\begin{equation}\label{eq-E1-nc}
E^{(nc)}_{\nu,\nu_C}=|\nu_C|N_M(U_1-\nu^2U_2)\ ,
\end{equation}
where we have defined 
\begin{equation}\label{eq-U1}
U_1=\frac{1}{2N_M\Omega_{\text{tot}}}\sum_{\kk,\qq,\GG} V(\qq+\GG)|F_{+1}(\kk,\qq+\GG)|^2\ ,
\end{equation}
and $U_2$ is from the second order perturbation of the nonchiral interaction term $O^1_{-\qq,-\GG}O^0_{\qq,\GG}$. $U_2$ is defined in Eq.~(\ref{seq:U2-def}) of App.~\ref{app:perturbation-energy-chern} and has a more complicated expression. Both $U_1$ and $U_2$ are proportional to $w_0^2$. In particular, one has $U_2=0$ if the FMC in Eq.~(\ref{eq-Mq=0}) holds. 
We note that the energy $U_1$ here is equivalent to the coupling $\lambda$ in Ref.~\cite{bultinck_ground_2020}. For $w_0\approx 0.8w_1$, our numerical calculation shows $U_1\approx 1.2$meV (which does not depend on whether FMC holds), and $U_2\approx0.10$meV without the FMC (see Fig. \ref{fig:U1U0}(b)). In particular, our numerical calculation shows that $U_1-\nu^2U_2>0$ for any $0\le w_0/w_1\le1$ and $|\nu|\le3$ (Fig. \ref{fig:U1U0}(b)). Therefore, the states with the smallest Chern number $|\nu_C|$ are prefered as the ground states. Note that for even fillings $\nu=0,\pm2$, the Chern number $\nu_C=0$ state $|\Psi_{\nu,0}\rangle$ is simply the exact ground state $|\Psi_{\nu}\rangle$ in Eq.~(\ref{eq-Psi-nu}). For odd fillings $\nu=\pm1,\pm3$, the Chern number $\nu_C=\pm1$ states $\Psi_{\nu,\pm1}$ give the perturbative ground states. 

For the exact nonchiral-flat ground states $|\Psi_{\nu}\rangle$ in Eq. (\ref{eq-Psi-nu}), we can calculate the Hartree-Fock Hamiltonian of state, which takes the form in the energy band basis
\begin{equation}
H_{HF}=\sum_{\kk}\sum_{\eta,s,m,n} \left[h_{HF,\nu}^{\eta,s} (\kk)\right]_{mn} c^\dagger_{\kk,m,\eta,s} c_{\kk,n,\eta,s}\ .
\end{equation}
The expression of $h_{HF,\nu}^{\eta,s}(\kk)$ is given in Eq. (\ref{seq:HF-nonchiral-flat}) of App.~\ref{app:HF-Hamiltonian}. The Hartree-Fock band energies reflect that of a branch of exact charge $\pm1$ excitations we derived in \cite{ourpaper5}.

\section{The (first) chiral-nonflat limit}\label{sec:chiralnonflat}

We now study the (first) chiral-nonflat limit, where the kinetic term $H_0\neq0$, and the U(4)$\times$U(4) symmetry in the (first) chiral-flat limit is broken down to a U(4) symmetry (different from the nonchiral-flat U(4)). The U(4) generators are $\widetilde{S}^{ab}=\sum_{\mathbf{k}} (\tilde{s}^{ab})_{m,\eta,s;n,\eta',s'}c_{\mathbf{k},m,\eta,s}^\dag c_{\mathbf{k},n,\eta',s'}$, where $\tilde{s}^{ab}=\zeta^0\tau^a s^b$ ($a,b=0,x,y,z$) (see derivation in Ref.~\cite{ourpaper3} and brief review in App.~\ref{app:U4-c-nf}). To see how $H_0$ perturbs the chiral-flat limit ground states $|\Psi_{\nu}^{\nu_+,\nu_-}\rangle$ in Eq.~(\ref{eq-Psi-nu-nup-num}), we note that $H_0$ in the chiral limit can be rewritten as
\begin{equation}\label{eq-H0+}
H_0=H_0^+=\sum_{\mathbf{k},e_Y,\eta,s} \epsilon_+(\mathbf{k}) d^\dag_{\mathbf{k},-e_Y,\eta,s} d_{\mathbf{k},e_Y,\eta,s}\ ,
\end{equation}
where $\epsilon_+(\mathbf{k})=\epsilon_{+}(-\mathbf{k})=[\epsilon_{+1,+}(\mathbf{k})- \epsilon_{-1,+}(\mathbf{k})]/2$ due to the chiral symmetry $C$. Since $H_0$ is off-diagonal in the Chern band basis, the first-order perturbation energy of states $|\Psi_{\nu}^{\nu_+,\nu_-}\rangle$ by $H_0$ is zero ($H_0$ has no matrix elements among different Chern insulator states, as this requires exciting every electron in one Chern band to another, which is $N_M$-th order). 
Note that $H_0$ excites $|\Psi_{\nu}^{\nu_+,\nu_-}\rangle$ into neutral excitations in all the half-filled valley-spin flavors $\{\eta,s\}$, while gives zero when acting on fully filled (empty) valley-spin flavors. Therefore, the non-positive second order perturbation energy due to $H_0$ favors as many half-filled valley-spin flavors $\{\eta,s\}$ as possible (see App.~\ref{app:kineticpert-u4u4} for details), as also shown in Ref.~\cite{bultinck_ground_2020}, which is opposite to the effect of $O^{1}_{\qq,\GG}$ in the nonchiral-flat limit. $H_0$ hence selects the following chiral-nonflat U(4) subset of the previously chiral flat U(4) $\times$ U(4) multiplet $|\Psi_{\nu}^{\nu_+,\nu_-}\rangle$ at filling $\nu=\nu_++\nu_--4$ and Chern number $\nu_C=\nu_+-\nu_-$ as the lowest states:
\begin{equation}\label{eq-Psi-nu-nuC2}
|\widetilde{\Psi}_{\nu,\nu_C}\rangle =\prod_{\mathbf{k}} \prod_{j=1}^{\nu_+}d^\dag_{\mathbf{k},+1,\eta_{j},s_{j}} \prod_{j=5-\nu_-}^{4}d^\dag_{\mathbf{k},-1,\eta_{j},s_{j}}|0\rangle\ ,
\end{equation}
where $\{\eta_j,s_j\}$ are the 4 valley-spin flavors arbitrarily sorted in $j$ ($1\le j\le 4$). This state has $4-|\nu|$ valley-spin flavors half-occupied, and has a second order perturbation energy
\begin{equation}\label{eq-E2-nf}
\widetilde{E}_{\nu,\nu_C}^{(2)}=-(4-|\nu|)N_M J_0\ ,
\end{equation}
Here the energy $J_0=N_M^{-1}\sum_{\ell} \frac{|Y_\ell|^2}{E_{\ell}-E_{0,\nu}}$, where $|Y_\ell|=|\langle \ell,\eta,s,\Psi_{\nu}^{\nu_+,\nu_-}|H_0|\Psi_{\nu}^{\nu_+,\nu_-}\rangle|$ are the amplitudes to neutral excitations $|\ell,e_Y,\eta,s,\Psi_{\nu}^{\nu_+,\nu_-}\rangle$ in a half-filled valley-spin flavor $\{\eta,s\}$ (which is independent of $\eta,s$, see App.~\ref{app:kineticpert-u4u4} for a short review and Ref.~\cite{ourpaper5} for a detailed calculation), $E_{\ell}$ are the unperturbed energies of the excited states $|\ell,e_Y,\eta,s,\Psi_{\nu}^{\nu_+,\nu_-}\rangle$, and $E_{0,\nu}$ is the unperturbed energy of state $|\Psi_{\nu}^{\nu_+,\nu_-}\rangle$ (which only depends on $\nu=\nu_++\nu_-=4$). We note that the energy $J_0$ here in Eq.~(\ref{eq-E2-nf}) is equivalent to the coupling $J$ in Ref.~\cite{bultinck_ground_2020}. Numerically, the coupling $J_0$ is given by $J_0=J_1=J_2$ in Tab. \ref{Tab-J123} at $w_0=0$.

\begin{figure}[tbp]
\begin{centering}
\includegraphics[width=\linewidth]{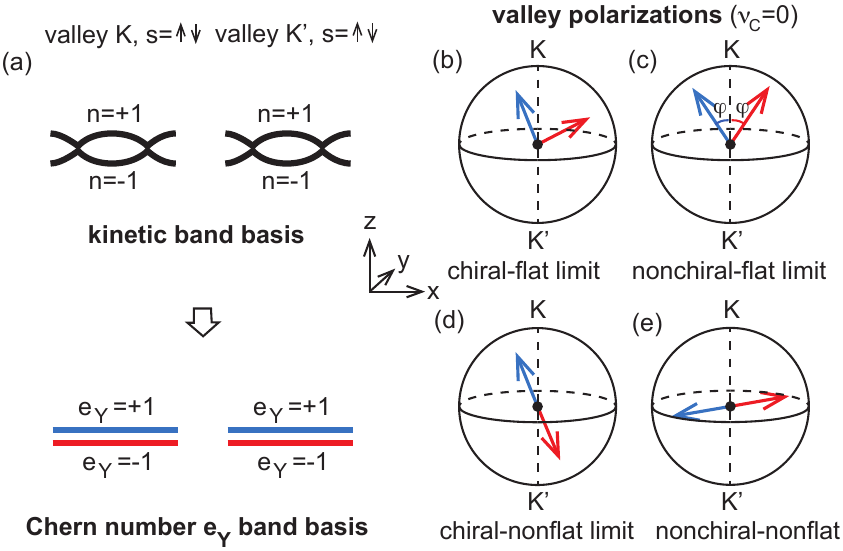}
\end{centering}
\caption{(a) Illustration of the $n=\pm1$ kinetic band basis and the Chern number $e_Y=\pm1$ Chern band basis. (b)-(e) The preferred valley polarization for a Chern number $\nu_C=0$ state in the (b) (first) chiral-flat limit, which has a U(4) FM within each Chern basis $e_Y$ space, while the polarization of $e_Y=\pm1$ subspaces are unrelated; (c) nonchiral-flat limit, where electrons occupying Chern basis $e_Y=\pm1$ have valley polarizations differ by a $\pi$ rotation about the $z$-axis; (d) (first) chiral-nonflat limit, where electrons in the Chern basis $e_Y=\pm1$ prefer opposite polarizations;  (e) nonchiral-nonflat case, where electrons in the $e_Y=\pm1$ Chern basis polarize in the valley $x$-$y$ plane and oppositely.}
\label{fig:valley}
\end{figure}

Since $\widetilde{E}_{\nu,\nu_C}^{(2)}$ in Eq.~(\ref{eq-E2-nf}) is independent of $\nu_C$, the chiral-nonflat U(4) multiplet states $|\widetilde{\Psi}_{\nu,\nu_C}\rangle$ (which are subsets of the chiral-flat multiplets $|\Psi_{\nu}^{\nu_+,\nu_-}\rangle$ in Eq.~(\ref{eq-Psi-nu-nup-num})) for a fixed $\nu$ with different Chern numbers $\nu_C=4-|\nu|,2-|\nu|,\cdots,|\nu|-4$ are degenerate up to the second order perturbation of $H_0$. We expect this degeneracy between different Chern number states to be broken by the 4-th order and higher order perturbations, since there is no symmetry protecting their degeneracy. Such higher order perturbations are difficult to be done analytically. However, as we will show numerically in a separate paper \cite{ourpaper6}, the states with the lowest possible Chern number $|\nu_C|$ wins and becomes the ground state in the chiral-nonflat limit. We will take this conclusion here, and leave the numerical verification to Ref. \cite{ourpaper6}.

In the state $|\widetilde{\Psi}_{\nu,\nu_C}\rangle$ in Eq.~(\ref{eq-Psi-nu-nuC2}), electrons of Chern basis $e_Y=\pm1$ tend to have distinct chiral-nonflat U(4) spin-valley polarizations (Fig.~\ref{fig:valley}(d)). The chiral-nonflat U(4) irrep of $|\widetilde{\Psi}_{\nu,\nu_C}\rangle$ is close to $[N_M^{4-|\nu|}]_4$, i.e., only differs from it by a few Young tableau boxes (the analogue for a SU(2) spin system would be an irrep with a total spin close but not equal to the maximal ferromagnetic value, see a discussion in App.~\ref{app:kineticpert-u4u4}). Therefore, one could view the state as having a chiral-nonflat U(4) AFM coupling between the two $e_Y=\pm1$ subspaces of the Chern basis.

\section{The nonchiral-nonflat case}\label{sec:nonchiralnonflat}

In realistic systems, the magic angle TBG has $w_0\approx 0.8 w_1$ \cite{Uchida_corrugation,Wijk_corrugation,dai_corrugation, jain_corrugation}, and the ratio between the energy scales of $H_0$ and $H_I$ is $\lesssim0.1$ \cite{xie2019spectroscopic,wong_cascade_2020}. If we view both $w_0$ and $H_0$ as perturbations to the first chiral-flat limit, our earlier analysis has shown that the nonchiral interaction due to $w_0$ and the kinetic term $H_0$ perturb the ground state energies at the first order (Eq.~(\ref{eq-E1-nc})) and the second order (Eq.~(\ref{eq-E2-nf})), respectively. Here we will first derive the nonchiral-nonflat ground states by perturbing the nonchiral-flat ground states in Eqs. (\ref{eq-Psi-nu}) and (\ref{eq-Psi-nu-nuC}) with the kinetic term $H_0$. We will then show that the same nonchiral-nonflat ground states can be obtained by perturbing the chiral-nonflat ground states in Eq. (\ref{eq-Psi-nu-nuC2}) with the nonchiral terms.

\subsection{Kinetic perturbation to the nonchiral-flat limit}

In this subsection,  we regard the kinetic term $H_0$ as a perturbation to the nonchiral-flat limit. In the nonchiral case without chiral symmetry $C$, we can rewrite $H_0$ as $H_0=H_0^++H_0^-$, where $H_0^+$ is given in Eq.~(\ref{eq-H0+}), and
\begin{equation}\label{eq-H0m}
H_0^-=\sum_{\mathbf{k},e_Y,\eta,s} \eta \epsilon_-(\mathbf{k}) d^\dag_{\mathbf{k},e_Y,\eta,s} d_{\mathbf{k},e_Y,\eta,s}\ .
\end{equation}
Here $\epsilon_-(\mathbf{k})=-\epsilon_{-}(-\mathbf{k})=[\epsilon_{+1,+}(\mathbf{k})+ \epsilon_{-1,+}(\mathbf{k})]/2$. Since $H_0$ breaks the nonchiral-flat U(4) symmetry down to U(2)$\times$U(2) of the spin-charge rotations of two valleys, the nonchiral-flat U(4) multiplet $|\Psi_{\nu,\nu_C}\rangle$ in Eq.~(\ref{eq-Psi-nu-nuC}) will no longer be degenerate, and a particular U(4) polarization will be favored. The first order perturbation of $H_0$ is zero (see App.~\ref{app:kineticpert-nonchiralflat}). The second order perturbation of $H_0=H_0^++H_0^-$ yields an energy dependence on the valley polarization of the state in the valley Bloch sphere. Here we consider a state obtained by applying a U(4) rotation 
\begin{equation}
U(\varphi_s)=e^{i S^{y0} (\varphi_\uparrow+\varphi_\downarrow) /4}e^{i S^{yz} (\varphi_\uparrow-\varphi_\downarrow) /4}
\end{equation} 
onto the state in Eq.~(\ref{eq-Psi-nu-nuC}), where $\varphi_s$ ($s=\uparrow,\downarrow$) gives a valley rotation in the spin $s$ sector, and we sort $\{\eta_j,s_j\}$ in Eq.~(\ref{eq-Psi-nu-nuC}) in the order $\{+,\uparrow\}$, $\{+,\downarrow\}$, $\{-,\downarrow\}$, $\{-,\uparrow\}$. $S^{ab}$ are the nonchiral-flat U(4) generators (see Eq.~(\ref{eq-SabeY})). Since the state in Eq.~(\ref{eq-Psi-nu-nuC}) is valley polarized in the $+z$ direction of the valley Bloch sphere, and since the $S^{y b}$ ($b=0,x,y,z$) representation matrices of the $e_Y=\pm1$ basis are opposite (proportional to $e_Y$, see Eq.~(\ref{eq-SabeY})), the valley polarization direction of spin $s$ electrons in the $e_Y=\pm1$ basis will be rotated oppositely away from the $z$ axis by angle $\varphi_s$ in the valley Bloch sphere, as illustrated by the blue ($e_Y=+1$) and red ($e_Y=-1$) arrows in Fig. \ref{fig:valley}(c). In App.~\ref{app:kineticpert-nonchiralflat}, we show that the second order perturbation energy of such a state is given by
\begin{equation}\label{eq-E2-nc-nf}
\begin{split}
&E^{(2)'}_{\nu,\nu_C}(\varphi_s)=-N_M\sum_{s=\uparrow,\downarrow}\Big[\nu_s^{(1)}J_1\cos^2\varphi_s \\
&\qquad\qquad\qquad  +(\nu_s^{(2)}J_2+\nu_s^{(3)}J_3)\sin^2\varphi_s\Big],
\end{split}
\end{equation}
where $J_i$ ($i=1,2,3$) are functions of $w_0$ (we assume $w_1=110$meV fixed) and the twist angle $\theta$ (which controls the bandwidth), while the three numbers $\nu_s^{(j)}$ ($j=1,2,3$) are defined in App. \ref{app:kineticpert-nonchiralflat} below Eq. (\ref{seq:approx-E-H0-J}) and given in Tab. \ref{Tab-nu123} (the left table). Their summations over spin are given by $\sum_s\nu_s^{(1)}=|\nu_C|$, $\sum_s\nu_s^{(2)}=4-|\nu|$, and $\sum_s\nu_s^{(3)}=4-|\nu_+-2|-|\nu_--2|$. The energies $J_1$ and $J_2$ come from the second order perturbation of $H_0^+$ in Eq.~(\ref{eq-H0+}) alone, while $J_3$ comes from the second order perturbation of $H_0^-$ in Eq.~(\ref{eq-H0m}) alone and the cross terms between $H_0^+$ and $H_0^-$ (see Eq.~(\ref{seq:approx-E-H0-J}) in App. \ref{app:kineticpert-nonchiralflat}). In general, one has $\nu_s^{(2)}\ge \nu_s^{(1)}$ and $\nu_s^{(2)}\ge \nu_s^{(3)}$ for any filling $\nu$ and any Chern number $\nu_C$. In particular, when the Chern number $|\nu_C|=4-|\nu|$, we have $\nu_s^{(1)}=\nu_s^{(2)}\ge \nu_s^{(3)}$. The numerical values of $J_i$ for $\theta=1.05^\circ$ and $w_0/w_1\in [0,0.8]$ with the FMC (Eq.~(\ref{eq-Mq=0})) imposed are given in Tab. \ref{Tab-J123}, which are independent of filling $\nu$. These values are almost the same as the values of $J_i$ without the FMC given in Tab. \ref{Tab-J123-noFMC} in App.~\ref{app:kineticpert-nonchiralflat}, which indicates the validity of the FMC. For small $w_0$ and single-particle bandwidth $t$, we have $J_1-J_2\propto w_0^2t^2$ and $J_3\propto w_0^2t^2$.

\begin{table}[tbp]
\centering
\begin{tabular}{c|c|c|c}
\hline
 $w_0/w_1$ & $J_1$ (meV) & $J_2$ (meV) & $J_3$ (meV)  \\
\hline
0 & 0.3018 & 0.3018 & 0\\
0.2 & 0.2650 & 0.2626 & 0.0004 \\
0.4 & 0.1735 & 0.1675 & 0.0013 \\
0.6 & 0.0751 & 0.0701 & 0.0020 \\
0.8 & 0.0174 & 0.0157 & 0.0012 \\
\hline
\end{tabular}
\caption{The numerically calculated perturbation energies $J_i$ ($i=1,2,3$) at twist angle $\theta=1.05^\circ$ with the FMC imposed (Eq.~(\ref{eq-Mq=0})), and Coulomb screening length $\xi=10$nm. The calculation is done in a $12\times12$ MBZ momentum lattice (large enough to simulate the thermodynamic limit). Their values depend on $w_0$ (we assume $w_1=110$meV is fixed), and are independent of filling $\nu$ with the FMC imposed. These values of $J_i$ are almost equal to that calculated without the FMC in Tab. \ref{Tab-J123-noFMC}. Note that $J_2\gg J_1-J_2>J_3>0$ for all $w_0>0$. Besides, when $w_0=0$, we have $J_1=J_2=J_0$, with $J_0$ defined in Eq.~(\ref{eq-E2-nf}).}\label{Tab-J123}
\end{table}

Generically, we always find $J_2\gg (J_1-J_2)> J_3>0$ when $w_0>0$. As a result, by minimizing the energy in Eq. (\ref{eq-E2-nc-nf}), we find the lowest insulator state at integer filling $\nu$ with Chern number $\nu_C$ favors $\varphi_s=\pi/2$ if $\nu_s^{(2)}>\nu_s^{(1)}$, and favors $\varphi_s=0$ if $\nu_s^{(2)}=\nu_s^{(1)}$, regardless of $\nu_s^{(3)}$ (see App. \ref{app:kineticpert-nonchiralflat} for details). The wave function of this lowest insulator state can be generically expressed as (see App. \ref{app:kineticpert-nonchiralflat} Eq. (\ref{seq-Psi-nc-nf0}))
\begin{equation}\label{eq-Psi-nc-nf}
\begin{split}
&|\Psi_{\nu,\nu_C}^{\text{nc-nf}}\rangle=\prod_{\mathbf{k}} \prod_{j=1}^{\nu_L} \Big( \prod_{e_Y=\pm} \frac{d^\dag_{\mathbf{k},e_{Y},\eta_j,s_j}+ e^{i\eta_j\gamma}\eta_j e_{Y}d^\dag_{\mathbf{k},e_{Y},-\eta_j,s_j}}{\sqrt{2}}\Big)\\
&\qquad\qquad \times\prod_{j=\nu_L+1}^{\nu_L+|\nu_C|} d^\dag_{\mathbf{k},\text{sgn}(\nu_C),\eta_j,s_j}|0\rangle\ ,
\end{split}
\end{equation}
where $\nu_C=\nu_+-\nu_-$ and $\nu=\nu_++\nu_--4$, and we have defined $\nu_L=\text{min}(\nu_+,\nu_-)$, while $\{\eta_j,s_j\}$ ($1\le j\le4$) are sorted in the order of $\{+,\uparrow\}$, $\{+,\downarrow\}$, $\{-,\downarrow\}$, $\{-,\uparrow\}$. $\gamma$ is an angle that can be chosen arbitrarily. More concretely, we can divide the insulator states in Eq. (\ref{eq-Psi-nc-nf}) into the following three classes:

(i) For zero Chern number $\nu_C=0$ states, which is only possible for even fillings $\nu$, we have $\varphi_\uparrow=\varphi_\downarrow=\pi/2$ (if the spin $\uparrow$/$\downarrow$ sector is not empty), and the insulator state is \emph{fully intervalley coherent}. All the electrons in such a state have a valley polarization in the $x$-$y$ plane of the valley Bloch sphere with an in-plane angle $\gamma$ (up to $\pi$) as shown in Fig. \ref{fig:valley}(e). The average number of electrons in valley $K$ and $K'$ are equal, which are coherent with each other. These zero Chern number intervalley coherent states agree with the K-IVC states at even fillings proposed in Ref. \cite{bultinck_ground_2020}.

(ii) For low Chern number $0<|\nu_C|<4-|\nu|$ states, we find $\varphi_\uparrow=\pi/2$ and $\varphi_\downarrow=0$. So the lowest insulator state is \emph{partially intervalley coherent}: it has intervalley coherence in the spin $\uparrow$ sector (where each valley before the U(4) rotation $U(\varphi_s)$ are fully occupied or empty), while it is valley polarized (in the $z$ direction of the valley Bloch sphere) in the spin $\downarrow$ sector (where at least one valley is half occupied). The average number of electrons in valleys $K$ and $K'$ are unequal.

(iii) For the highest Chern number $|\nu_C|=4-|\nu|$ states, we have $\varphi_\uparrow=\varphi_\downarrow=0$, and $\nu_L=0$ in Eq. (\ref{eq-Psi-nc-nf}). The lowest insulator state is then \emph{fully valley polarized} (in the $z$-direction of the valley Bloch sphere). In this case, the number of electrons in valleys $K$ and $K'$ are maximally imbalanced.

All the U(2)$\times$U(2) rotations of state $|\Psi_{\nu,\nu_C}^{\text{nc-nf}}\rangle$ in Eq. (\ref{eq-Psi-nc-nf}) form a U(2)$\times$U(2) multiplet of degenerate states (see App. \ref{app:kineticpert-nonchiralflat}). The $|\nu_C|=4$ state at $\nu=0$ and the $|\nu_C|=2$ state at $\nu=\pm2$ in Eq. (\ref{eq-Psi-nc-nf}) are singlets of U(2)$\times$U(2). In all the other cases, the state $|\Psi_{\nu,\nu_C}^{\text{nc-nf}}\rangle$ in Eq. (\ref{eq-Psi-nc-nf}) spontaneously breaks the nonchiral-nonflat U(2)$\times$U(2) symmetry, and the remaining symmetry little group for different $\nu,\nu_C$ is given in Tab. \ref{Tab-little} in App. \ref{app:kineticpert-nonchiralflat}. We can decompose the nonchiral-nonflat U(2)$\times$U(2) symmetry into SU(2)$_K\times$SU(2)$_{K'}\times$U(1)$_C\times$U(1)$_V$, where SU(2)$_\eta$ is the spin rotation symmetry of valley $\eta$ (generated by $(\tau^0+\eta\tau_z)s^a/2$), U(1)$_C$ is the global charge U(1) symmetry (generated by $\tau^0s^0$), and U(1)$_V$ is the valley U(1) symmetry (generated by $\tau^zs^0$). In this notation, the U(1)$_C$ symmetry is always unbroken due to the charge conservation. The valley U(1)$_V$ symmetry is unbroken when $|\nu_C|=4-|\nu|$, and is broken when $|\nu_C|<4-|\nu|$. 

\subsection{Nonchiral perturbation to the chiral-nonflat limit}

An alternative approach is to treat the nonchiral terms of Eqs.~(\ref{eq-OqG1}) and (\ref{eq-H0m}) as perturbations to the chiral-nonflat U(4) multiplet $|\widetilde{\Psi}_{\nu,\nu_C}\rangle$ in Eq.~(\ref{eq-Psi-nu-nuC2}). This yields the same ground state as given in Eq.~(\ref{eq-Psi-nc-nf}), which is shown in details in App.~\ref{app:ncpert-chiralnonflat}). In fact, one could see this most easily by noting that the U(4) rotated state $|\Psi_{\nu,\nu_C}^{\text{nc-nf}}\rangle$ in Eq.~(\ref{eq-Psi-nc-nf}) (and its U(2)$\times$U(2) rotations) is a state simultaneously in the nonchiral-flat U(4) multiplet of state $|\Psi_{\nu,\nu_C}\rangle$ in Eq.~(\ref{eq-Psi-nu-nuC}) and the chiral-nonflat U(4) multiplet $|\widetilde{\Psi}_{\nu,\nu_C}\rangle$ in Eq.~(\ref{eq-Psi-nu-nuC2}) (see proof in App.~\ref{app:ncpert-chiralnonflat}). Therefore, for a fixed filling $\nu$ and Chern number $\nu_C$, state $|\Psi_{\nu,\nu_C}^{\text{nc-nf}}\rangle$ in Eq.~(\ref{eq-Psi-nc-nf}) simultaneously minimizes the nonchiral interaction energy and the kinetic energy, thus is favored as a candidate of the lowest state. We note that, however, for $|\nu_C|=4-|\nu|$, one has $\nu_+=0$ or $\nu_-=0$ (if $\nu\le4$), and the entire nonchiral-flat U(4) multiplet~(\ref{eq-Psi-nu-nuC}) is the same as the chiral-nonflat U(4) multiplet~(\ref{eq-Psi-nu-nuC2}). In this case, we cannot easily determine the favored U(4) polarization, and a careful higher order energy calculation is needed to see that the valley polarized state is favored (App. \ref{app:kineticpert-nonchiralflat}). For a similar reason, for $0<|\nu_C|<4-|\nu|$, the valley polarization of the spin $\downarrow$ sector has to rely on a higher order energy calculation (App. \ref{app:kineticpert-nonchiralflat}).

\subsection{An intuitive picture}

For insulator states with Chern number $\nu_C=0$ which have electrons equally occupying the two Chern basis $e_Y=\pm1$, their intervalley coherence can be more intuitively understood with the valley Bloch sphere. For example, consider the insulator state with Chern number $\nu_C=0$ at filling $\nu=0$, which has 2 $e_Y=+1$ Chern bands and 2 $e_Y=-1$ Chern bands fully occupied. We assume the electrons within each Chern basis $e_Y=\pm1$ subspace form a spin singlet with maximal valley polarization along certain direction of the valley Bloch sphere, as illustrated by the blue and red arrows in Fig. \ref{fig:valley}(b)-(e) (the north/south pole of the Bloch sphere represent valley $K$ and $K'$, respectively). In the chiral-flat limit, the valley polarizations of the occupied Chern $+1$ basis and Chern $-1$ basis are unrelated due to the U(4)$\times$U(4) symmetry (see Sec.~\ref{sec:chernchiralflatband}), as shown in Fig. \ref{fig:valley}(b). 
When reduced to the nonchiral-flat limit with a nonchiral-flat U(4) symmetry, we have shown in Sec.~\ref{sec:gsnonchiralflat} that the coupling between the U(4) polarizations of electrons in the Chern $e_Y=\pm1$ subspaces is ferromagnetic. Because of the relative unitary transformation $\tau_z$ between the Chern $e_Y=\pm1$ basis irreps (Eq.~(\ref{eq-SabeY})), the physical valley polarizations of the $e_Y=\pm1$ Chern basis differ by a valley $z$-axis $\pi$ rotation, as illustrated by Fig. \ref{fig:valley}(c). In contrast, when reduced to the chiral-nonflat limit which has a chiral-nonflat U(4) symmetry, we have shown in Sec.~\ref{sec:chiralnonflat} that the two $e_Y=\pm1$ subspaces have a chiral-nonflat U(4) AFM coupling in between. Since the  chiral-nonflat U(4) irreps of the $e_Y=\pm1$ Chern basis are identical (without differing by a unitary transformation), the valley polarizations of the $e_Y=\pm1$ subspaces are opposite (AFM) to each other in the valley Bloch sphere, as illustrated by Fig. \ref{fig:valley}(d). It is then straightforward to see that, in the nonchiral-nonflat case, to compromise between the valley polarization configurations in Figs.~\ref{fig:valley}(c) and (d), the valley polarizations of the electrons in the $e_Y=\pm1$ Chern basis will be pinned in the $x$-$y$ plane of valley Bloch sphere and opposite to each other, as illustrated by Fig. \ref{fig:valley}(e). Thus the state is intervalley coherent.

The same argument can be made for the intervalley coherence in the spin $\uparrow$ sector of the insulator states with Chern number $0<|\nu_C|<4-|\nu|$, which has equal number of electrons occupying the spin $\uparrow$ $e_Y=\pm1$ Chern basis. However, for the states with $|\nu_C|=4-|\nu|$, or the spin down sector of the states with $|\nu_C|<4-|\nu|$, only one of the Chern basis $e_Y=\pm1$ subspaces is occupied (when $\nu\le0$). The valley polarization configurations of the nonchiral-flat limit and the chiral-nonflat limit are then no different, and the above argument fails. Higher order calculations are therefore necessary to show that the polarization along $z$-direction of the valley Bloch sphere is favored (i.e., valley polarized).

\subsection{The ground states}

\begin{figure}[tbp]
\begin{centering}
\includegraphics[width=\linewidth]{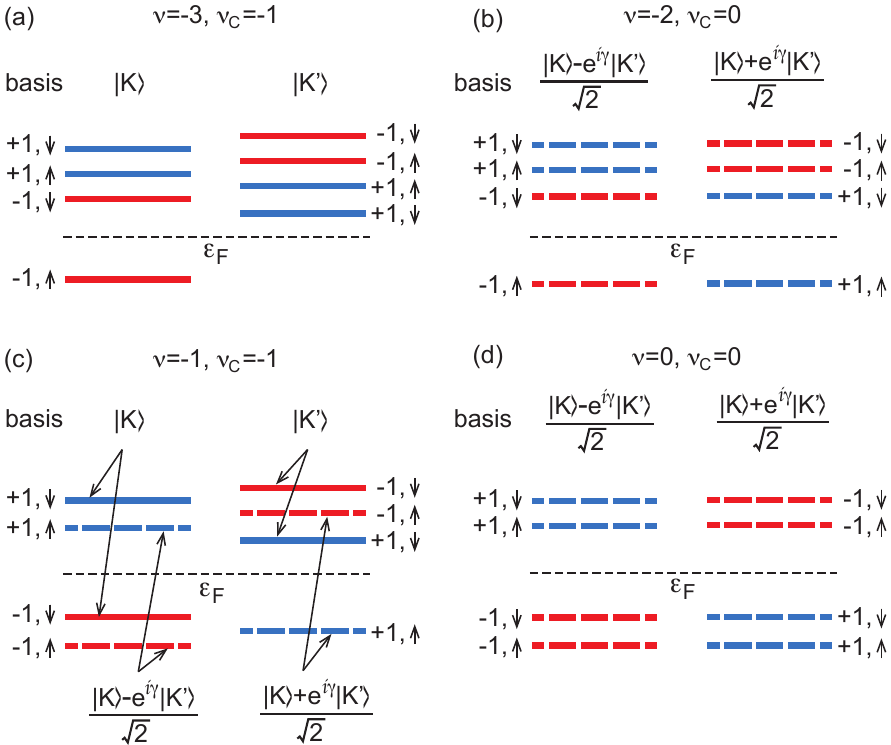}
\end{centering}
\caption{Illustration of filling of the band occupation of the nonchiral-nonflat ground states at integer fillings (a) $\nu=-3$, (b) $\nu=-2$, (c) $\nu=-1$ and (d) $\nu=0$.  $|K\rangle$, $|K'\rangle$ stand for the single-particle basis at valley $K$ and $K'$, and $\gamma$ is an arbitrary phase. The solid lines stand for valley polarized basis (eigenbasis of $\tau_z$, $z$-direction polarized in the valley Bloch sphere), while dashed lines stand for the intervalley coherent basis (in-plane polarized in the valley Bloch sphere). Each band is labeled by its Chern number $e_Y=\pm1$ (blue and red for Chern numbers $\pm1$, respectively) and spin $s=\uparrow,\downarrow$, and the dashed line represents the Fermi level. As we have shown, the $\nu=-3$ ground state has $|\nu_C|=4-|\nu|$, thus is valley polarized; the $\nu=-2,0$ ground states have $\nu_C=0$ and are fully intervalley coherent, while the $\nu=-1$ ground state has $0<|\nu_C|<4-|\nu|$ and is partially intervalley coherent.}
\label{fig:filling}
\end{figure}

The total perturbation energy of the state $|\Psi_{\nu,\nu_C}^{\text{nc-nf}}\rangle$ in Eq. (\ref{eq-Psi-nc-nf}) can be calculated by $E_{\nu,\nu_C}(\varphi_s)=E_{\nu,\nu_C}^{(1)}+E_{\nu,\nu_C}^{(2)'}(\varphi_s)$ with $\varphi_s=0$ (valley polarized) or $\pi/2$ (intervalley coherent), where $E_{\nu,\nu_C}^{(1)}$ and $E_{\nu,\nu_C}^{(2)'}(\varphi_s)$ are defined by Eq.~(\ref{eq-E1-nc}) and (\ref{eq-E2-nc-nf}), respectively. We thus find the ground states in the nonchiral-nonflat case are the insulator state $|\Psi_{\nu,\nu_C}^{\text{nc-nf}}\rangle$ in Eq.~(\ref{eq-Psi-nc-nf}) with Chern number $\nu_C=0$ ($\nu_C=\pm1$) for even (odd) filling $\nu$. In particular, (i) at $\nu=0,\pm2$, the ground states have Chern number $\nu_C=0$ and are fully intervalley coherent, which are about $0.05\sim 0.5$meV (of order $J_1\approx J_2$, thus depending on $w_0$) per electron lower than the valley polarized state with the same Chern number at $\theta=1.05^\circ$. (ii) At $\nu=\pm1$, the ground states with Chern number $\nu_C=\pm1$ are partially intervalley coherent, where the spin $\uparrow$ sector is intervalley coherent, while the spin $\downarrow$ sector is valley polarized. At $\theta=1.05^\circ$, the intervalley coherent spin sector is $0.05\sim 0.5$meV (around $J_2$) per electron lower than its valley polarized counterpart, while the valley polarized spin sector is only about $J_1-J_2-J_3=0.001\sim 0.005$meV (depending on $w_0$) per electron lower than its intervalley coherent counterpart. This means the partially intervalley coherent ground state at $\nu=\pm1$ is $0.05\sim 0.5$meV per electron lower than a fully valley polarized state, and is only $\sim 0.005$meV per electron lower than a fully intervalley coherent state. (iii) Lastly, at $\nu=\pm3$, the Chern number $\nu_C=\pm1$ ground state is valley polarized, which is only about $J_1-J_2-J_3=0.001\sim 0.005$meV per moir\'e unit cell lower than the intervalley coherent state with Chern number $\nu_C=\pm1$ at $\theta=1.05^\circ$. All of these energy differences are expected to be proportional to $t^2$, with $t$ being the single-particle bandwidth. The occupied bands and valley polarization of the ground states at integer fillings $\nu\le 0$ are illustrated in Fig. \ref{fig:filling}.

The ground state we find in Eq.~(\ref{eq-Psi-nc-nf}) at $\nu=0$ with Chern number $\nu_C=0$ is a spin-singlet, and exactly agrees with the $\nu=0$ K-IVC state found in Ref.~\cite{bultinck_ground_2020}. In Ref.~\cite{bultinck_ground_2020}, the K-IVC state is shown to preserve an anti-unitary Kramers time-reversal symmetry $T'=i\tau_yT$, which is the spinless time-reversal multiplied by a valley rotation $i\tau_y$, and satisfies $T'^2=-1$ (in contrast to $T^2=1$ of the physical spinless time-reversal $T$). By noting that the physical time-reversal $T$ flips $e_Y\rightarrow -e_Y$ and valley $\eta\rightarrow -\eta$, it is easy to verify that the state $|\Psi_{0,0}^{\text{nc-nf}}\rangle$ in Eq.~(\ref{eq-Psi-nc-nf}) at $\nu=0$ satisfies 
\begin{equation}
T'|\Psi_{0,0}^{\text{nc-nf}}\rangle =(-e^{-i\gamma})^{4N_M} |\Psi_{0,0}^{\text{nc-nf}}\rangle\ ,
\end{equation}
thus is invariant under the Kramers time-reversal $T'$. The $\nu=-2$ Chern number $0$ state $|\Psi_{-2,0}^{\text{nc-nf}}\rangle$ we found also agrees with the K-IVC state suggested for $\nu=-2$ in Ref.~\cite{bultinck_ground_2020}, while we have further identified the FMC Eq.~(\ref{eq-Mq=0}) as the sufficient condition for it to be the ground state. The state $|\Psi_{-2,0}^{\text{nc-nf}}\rangle$ is also similar to the $\nu=-2$ ground state found by Ref.~\cite{kang_strong_2019}, but our Hamiltonians are different (see discussions in Ref.~\cite{ourpaper3}). The valley coherence/polarization of the ground states at odd integer fillings and the higher Chern number low-lying states at all integer fillings that we have identified in Eqs.~(\ref{eq-Psi-nc-nf}) have not been analytically studied before. The valley polarized Chern number $\pm1$ state at $\nu=-3$ we identified here is also verified in our exact diagonalization study in Ref.~\cite{ourpaper6}. Besides, we note that the ground state valley coherence/polarization at integer fillings $\nu$ we found here (fully/partially intervalley coherent at $\nu=0,\pm1,\pm2$ and valley polarized at $\nu=\pm3$) are in agreement with the Hartree-Fock calculation in Ref.~\cite{zhang_HF_2020}. However, the ground state Chern numbers are not studied in Ref.~\cite{zhang_HF_2020}.

\section{First-order phase transitions in magnetic field}\label{sec:phasetransitionbfield}

\begin{figure}[tbp]
\begin{centering}
\includegraphics[width=\linewidth]{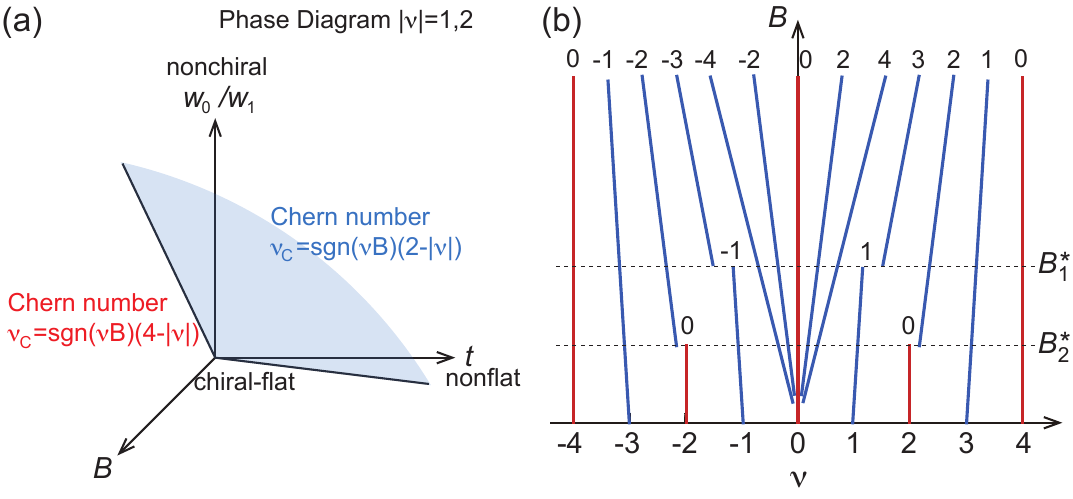}
\end{centering}
\caption{(a) The phase diagram near integer fillings $\nu=\pm1,\pm2$ with respect to the nonchiral interaction strength $w_0/w_1$, single-particle bandwidth $t$ and out-of-plane magnetic field $B$. A first-order phase transition from an insulator with Chern number $\nu_C=\text{sgn}(\nu B)(2-|\nu|)$ (fully/partially intervalley coherent) to $\nu_C=\text{sgn}(\nu B)(4-|\nu|)$ (valley polarized) happens with respect to $B$. (b) The expected dominant Landau fan diagram with respect to filling $\nu$ and out-of-plane magnetic field $B$ based on our theory, where the lines indicate the positions of Chern gaps with their Chern numbers labeled (red lines stand for Chern number zero). Interaction-driven first-order transitions are expected at certain magnetic field $B_\nu^*$ for fillings $\nu=\pm1,\pm2$, and our theory gives $B_1^*=2B_2^*$. For $\theta=1.05^\circ$, $w_0=0.8w_1$ and top/bottom screening lengths $\xi\approx10$nm, our numerical calculation estimates $B_1^*\approx0.5$T and $B_2^*\approx0.2$T.}
\label{fig:PD}
\end{figure}

We now discuss the effect of an out-of-plane magnetic field $B$ on the TBG insulator ground states in the nonchiral-nonflat case by examining their free energies. As we will show, the magnetic field $B$ can drive phase transitions between insulator states with different Chern numbers.

By Eq.~(\ref{eq-HIrewritten}), we find the chemical potential for realizing the insulator states $|\Psi_{\nu,\nu_C}^{\text{nc-nf}}\rangle$ is approximately $\mu_\nu=\nu U_0$, where 
\begin{equation}\label{eq-U0}
U_0=\frac{1}{N_M\Omega_{\text{tot}}}\sum_{\GG}V(\GG)\left(\sum_\kk \alpha_0(\mathbf{k,G})\right)^2. 
\end{equation}
In the nonchiral case, we have shown that states with the lowest Chern numbers $|\nu_C|$ are preferred at all integer fillings $\nu$ (Eq.~(\ref{eq-E1-nc})).  When an out-of-plane magnetic field $B$ is added, the Streda formula \cite{streda1982} implies that the number of occupied electrons $N$ adiabatically change by $\Delta N(B)=\nu_CN_M\Phi/\Phi_0$, where $\Phi=B\Omega_M$ is the magnetic flux per moir\'e unit cell area $\Omega_M$, and $\Phi_0=h/e$ is the flux quanta. Since the orbital magnetic moments \cite{shi_orbmag_2007} of the flat bands are zero in the flat limit $H_0=0$ (App.~\ref{app:magnetic}), we expect the interaction energy to be roughly unchanged at small $B$. 
Therefore, the change of the free energy $F=\langle H_I-\mu_\nu N\rangle$ is approximately $\Delta F(B)=-\mu_\nu \Delta N (B)=-\nu\nu_CN_MU_0 \Phi/\Phi_0$. Thus, with the nonchiral perturbation energy in Eq.~(\ref{eq-E1-nc}), and neglecting the kinetic perturbation energy in Eq.~(\ref{eq-E2-nc-nf}) which has a negligible $\nu_C$ dependence (see App.~\ref{app:magnetic}), we find the state $|\Psi_{\nu,\nu_C}^{\text{nc-nf}}\rangle$ has a free energy (up to $\nu_C$ independent terms)
\begin{equation}\label{eq:free-energy}
F_{\nu,\nu_C}(B)\approx N_M\Big[|\nu_C|(U_1-\nu^2U_2)-\nu\nu_CU_0\frac{e\Omega_M}{h} B\Big],
\end{equation}
where $U_0$, $U_1$ and $U_2$ are defined by Eqs.~(\ref{eq-U0}) and~(\ref{eq-E1-nc}). When $\nu=\pm1,\pm2$, and as $B$ increases to a magnitude
\begin{equation}
B_{\nu}^*=\frac{U_1-\nu^2U_2}{|\nu|U_0}\frac{h}{e\Omega_M}\ ,
\end{equation}
we find the ground state at filling $\nu$ undergoes an interaction-driven first order phase transition from the lowest Chern number $\nu_C=\text{sgn}(\nu B)(2-|\nu|)$ (which is fully or partially intervalley coherent) to the highest Chern number $\nu_C=\text{sgn}(\nu B)(4-|\nu|)$ (which is valley polarized), as illustrated in Fig. \ref{fig:PD}(a). Note that we numerically find $U_1-\nu^2U_2>0$ for any $|\nu|\le3$ and $0< w_0/w_1\le 1$ without the FMC, so $B_{\nu}^*>0$. For filling $\nu=\pm3$, the ground state always have a Chern number $\nu_C=\text{sgn}(\nu B)$. For $\nu=0$, the magnetic field $B$ has no contribution in Eq.~(\ref{eq:free-energy}), and we expect the Chern number $\nu_C=0$ state to stay robust. This leads to a predicted dominant Landau fan diagram as shown in Fig. \ref{fig:PD}(b), with interaction-driven first order transitions between different Chern numbers at finite $B$ near $\nu=\pm1,\pm2$. Near $\nu=0$, we expect no interaction-driven phase transition, but has the Landau fan contributed by Landau levels at the CNP (which we expect to be spin 2-fold degenerate, since the ground state $|\Psi_{0,0}^{\text{nc-nf}}\rangle$ we found in Eq.~(\ref{eq-Psi-nc-nf}) is a spin-singlet thus spin degenerate, but breaks valley U(1) symmetry thus valley nondegenerate). For $w_0\approx 0.8w_1$ near the magic angle \cite{Uchida_corrugation,Wijk_corrugation,dai_corrugation, jain_corrugation}, and top/bottom gate screening length $\xi\approx 10$nm, we estimated that $U_1/U_0\approx0.02$ and $U_2/U_0\approx 0.0015$ without the FMC (see App. \ref{app:magnetic} Fig. \ref{fig:U1U0}(b)), which gives a critical field $B_{1}^*\approx 0.5$T at fillings $\nu=\pm1$, and $B_{2}^*\approx 0.2$T at fillings $\nu=\pm2$. 

Remarkably, our prediction (Fig. \ref{fig:PD}(b)) is well supported by the recent experimental discoveries by scanning tunneling spectroscopy \cite{nuckolls_chern_2020,choi2020tracing} as well as transport experiments \cite{saito2020,das2020symmetry, wu_chern_2020}, where Chern number $\nu_C=\text{sgn}(\nu B)(4-|\nu|)$ interacting gaps are found to arise above a certain magnetic field $B$ near all integer fillings $\nu\neq 0$. The hysteresis loop \cite{lu2019superconductors,das2020symmetry} and Pomeranchuk effect \cite{saito2020isospin,rozen2020entropic} observed in transport in magnetic field $B>0$ near $\nu=\pm1$ \cite{das2020symmetry} also suggest the presence of first-order phase transitions therein, thus supporting our prediction of the nonzero $B$ field first-order phase transitions. In particular, the hysteresis near $\nu=\pm1$ in transport experiments is observed around a magnetic field $1$T in Ref.~\cite{lu2019superconductors} and around $3$T in Ref.~\cite{das2020symmetry}, which have the same order of magnitude as our estimations ($\sim 0.5$T), considering that unknown realistic complications (sample strain, etc.) are not taken into account in our calculations.

\section{The stabilizer code limit}\label{sec:stabilizercode}

Lastly, we study the many-body states in a stabilizer code limit revealed in Ref.~\cite{ourpaper3} (see also Sec.~\ref{app:stabilizer}). The stabilizer code limit is defined as the chiral-flat limit plus the condition that the form factors $M_{mn}^{(\eta)}\left(\mathbf{k},\mathbf{q}+\mathbf{G}\right)$ in Eq.~(\ref{eq-Mmn}) are independent of $\mathbf{k}$ for all $\qq,\GG$. As a result, one will have $[O_{\qq,\GG},O_{\qq',\GG'}]=0$, and thus all the terms of the Hamiltonian $H=H_I$ in Eq.~(\ref{eq-pHI}) commute with each other (see Ref.~\cite{ourpaper3}):
\begin{equation}
[O_{-\qq,-\GG}O_{\qq,\GG},O_{-\qq',-\GG'}O_{\qq',\GG'}]=0\ .
\end{equation}
By a Fourier transformation into the real space, the Hamiltonian $H_I$ can be rewritten into an extended Hubbard model (App.~\ref{app:stabilizer}):
\begin{equation}
H_I=\sum_{e_Y,s,\eta, e_Y',s',\eta'}\sum_{\RR_M,\RR_M'} \frac{U^{e_Y,e_Y'}_{\RR_M-\RR_M'}}{2} n_{e_Y,\mathbf{R}_M}^{\eta,s} n_{e_Y',\mathbf{R}_M'}^{\eta',s'},
\end{equation}
where $\RR_M$ are the AA stacking center sites of TBG, $n_{e_Y,\mathbf{R}_M}^{\eta,s}=d_{e_Y,\eta,s,\RR_M}^\dag d_{e_Y,\eta,s,\RR_M}-\frac{1}{2}$, and we have defined $d_{e_Y,\eta,s,\RR_M}^\dag=\frac{1}{\sqrt{N_M}}\sum_{\mathbf{k}}e^{i\mathbf{k\cdot R}_M}d^\dag_{\mathbf{k},e_Y,\eta,s}$. The extended Hubbard interaction is given by (see App.~\ref{app:stabilizer})
\begin{equation}
\begin{split}
U^{e_Y,e_Y'}_{\RR_M-\RR_M'}=&\frac{1}{\Omega_{\text{tot}}}\sum_{\qq,\GG} e^{i(\qq+\GG)\cdot(\RR_M-\RR_M')}\\ 
&\times \beta_{e_Y}(\mathbf{q+G}) \beta_{e_Y'}(\mathbf{-q-G})\ ,
\end{split}
\end{equation}
where $\beta_{e_Y}(\mathbf{q+G})=\sqrt{V(\mathbf{G}+\mathbf{q})} M_{e_Y}(\kk,\qq+\GG)$ is $\kk$-independent in this limit. 
Due to the  Wannier obstruction of a Chern band, one expects $U^{e_Y,e_Y'}_{\RR_M-\RR_M'}$ to be long range. 

Since $[n_{e_Y,\mathbf{R}_M}^{\eta,s}, n_{e_Y',\mathbf{R}_M'}^{\eta',s'}]=0$, the many-body eigenstates of $H_I$ is simply given by the Fock states of all the on-site electron occupation configurations, where $n_{e_Y,\mathbf{R}_M}^{\eta,s}=\pm 1/2$.

Generically, this stabilizer code limit cannot be reached by realistic TBG parameters. However, it provides us a rough understanding of the TBG physics in terms of Hubbard subbands, as suggested by the recent scanning tunneling spectroscopy experiments \cite{xie2019spectroscopic,wong_cascade_2020,nuckolls_chern_2020}.

\section{Conclusion and Discussion}\label{sec:discussion}

Under the influence of  Coulomb interactions, we have shown that under the FMC in Eq.~(\ref{eq-Mq=0}), or if the FMC is not strongly violated, exact Chern insulator Fock ground states of Chern number $\nu_C=4-|\nu|,2-|\nu|,\cdots,|\nu|-4$ can be obtained at all integer fillings $\nu$ of TBG in the (first) chiral-flat limit (defined by $w_0=0$) with a U(4)$\times$U(4) symmetry. In Refs.~\cite{ourpaper1,ourpaper5,ourpaper6}, the validity of the FMC (that it is not strongly violated) is justified by both analytical and numerical calculations. Exact Chern number $0$ Fock ground states can also be derived at even fillings $\nu=0,\pm2$ in the nonchiral-flat limit with a U(4) symmetry, which are similar to the exact ground state of Kang and Vafek at $\nu=-2$ in Ref.~\cite{kang_strong_2019}. At odd fillings $\nu=\pm1,\pm3$ in the nonchiral-flat limit, we find perturbative Chern insulator ground states with Chern number $\nu_C=\pm1$. In the chiral-nonflat limit, we find different Chern number states at each filling $\nu$ are degenerate up to the second order perturbation; while our exact diagonalization calculation in Ref.~\cite{ourpaper6} suggests that the higher order perturbations favor the lowest Chern number state. The expressions of these exact/perturbative insulator states allow us to further calculate the charge excitations and neutral Goldstone collective modes of TBG, which is studied in Ref.~\cite{ourpaper5}. The charge gaps of the insulating states in this paper will also be studied in Ref.~\cite{ourpaper5,ourpaper6}. In particular, the exact charge $\pm1$ excitations derived in Ref.~\cite{ourpaper5} for the exact insulator states here are equivalent to the Hartree-Fock bands of these insulator states (see App. \ref{app:HF-Hamiltonian}).

In the perspective of the $e_Y=\pm1$ Chern basis, all these low-energy insulator states we have found can be viewed as U(4)$\times$U(4) FM in the (first) chiral-flat limit, where the spin-valley U(4) polarizations of electrons in the $e_Y=\pm1$ Chern basis are unrelated to each other. In the nonchiral-flat limit, the nonchiral-flat U(4) polarizations of the $e_Y=\pm1$ Chern basis have a FM coupling between each other. In contrast, in the (first) chiral-nonflat limit, the chiral-nonflat U(4) (different from the nonchiral-flat U(4)) polarizations of the $e_Y=\pm1$ Chern basis effectively have an AFM coupling in between. We note that all of these spin-valley magnetizations (polarizations) have an orbital region, because of the absence of spin-orbital couplings in graphene.

In the nonchiral-nonflat case which corresponds to the experimental reality, due to the kinetic energy, the insulating states with zero Chern number $\nu_C=0$ (e.g., the ground states at $\nu=0,\pm2$) are further aligned into fully intervalley coherent, where electrons occupying the Chern basis $e_Y=\pm1$ have opposite in-plane valley Bloch sphere polarizations. The insulating states with Chern number $\nu_C=4-|\nu|$ (e.g., the $\nu=\pm3$ ground states with the highest Chern number $\nu_C=\pm1$) are pinned to be valley polarized, with a maximal number of electrons imbalance between the two valleys. Besides, the states with low Chern numbers $0<|\nu_C|<4-|\nu|$ (e.g., the ground state at $\nu=\pm1$ with Chern number $\nu_C=\pm1$) are found to be partially intervalley coherent. However, for $0<|\nu_C|\le4-|\nu|$, the lowest state (valley polarized if $|\nu_C|=4-|\nu|$ and partially valley coherent if $0<|\nu_C|<4-|\nu|$) is only $\sim0.005$meV per electron lower than the fully intervalley coherent state, making the latter still a competitive state. The lowest Chern number state at each filling $\nu$ is generically favored, while the higher Chern number states are competing low-lying states. 
In particular, the fully intervalley coherent states with Chern number $\nu_C=0$ at even fillings we found agree exactly with the K-IVC states studied in Ref. \cite{bultinck_ground_2020}. We also note that the ground state valley coherence/polarization we found at all integer fillings $\nu$ agrees with that from the Hartree-Fock calculations in Ref.~\cite{zhang_HF_2020}, but the ground state Chern numbers are not discussed in Ref.~\cite{zhang_HF_2020}.

Further, we showed that for $\nu=\pm1,\pm2$, the TBG ground state undergoes first order transitions from Chern number $\nu_C=\text{sgn}(\nu B)(2-|\nu|)$ (intervalley coherent) to $\nu_C=\text{sgn}(\nu B)(4-|\nu|)$ (valley polarized) can be driven by an out-of-plane magnetic field around a nonzero critical field $B_\nu^*$ ($B_1^*\approx 0.5$T, $B_2^*\approx 0.2$T in our numerical calculations). This explains the Chern number $\nu_C=\text{sgn}(\nu B)(4-|\nu|)$ insulating states arising in magnetic fields as observed by recent scanning tunneling spectroscopy experiments \cite{nuckolls_chern_2020}, and by transport experiments \cite{saito2020,das2020symmetry,saito2020isospin, wu_chern_2020,rozen2020entropic}.

When the nonchiral interaction terms are large (i.e., large $w_0/w_1$), our perturbative treatment for odd fillings $\nu=\pm1,\pm3$ may become invalid, in which case the Chern number $\nu_C=\pm1$ ground states at these odd fillings $\nu$ may give way to an unpolarized metallic state, or translation and/or rotational symmetry broken phases as proposed in Refs.~\cite{kang_strong_2019,dodaro2018phases}. Further, if the bandwidths of the active bands become large (e.g., away from the magic angle), the insulator ground states at all integer fillings $\nu$ we discussed in this paper will eventually give way to weakly interacting unpolarized metallic phases. We leave the studies of these situations for our numerical paper \cite{ourpaper6} as well as for future theoretical analysis.

\ \\

\begin{acknowledgments}
We thank Aditya Cowsik and Fang Xie for valuable discussions. We are also grateful to Michael Zaletel for helpful comments and discussions on our results. B.A.B thanks Oskar Vafek for fruitful discussions, and for sharing their similar results on this problem before publication \cite{kang_nonabelian_2020}. This work was supported by the DOE Grant No. DE-SC0016239, the Schmidt Fund for Innovative Research, Simons Investigator Grant No. 404513, the Packard Foundation, the Gordon and Betty Moore Foundation through Grant No. GBMF8685 towards the Princeton theory program, and a Guggenheim Fellowship from the John Simon Guggenheim Memorial Foundation. Further support was provided by the NSF-EAGER No. DMR 1643312, NSF-MRSEC No. DMR-1420541 and DMR-2011750, ONR No. N00014-20-1-2303, Gordon and Betty Moore Foundation through Grant GBMF8685 towards the Princeton theory program, BSF Israel US foundation No. 2018226, and the Princeton Global Network Funds. B.L. acknowledge the support of Princeton Center for Theoretical Science at Princeton University during the early stage of this work. D.K.E. acknowledges support from the Ministry of Economy and Competitiveness of Spain through the “Severo Ochoa” program for Centres of Excellence in R\&D (SE5-0522), Fundació Privada Cellex, Fundació Privada Mir-Puig, the Generalitat de Catalunya through the CERCA program, funding from the European Research Council (ERC) under the European Union’s Horizon 2020 research and innovation programme (grant agreement No. 852927) and the La Caixa Foundation. AY is supported by the Gordon and Betty Moore Foundation’s EPiQS initiative grants GBMF9469, DOE-BES grant DE-FG02-07ER46419, NSF-MRSEC through the Princeton Center for Complex Materials NSF-DMR-1420541, and NSF-DMR-1904442.
\end{acknowledgments}

\bibliography{TBLGHexalogy,HexalogyInternalRefs}

\appendix

\onecolumngrid
\tableofcontents

\section{Review of Notations: Single-Particle and Interaction Hamiltonian}\label{app:notations}

This appendix is devoted to a very brief review - for self-completeness-  of the single-particle Hamiltonian and interaction Hamiltonian of TBG. The detailed discussion of the TBG Hamiltonian can be found in Ref.~\cite{ourpaper3}.

\subsection{Single-particle Hamiltoian: short review}\label{app:notationsonebody}

\subsubsection{Continuum model Hamiltonian}

The single-particle Hamiltonian of TBG for small twist angle $\theta$ known as the Bistritzer-Macdonald continuum model  \cite{bistritzer_moire_2011} takes the form (see e.g. Refs.~\cite{ourpaper1,ourpaper2,ourpaper3})]
\begin{equation}\label{eq:H0}
\hat{H}_{0}=\sum_{\mathbf{k}\in\text{MBZ}}\sum_{\eta\alpha\beta s}\sum_{\mathbf{Q}\mathbf{Q}^{\prime}}\left[ h_{\mathbf{Q},\mathbf{Q}^{\prime}}^{\left(\eta\right)}\left(\mathbf{k}\right)\right]_{\alpha\beta} c_{\mathbf{k},\mathbf{Q},\eta,\alpha, s}^{\dagger}c_{\mathbf{k},\mathbf{Q}^{\prime},\eta,\beta, s}\ ,
\end{equation}
where $\mathbf{k}$ takes value in the moir\'e Brillouin zone (MBZ), and $\mathbf{k=0}$ is chosen at the center ($\Gamma_M$ point) of the MBZ. The momenta $\mathbf{Q}\in\mathcal{Q}_\pm$ runs over the sets $\mathcal{Q}_\pm=\pm\mathbf{q}_1+\mathcal{Q}_0$, where we have defined momenta $\mathbf{q}_j=k_\theta C_{3z}^{j-1}(0,1)^T$ ($j=1,2,3$), with $k_\theta=8\pi\sin(\theta/2) /3a_0$ for graphene lattice constant $a_0=0.246$nm, and $C_{3z}$ being 3-fold rotation about z axis. $\mathcal{Q}_0$ is the triangular reciprocal lattice generated by the moir\'e reciprocal vectors $\widetilde{\mathbf{b}}_1=\mathbf{q}_3-\mathbf{q}_1$ and $\widetilde{\mathbf{b}}_2=\mathbf{q}_3-\mathbf{q}_2$. $\eta=\pm$ is the graphene valley index for valley K and K', respectively, $s$ is electron spin, and $\alpha,\beta=A, B$ are indices for the graphene A and B sublattices. $h_{\mathbf{Q},\mathbf{Q}^{\prime}}^{\left(\eta\right)}\left(\mathbf{k}\right)$ is the first-quantized momentum space Hamiltonian at valley $\eta$ in the sublattice space, and $\QQ,\QQ'\in \mathcal{Q}_\pm$, which is independent of spin $s$ due to the absence of spin-orbital coupling (SOC). At valley $\eta=\pm$, they take the form
\begin{equation}
h_{\mathbf{Q},\mathbf{Q}^{\prime}}^{\left(+\right)}\left(\mathbf{k}\right) =v_F(\mathbf{k-Q})\cdot\bm{\sigma}\delta_{\mathbf{Q,Q'}}+\sum_{j=1}^3 T_j\delta_{\mathbf{Q,Q'\pm q}_j}\ ,
\end{equation}
\begin{equation}
h_{\mathbf{Q},\mathbf{Q}^{\prime}}^{\left(-\right)}\left(\mathbf{k}\right) 
=-v_F(\mathbf{k-Q})\cdot\bm{\sigma}^*\delta_{\mathbf{Q,Q'}}+\sum_{j=1}^3 (\sigma_xT_j\sigma_x)\delta_{\mathbf{Q,Q'\pm q}_j}\ ,
\end{equation}
where $v_F$ is the graphene Fermi velocity,  $\bm{\sigma}=(\sigma_x,\sigma_y)$,  $\bm{\sigma}^*=(\sigma_x,-\sigma_y)$, and the matrices
\begin{equation}\label{seq-Tj}
T_j=w_0\sigma_0+w_1\Big[\sigma_x\cos\frac{2\pi(j-1)}{3}+\sigma_y\sin\frac{2\pi(j-1)}{3}\Big]\ .
\end{equation}
Here $\sigma_0,\sigma_x,\sigma_y,\sigma_z$ are the $2\times2$ identity matrix and Pauli matrices in the space of sublattice indices, while $w_0\ge0$ and $w_1\ge0$ are the interlayer hoppings at the AA and AB stacking centers of TBG, respectively. Generically, in realistic systems $w_0<w_1$ due to the lattice relaxation. In the absence of lattice relaxation, one has $w_0=w_1$ \cite{bistritzer_moire_2011}.

The single-particle Hamiltonian can be diagonalized into
\beq
\hat{H}_0 = \sum_{\kk} \sum_{n \eta s} \epsilon_{n, \eta}(\kk) c_{\kk n \eta s}^\dagger c_{\kk n\eta s} ,\label{eq:H0-energyband}
\eeq
where
\begin{equation}
c_{\mathbf{k},n,\eta, s}^{\dagger}=\sum_{\mathbf{Q}\alpha}u_{\mathbf{Q}\alpha;n\eta}\left(\mathbf{k}\right)c_{\mathbf{k},\mathbf{Q},\eta,\alpha s}^{\dagger}\ ,\label{eq:solution}
\end{equation}
is the energy band electron basis, while $\epsilon_{n, \eta}(\kk)$ and $u_{\mathbf{Q}\alpha;n\eta}(\kk)$ are the eigen-energy and eigenstate wavefunction of band $n$ of the first quantized Hamiltonian $h_{\mathbf{Q},\mathbf{Q}^{\prime}}^{\left(\eta\right)}\left(\mathbf{k}\right)$ in valley $\eta$. The wavefunction satisfies the Bloch periodicity with  unit embedding matrix $u_{\mathbf{Q}\alpha; n\eta}\left(\mathbf{k}+\tilde{\mathbf{b}}_{i}\right)=u_{\mathbf{Q}-\tilde{\mathbf{b}}_{i}\alpha;n\eta}\left(\mathbf{k}\right)$. In each valley and spin, we shall use integers $n>0$ to label the $n$-th conduction band, and use integer $n<0$ to label the $|n|$-th valence band (thus $n\neq0$). The lowest conduction and valence bands in each valley-spin flavor is thus labeled by $n=\pm1$. 

\subsubsection{Discrete symmetries}\label{app:discretesymmetries}

The discrete symmetries of TBG include spinless (due to absence of SOC) unitary discrete rotational symmetries $C_{2z}$, $C_{3z}$ and $C_{2x}$, and the spinless anti-unitary time-reversal symmetry $T$ (see Refs.~\cite{ourpaper1,ourpaper2} for more details). Furthermore, there is a unitary particle-hole transformation $P$ which anti-commutes with the single-particle Hamiltonian, namely, $\{P,\hat{H}_0\}=0$ \cite{song_all_2019,ourpaper2}. 
Lastly, in the (first) chiral limit when $w_0=0<w_1$, there is another chiral transformation $C$ which anti-commutes with the single-particle Hamiltonian, $\{C,\hat{H}_0\}=0$ \cite{tarnopolsky_origin_2019,ourpaper2}.

The operations of a symmetry operator $g$ can be generically represented by
\begin{equation}
g c_{\mathbf{k},\mathbf{Q},\eta,\alpha,s}^{\dagger} g^{-1}=\sum_{\QQ'\eta'\beta} [D(g)]_{\QQ'\eta'\beta,\QQ\eta\alpha} c_{g\mathbf{k},\mathbf{Q}',\eta',\beta,s}^{\dagger}\ ,
\end{equation}
where $D(g)$ is the representation matrix of the symmetry operation $g$ in the space of indices $\{\QQ,\eta,\alpha\}$, and $g\kk$ is the momentum after acting $g$ on momentum $\kk$. In particular, $C_{2z}\kk=T\kk=P\kk=-\kk$, while $C\kk=\kk$. The representation matrices for the discrete symmetries of TBG that will be used in this paper are given by
\beq
[D(C_{2z})]_{\QQ^\pr \eta^\pr \beta, \QQ\eta \alpha} = \delta_{\QQ^\pr,- \QQ} \delta_{\eta^\pr,-\eta} (\sigma_x)_{\beta\alpha}, \qquad [D(T)]_{\QQ^\pr \eta^\pr \beta, \QQ \eta \alpha}= \delta_{\QQ^\pr,-\QQ} \delta_{\eta^\pr,-\eta} \delta_{\beta,\alpha, \label{eq:C2}
}\eeq
\beq
\quad[D(P)]_{\QQ^\pr \eta^\pr \beta, \QQ \eta \alpha} = \delta_{\QQ^\pr,-\QQ} \delta_{\eta^\pr, \eta} \delta_{\beta,\alpha} \zeta_{\QQ}\ ,\qquad \quad[D(C)]_{\QQ^\pr \eta^\pr \beta, \QQ \eta \alpha} = \delta_{\QQ^\pr,\QQ} \delta_{\eta^\pr, \eta} (\sigma_z)_{\beta,\alpha}\ . \label{eq:P}
\eeq
The $C_{2z}$, $T$ and $P$ symmetries imply that
\beq
\epsilon_{n,\eta}(\kk) = \epsilon_{n,-\eta}(-\kk)\ ,\qquad
\epsilon_{n,\eta}(\kk) =-\epsilon_{-n,\eta}(-\kk)\ .
\eeq
Furthermore, in the (first) chiral limit $w_0=0$, we further have
\beq
\epsilon_{n,\eta}(\kk) =-\epsilon_{-n,\eta}(\kk)\ .
\eeq

The detailed properties of these symmetries are given in Refs.~\cite{ourpaper2,ourpaper3}.

\subsection{Interacting Hamiltonian: short review}\label{app:interactinghamiltonian}

Here we review the full Hamiltonian of TBG under Coulomb interaction, and the symmetries of the Hamiltonian in the nonchiral-flat, (first) chiral-flat and (first) chiral-nonflat limits. The detailed full discussion has been given in Ref.~\cite{ourpaper3}.

The Coulomb interaction is assumed to be screened by a top gate plate and bottom gate plate at distances $\xi$ away from TBG, which takes the form of $\widetilde{V}\left(\mathbf{r}\right)=U_{\xi}\sum_{n=-\infty}^{\infty}\frac{\left(-1\right)^{n}}{\sqrt{\left(r/\xi\right)^{2}+n^{2}}}$, with $U_{\xi}=e^2/(\epsilon\xi)$. As an approximate estimation, we choose $\xi=10$nm, and $\epsilon\approx 6$, which yields $U_{\xi}=24$meV. This agrees with the experimental observations of Ref.~\cite{xie2019spectroscopic}, which suggest an on-site Hubbard interaction $\sim 25$meV in magic angle TBG.

The full Coulomb interaction Hamiltonian can be written in the momentum space as
\begin{equation}\label{seq-HI}
\hat{H}_I=\frac{1}{2\Omega_{\text{tot}}}\sum_{\mathbf{G}\in\mathcal{Q}_0}\sum_{\mathbf{q}\in \text{MBZ}}V(\mathbf{q+G})\delta\rho_{\mathbf{-q-G}}\delta\rho_{\mathbf{q+G}}\ ,
\end{equation}
where
\begin{equation}\label{seq-Vq}
V\left(\mathbf{q}\right)=\frac{2\pi e^2}{\epsilon}\frac{\tanh\left(\xi q/2\right)}{q}
\end{equation}
is the Fourier transform of $\widetilde{V}\left(\mathbf{r}\right)$, and
\begin{equation}\label{eq:drho-0}
\delta\rho_{\mathbf{q+G}}=\sum_{\eta,\alpha,s}\sum_{\mathbf{k}\in\text{MBZ}}\sum_{\mathbf{Q}\in\mathcal{Q}_\pm} \left(c_{\mathbf{k+q},\mathbf{Q-G},\eta,\alpha, s}^\dag c_{\mathbf{k},\mathbf{Q},\eta,\alpha, s}-\frac{1}{2}\delta_{\mathbf{q,0}}\delta_{\mathbf{G,0}}\right)\ .
\end{equation} 
is the Fourier transform of the total electron density at momentum $\mathbf{q+G}$ relative to the filling of the graphene charge neutral point (CNP).

\subsubsection{Projected Hamiltonian and gauge fixing}\label{app:ihamiltoniangaugefixing}

Near the magic angle, since the lowest two bands $n=\pm1$ per spin-valley are almost flat and away from higher bands, we can project the TBG Hamiltonian into the 8 flat bands (2 per spin-valley). The projected Hamiltonian $H$ can be written as two terms $H=H_0+H_I$ (note that here we denote projected Hamiltonian in notations without hat, to distinguish with the full Hamiltonian $\hat{H}_0$ and $\hat{H}_I$ which have a hat). The kinetic Hamiltonian is simply given by
\beq
H_0 =\sum_{n=\pm1}\sum_{\eta s} \sum_{\kk\in\text{MBZ}}  \epsilon_{n, \eta}(\kk) c_{\kk n \eta s}^\dagger c_{\kk n\eta s}\ .\label{eq:H0-proj}
\eeq
As derived in detail in Ref.~\cite{ourpaper3}, the projected interaction Hamiltonian can be written as
\begin{equation}\label{seq-pHI}
H_I=\frac{1}{2\Omega_{\text{tot}}}\sum_{\mathbf{q}\in\text{MBZ}}\sum_{\mathbf{G}\in\mathcal{Q}_0} O_{\mathbf{-q,-G}} O_{\mathbf{q,G}} \ ,
\end{equation}
where we have defined a set of operators
\begin{equation}\label{seq-OqG}
O_{\mathbf{q,G}} =\sum_{\mathbf{k}\eta s}\sum_{m,n=\pm1} \sqrt{V(\mathbf{q+G})} M_{m,n}^{\left(\eta\right)} \left(\mathbf{k},\mathbf{q}+\mathbf{G}\right) \left(\rho_{\mathbf{k,q},m,n,s}^\eta-\frac{1}{2}\delta_{\mathbf{q,0}}\delta_{m,n}\right)\ .
\end{equation}
Here $\rho_{\mathbf{k,q},m,n,s}^\eta=c^\dagger_{\kk+\qq,m,\eta,s} c_{\kk,n,\eta,s}$ is the density operator within the flat bands, and we have defined the wavefunction overlap matrix
\begin{equation}
M_{m,n}^{\left(\eta\right)}\left(\mathbf{k},\mathbf{q}+\mathbf{G}\right)=\sum_{\alpha}\sum_{\substack{\mathbf{Q}\in\mathcal{Q}_{\pm}}
}u_{\mathbf{Q}-\mathbf{G},\alpha;m\eta}^{*}\left(\mathbf{k}+\mathbf{q}\right)u_{\mathbf{Q},\alpha;n\eta}\left(\mathbf{k}\right) . \label{eq:M-def}
\end{equation}
In particular, one has \cite{ourpaper3} $[O_{\mathbf{q,G}},O_{\mathbf{q',G'}}]\ne 0$ 
unless $\qq = \qq'$ or $\GG =\GG'$. Therefore, different terms in the interaction Hamiltonian $H_I$ do not commute, unless certain special conditions are satisfied (see the stabilizer code limit defined in Ref.~\cite{ourpaper3}, and the exact solutions in App.~\ref{app:stabilizer}) 

For convenience, we gauge fix the wavefunctions by the TBG symmetries $C_{2z}$, $T$ and $P$, which are discussed in details in Ref.~\cite{ourpaper3}. In the energy band basis, a symmetry $g$ acts as
\begin{equation}
g c_{\mathbf{k},n,\eta' ,s}^{\dagger} g^{-1}= \sum_{m\eta} [B^g(\kk)]_{m\eta,n\eta'} c_{g\mathbf{k},m,\eta, s}^{\dagger}\ ,
\end{equation}
where $B^g(\kk)$ is called the sewing matrix of $g$. Hereafter we use $\zeta^a$ and $\tau^a$ ($a=0,x,y,z$) to denote the identity and Pauli matrices in the energy band $n=\pm1$ space and the valley space, respectively. Throughout this paper, we choose the following $\kk$-independent gauge fixings:
\begin{equation}\label{eq:gauge-0}
B^{C_{2z}T}(\kk)=\zeta^0\tau^0\ ,\qquad B^{C_{2z}P}(\kk)=\zeta^y\tau^y\ ,\qquad B^{C_{2z}}(\kk)=\zeta^0\tau^x\ ,\qquad B^{P}(\kk)=-i\zeta^y\tau^z\ ,
\end{equation}
Furthermore, we require the $\kk$-space continuous condition for the single-particle wavefunctions
\begin{equation}\label{seq:c-continuous}
\lim_{\qq\rightarrow \mathbf{0}}\left|u^\dag_{n,\eta}(\kk+\qq) u_{n,\eta}(\kk)- u^\dag_{-n,\eta}(\kk+\qq) u_{-n,\eta}(\kk)\right|=0
\end{equation} 
for any $\kk$ and $\qq$. Eqs.~(\ref{eq:gauge-0}) and~(\ref{seq:c-continuous}) are shown to be consistent with each other in Ref.~\cite{ourpaper3}. Under the gauge fixing of Eq.~(\ref{eq:gauge-0}), the overlap matrix in Eq.~(\ref{eq:M-def}) in the band-valley space is fixed into the form
\beq 
M(\kk,\qq+\GG) = \zeta^0\tau^0 \alpha_0(\kk,\qq+\GG) + \zeta^x\tau^z \alpha_1(\kk,\qq+\GG) + i\zeta^y\tau^0 \alpha_2(\kk,\qq+\GG) + \zeta^z\tau^z \alpha_3 (\kk,\qq+\GG).  \label{eq:M-para}
\eeq 
where $\alpha_{0,1,2,3}(\kk,\qq+\GG)$ are all real functions. We denote the matrix coefficient of $\alpha_j(\kk,\qq+\GG)$ in Eq.~(\ref{eq:M-para}) as $M_j$. These real functions satisfy the following conditions:
\beq
\alpha_a(\kk,\qq+\GG) = \alpha_a(\kk+\qq,-\qq-\GG)\quad \text{for }a=0,1,3,\qquad
\alpha_2(\kk,\qq+\GG) =-\alpha_2(\kk+\qq,-\qq-\GG), \label{eq:alpha-cond1}
\eeq
\beq
\alpha_a(\kk,\qq+\GG) = \alpha_a(-\kk,-\qq-\GG)\quad \text{for }a=0,2,\qquad
\alpha_a(\kk,\qq+\GG) =-\alpha_a(-\kk,-\qq-\GG)\quad \text{for }a=1,3.  \label{eq:alpha-cond2}
\eeq
In particular, the combination of Eqs.~(\ref{eq:alpha-cond1}) and~(\ref{eq:alpha-cond2}) implies that at $\qq=\mathbf{0}$, we have
\begin{equation}\label{eq:alpha-cond3}
\alpha_0(\kk,\GG)=\alpha_0(-\kk,\GG)\ ,\qquad \alpha_j(\kk,\GG)=-\alpha_j(-\kk,\GG),\quad (j=1,2,3).
\end{equation}

Furthermore, in the (first) chiral limit $w_0=0$ which has the chiral symmetry $C$, given the gauge fixings in Eq.~(\ref{eq:gauge-0}), we can fix the sewing matrix of $C$ into a $\kk$-independent form
\begin{equation}\label{seq:sewing-C}
B^C(\kk)=\zeta^y\tau^z\ .
\end{equation}
It was proven in Ref.~\cite{ourpaper3} that the chiral symmetry $C$ restricts the functions $\alpha_1(\kk,\qq+\GG)=\alpha_3(\kk,\qq+\GG)=0$ in Eq.~(\ref{eq:M-para}), and thus the $M$ matrix coefficient will become
\begin{equation}
M(\kk,\qq+\GG) = \zeta^0\tau^0 \alpha_0(\kk,\qq+\GG) + i\zeta^y\tau^0 \alpha_2(\kk,\qq+\GG).  \label{eq:M-para-chiral}
\end{equation}
We will use this result in the discussions of many-body states in the chiral limit.

\subsubsection{Many-body charge conjugation symmetry of the projected Hamiltonian}

It was also shown in Ref.~\cite{ourpaper3} that the full projected Hamiltonian $H=H_0+H_I$ has a many-body charge-conjugation symmetry $\mathcal{P}_c$, which ensures that all the physical phenomena is PH symmetric about the filling of the charge neutrality point (CNP) at $\nu=0$. The \textit{many-body charge conjugation} $\CC_c$ is defined as the single-particle transformation $C_{2z}TP$ followed by an interchange between electron annihilation operators $c$ and creation operators $c^\dag$, namely,
\beq
\CC_c c_{\kk,n,\eta,s}^\dg \CC_c^{-1}= c_{-\kk,m,\eta^\pr,s} [B^{C_{2z}TP}(\kk)]_{m\eta^\pr,n\eta}(\kk), \qquad
\CC_c c_{\kk,n,\eta,s} \CC_c^{-1}= c_{-\kk,m,\eta^\pr,s}^\dg [B^{C_{2z}TP*}(\kk)]_{m\eta^\pr,n\eta}\ .
\eeq
Under the gauge fixings of \cref{eq:gauge-0}, one has $B^{C_{2z}TP}_{m\eta^\pr, n\eta} = B^{P}_{m\eta^\pr, n\eta} = (-i\zeta^y \tau^z)_{m \eta^\pr, n\eta}$ (Eq.~\ref{eq:gauge-0}). It can then be proved that $\CC_c H_0 \CC_c^{-1}=H_0+\text{const}.$, and $\CC_c O_{\qq,\GG}\CC_c^{-1}=-O_{\qq,\GG}$, which indicates the projected interaction in Eq.~(\ref{seq-pHI}) satisfies $[\CC_c,H_I]=0$. In total, one has the $\CC_c$ symmetry
\beq
\CC_cH\CC_c^{-1}=H+\text{const}.
\eeq
for $H=H_0+H_I$.  $\CC_c$ maps a many-body state at filling $\nu$ to filling $-\nu$, where $\nu$ is the number of electrons per moir\'e unit cell relative to the CNP. This ensures the TBG ground states at $\nu$ and $-\nu$ are PH symmetric.

\subsection{The Chern band basis}\label{app:chernbasis}

A useful electron basis in the discussion of many-body states in this paper is the Chern (band) basis within the lowest two bands (in each valley-spin flavor) $n=\pm1$ defined in Refs.~\cite{ourpaper2,ourpaper3}. Under the gauge fixings of Eqs.~(\ref{eq:gauge-0}) and~(\ref{seq:c-continuous}), it is given by
\beq
d^\dagger_{\kk,e_Y,\eta,s} = \frac{1}{\sqrt2} ( c^\dg_{\kk,+1,\eta,s} + i e_Y c^\dg_{\kk,-1,\eta,s})\ ,
\label{eq:Chern-band}
\eeq
where $e_Y=\pm1$. As proved in Ref.~\cite{ourpaper3}, the Chern basis $d^\dagger_{\kk,e_Y,\eta,s}$ of all $\kk$ for a fixed $e_Y,\eta,s$ correspond to a Chern band carrying Chern number $e_Y$. In total, the projected Hilbert space contains $4$ Chern number $e_Y=+1$ bands, and $4$ Chern number $e_Y=-1$ bands (which are not single-particle energy eigenstates).

\subsection{Symmetry review}\label{app:symmetryreview}
We now review the enhanced continuous symmetries of the TBG Hamiltonian in various limits, which have been proved in Ref.~\cite{ourpaper3}. Hereafter, with the understanding that we assume the gauge fixing given by Eq.~(\ref{eq:gauge-0}), we shall use $\zeta^a$, $\tau^a$, $s^a$ to denote the identity matrix ($a=0$) and Pauli matrices ($a=x,y,z$) in the band $n=\pm1$, valley $\eta=\pm$ and spin $s=\uparrow,\downarrow$ bases, respectively.

\subsubsection{U(2)$\times$U(2) symmetry in the nonchiral-nonflat case}

The total projected Hamiltonian $H=H_0+H_I$ in bands $n=\pm1$ enjoys a U(2)$\times$U(2) symmetry of the spin-charge rotations in each valley. The 8 generators $S^{ab}$ ($a=0,z$, $b=0,x,y,z$) of the U(2)$\times$U(2) symmetry take the form
\begin{equation}\label{seq-U2U2-Sab}
S^{ a b}=\sum_{\mathbf{k},m,\eta,s;n,\eta',s'} (s^{ab})_{m,\eta,s;n,\eta',s'}c_{\mathbf{k},m,\eta,s}^\dag c_{\mathbf{k},n,\eta',s'}\ ,\qquad (a=0,z,\quad b=0,x,y,z)\ ,
\end{equation}
where the matrices
\begin{equation}\label{seq-s0zb}
s^{0b}=\zeta^0\tau^0 s^b,\qquad s^{zb}=\zeta^0\tau^z s^b, \qquad (b=0,x,y,z).
\end{equation}
In particular, $S^{0b}$ and $S^{zb}$ give the global spin-charge U(2) rotations and the valley spin-charge U(2) rotations, respectively.

\subsubsection{U(4) symmetry in the nonchiral-flat limit}\label{app:U4-nc-f}

The nonchiral-flat limit is defined as the limit where the projected kinetic Hamiltonian in Eq.~(\ref{eq:H0-proj}) becomes exactly $H_0=0$, while both $w_0>0$ and $w_1>0$ in Eq.~(\ref{seq-Tj}). In this case, the total projected Hamiltonian is $H=H_I$, and $C_{2z}P$ becomes a symmetry of the system, namely, $[C_{2z}P,H_I]=0$. Note that $C_{2z}P$ preserves the electron momentum $\kk$. This enhances the U(2)$\times$U(2) symmetry in Eq.~(\ref{seq-s0zb}) into a U(4) symmetry. The 16 generators of this U(4) symmetry are
\begin{equation}\label{eq:U4-generator0}
S^{ a b}=\sum_{\mathbf{k},m,\eta,s;n,\eta',s'} (s^{ab})_{m,\eta,s;n,\eta',s'}c_{\mathbf{k},m,\eta,s}^\dag c_{\mathbf{k},n,\eta',s'}\ ,\qquad (a, b=0,x,y,z)\ ,
\end{equation}
where
\begin{equation}\label{eq:U4-generator}
s^{ab}=\{\zeta^0\tau^0 s^b,\ \zeta^y\tau^x s^b,\ \zeta^y\tau^y s^b,\ \zeta^0\tau^z s^b\}, \qquad (a, b=0,x,y,z)\ .
\end{equation}

The U(4) single-electron irreducible representations (irreps) in the nonchiral-flat limit are given by the Chern band basis $d^\dag_{\kk,e_Y,\eta,s}$ at a fixed $\kk$ and $e_Y=\pm1$, where the representation matrices of the U(4) generators when acting on the space of single-electron states $d^\dag_{\kk,e_Y,\eta,s}|0\rangle$ are \begin{equation}\label{seq-SabeY}
s^{ab}(e_Y)=\{ \tau^0s^b,\ e_Y\tau^x s^b,\ e_Y\tau^y s^b,\ \tau^zs^b \}. 
\end{equation}
The irrep occupied by $d^\dag_{\kk,e_Y,\eta,s}$ at a fixed $\kk$ and $e_Y$ is the fundamental irrep $[1]_4$ of the U(4) group (the notation $[1]_4$ will be explained in App.~\ref{app:U4group}). However, we note that the $e_Y=+1$ and $e_Y=-1$ irreps differ by a $\pi$ valley rotation $e^{i\pi\tau^z/2}$ about the $z$ axis.

\subsubsection{U(4)$\times$U(4) symmetry in the (first) chiral-flat limit}\label{app:U(4)U(4)-cf}

In the (first) chiral-flat limit where one has both flat bands $H_0=0$ and the chiral condition that $w_0=0<w_1$, the symmetry of TBG is enhanced into U(4)$\times$U(4). The detailed proof is given in Refs.~\cite{ourpaper3, bultinck_ground_2020}, and we summarize the conclusions here.

At $w_0=0$, the interaction Hamiltonian acquires an additional chiral symmetry $C$, namely, $[C,H_I]=0$. Note that $C$ preserves the electron momentum $\kk$. The symmetries $C$ further enhances the U(4) symmetry in the nonchiral-flat limit (Eq.~(\ref{eq:U4-generator})) into a U(4)$\times$U(4) symmetry. The generators of this U(4)$\times$U(4) group is given by the 16 operators $S^{ab}$ in Eq.~(\ref{eq:U4-generator0}) and the 16 operators $S'^{ab}$ defined by
\begin{equation}\label{eq:U4U4-generator0}
S'^{ a b}=\sum_{\mathbf{k},m,\eta,s;n,\eta',s'} (s'^{ab})_{m,\eta,s;n,\eta',s'}c_{\mathbf{k},m,\eta,s}^\dag c_{\mathbf{k},n,\eta',s'}\ ,\qquad (a, b=0,x,y,z)\ ,
\end{equation}
where
\begin{equation}\label{eq:U4U4-generator1}
s'^{ab}=\{\zeta^y\tau^0 s^b,\ \zeta^0\tau^x s^b,\ \zeta^0\tau^y s^b,\ \zeta^y\tau^z s^b\}, \qquad (a, b=0,x,y,z)\ .
\end{equation}
It is more useful to linear combine the generators into
\begin{equation}
S^{ab}_{\pm}=\sum_{\mathbf{k},m,\eta,s;n,\eta',s'} (s_\pm^{ab})_{m,\eta,s;n,\eta',s'}c_{\mathbf{k},m,\eta,s}^\dag c_{\mathbf{k},n,\eta',s'}\ ,
\end{equation}
where
\begin{equation} 
s^{ab}_\pm=\frac{1}{2}(\zeta^0\pm\zeta^y) \tau^a s^b\ , \qquad
(a,b=0,x,y,z). \label{eq:U4U4-generator}
\end{equation}
In this form, the 16 generators $S_+^{ab}$ generates the first U(4), and the 16 generators $S_-^{ab}$ generates the second U(4), and $[S_+^{ab},S_-^{cd}]=0$. Therefore, in total they give a U(4)$\times$U(4) symmetry in the (first) chiral-flat limit.

The U(4)$\times$U(4) single-electron irreps in the chiral-flat limit are given by the Chern band basis $d^\dag_{\kk,e_Y,\eta,s}$ at a fixed $\kk$ and $e_Y=\pm1$, for which the representation matrices of the U(4)$\times$U(4) generators are
\begin{equation}\label{seq-Sabpm}
 s_\pm^{ab}=\frac{1}{2}\left(1\pm e_Y\right)\tau^a s^b\ . 
\end{equation}
At a fixed $\kk$, the irrep U(4)$\times$U(4) occupied by $d^\dag_{\kk,+1,\eta,s}$ is $([1]_4,[0]_4)$, while the irrep U(4)$\times$U(4) occupied by $d^\dag_{\kk,+1,\eta,s}$ is $([0]_4,[1]_4)$. Here $([\lambda_1]_4,[\lambda_2]_4) $ stands for the U(4)$\times$U(4) irrep given by the tensor product of an irrep $[\lambda_1]_4$ of the first U(4) and an irrep $[\lambda_2]_4$ of the second U(4). More detailed explanations of the irrep notations will be given in App.~\ref{app:U4group}. We note that these two different irreps $([1]_4,[0]_4)$ and $([0]_4,[1]_4)$ subduce into the same nonchiral-flat irrep (but differ by a unitary transformation, see Eq.~(\ref{seq-SabeY})).

\subsubsection{U(4) symmetry in the (first) chiral-nonflat limit}\label{app:U4-c-nf}

In Ref.~\cite{ourpaper3} it was proved that, in the (first) chiral-nonflat limit where $w_0=0<w_1$ but the bands are no longer exactly flat, namely $H_0\neq0$, the Hamiltonian still has a remaining U(4) symmetry. This is due to the combined symmetry $CC_{2z}P$ of the Hamiltonian. The 16 generators of this U(4) group are given by a subset of the U(4)$\times$U(4) generators in Eq.~(\ref{eq:U4U4-generator}), and we redefine their notations as
\begin{equation}
\widetilde{S}^{ab}=\sum_{\mathbf{k},m,\eta,s;n,\eta',s'} (\tilde{s}^{ab})_{m,\eta,s;n,\eta',s'}c_{\mathbf{k},m,\eta,s}^\dag c_{\mathbf{k},n,\eta',s'}\ ,
\end{equation}
where
\begin{equation}\label{eq:U4-generator-chiral-nonflat}
\tilde{s}^{ab}= \zeta^0 \tau^a s^b,\qquad (a,b=0,x,y,z) \ .
\end{equation}
Note that the chiral-nonflat U(4) symmetry here is simply the valley-spin rotation symmetry without transformations in the band basis, which is different from the nonchiral-flat U(4) symmetry in Eq.~(\ref{eq:U4-generator}).

Since the generators are all proportional to $\zeta^0$, either the energy band basis $c^\dag_{\kk,n,\eta,s}$ of a fixed band $n=\pm1$ or the Chern band basis $d^\dag_{\kk,e_Y,\eta,s}$ of a fixed $e_Y=\pm1$ at certain momentum $\kk$ is occupying a single-electron fundamental U(4) irrep $[1]_4$ in the chiral-nonflat limit. We note that these U(4) irreps are simply the subduction of the U(4)$\times$U(4) irreps into its chiral-nonflat U(4) subgroup. 
The representation matrices of the U(4) generators are
\begin{equation}
\tau^as^b \ ,\qquad (a,b=0,x,y,z)\ .
\end{equation}

\section{Brief Review of the U(4) group}\label{app:U4group}

In this appendix, we briefly review the group representations of the U(4) and U(4)$\times$U(4) groups, which will be useful for our discussions of many-body states in this paper.

\subsection{The U(4) Irreps}\label{app:U4irrep}

The U($N$) group is defined by all the $N\times N$ unitary matrices $U$ satisfying $U^\dag U=I_{N}$, where $I_N$ is the identity matrix. The matrices $U$ are generated by all the linearly independent $N\times N$ Hermitian matrices, thus the total number of generators is $N^2$. In particular, for the U(4) group, the 16 generators can be represented by the tensor product of two sets of $2\times2$ identity and Pauli matrices $\tau^a$ and $s^a$ ($a=0,x,y,z$) as
\begin{equation}\label{seq:U(4)-fundamental}
s_0^{ab}=\tau^a s^b\ ,\qquad (a,b=0,x,y,z)\ .
\end{equation}
We denote their commutation relations as
\begin{equation}\label{seq:U4-structure-const}
[s_0^{ab},s_0^{cd}]=f^{ab,cd}_{ef} s_0^{ef}\ .
\end{equation}
Then $f^{ab,cd}_{ef}$ are the group structure constants, which are the same for all representations of U(4) group.

The representations of the U($N$) group are the same as that of the SU($N$) group plus a U(1) generator which is proportional to the identity matrix. The basis for the irreps are the same for U($N$) and SU($N$). Therefore, it is sufficient to discuss the SU($N$) irreps, which we will briefly review (in particular for $N=4$) in this appendix.

The set of all the $N\times N$ traceless matrices $U$ defines the $N$-dimensional fundamental irreducible representation (irrep) of the SU($N$) group, and the representation matrices of the SU($N$) generators are given by all the linearly independent traceless Hermitian $N\times N$ matrices. These matrices act on an $N$-dimensional complex vector basis $V_a$ ($1\le a\le N$). For the SU(4) group, the $15$ generators in the fundamental irrep are exactly given by Eq.~(\ref{seq:U(4)-fundamental}) with $ab\neq0$. There is also a 1-dimensional trivial identity irrep for group SU($N$), in which the representation matrices of all the SU($N$) generators are given by $0$. 

In this paper, we shall use the following notations to denote the fundamental irrep and trivial identity irrep of the U(4) group (which is the same as that of SU(4), except that there is an additional U(1) generator):
\begin{equation}
\text{U(4) fundamental irrep:}\qquad [1]_4\ ,\qquad \qquad  \text{U(4) trivial identity irrep:}\qquad [0]_4\ ,
\end{equation}
which will be explained below. In particular, we assume the additional U(1) generator $S^{00}$ (compared to SU(4)) has a representation matrix $s^{00}_0$ given by Eq.~(\ref{seq:U(4)-fundamental}) in the fundamental U(4) irrep $[1]_4$, while its representation matrix is simply $0$ in the trivial identity irrep $[0]_4$.

\begin{figure}[htbp]
\begin{centering}
\includegraphics[width=0.7\linewidth]{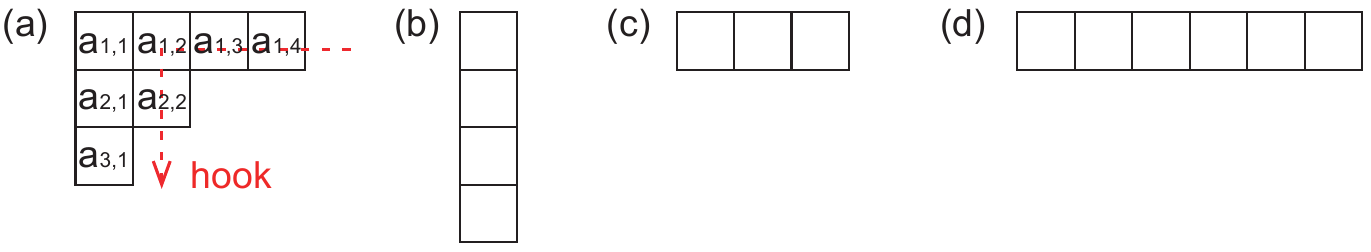}
\end{centering}
\caption{Young tableau.}
\label{fig:young}
\end{figure}

The irreducible representations (irreps) of the SU($N$) group can be labeled by the Young tableau as shown in Fig.~\ref{fig:young}. All the irreps of SU($N$) can be obtained by decomposition of the tensor product of the fundamental $N$-dimensional SU($N$) representation. Assume the fundamental SU($N$) representation acts on the basis of an $N$-dimensional complex vector $V_a$ with component index $1\le a\le N$. A rank-$m$ tensor product representation then has a tensor basis
\begin{equation}\label{eq:tensorrep}
V_{a_1 a_2\cdots a_m}=V_{a_1}V_{a_2}\cdots V_{a_m}\ .
\end{equation}
These basis form a reducible representation. It can be reduced into an irrep by symmetrization among a subset of indices and anti-symmetrization among the rest indices, which can be conveniently represented by a Young tableau. As shown in Fig.~\ref{fig:young}(a), a Young tableau consists of rows of boxes, where the number of boxes in the $i$-th row is no smaller than that in the $(i+1)$-th row. It can be conveniently denoted by 
\begin{equation}\label{eq:UNirrep}
[\lambda_1,\lambda_2,\cdots]_N, 
\end{equation}
where $\lambda_i$ is the number of boxes in row $i$ ($\lambda_i\ge\lambda_{i+1}$). Note that in general, only the first $N-1$ rows are provided in such a notation. For convenience, we will add the $N$-th row (the number of boxes will then match the number of particles, i.e. the U(1) charge) and we will identify $[\lambda_1,\lambda_2,\cdots,\lambda_N]_N$ with $[\lambda_1-\lambda_N,\lambda_2-\lambda_N,\cdots,\lambda_{N-1}-\lambda_N]_N$. By putting a number $a_{i,j}$ in each box at row $i$ and column $j$, we obtain a basis 
\begin{equation}\label{eq:young-basis}
V^{[\lambda_1,\lambda_2,\cdots]_N}_{a_{11},a_{12}\cdots}
\end{equation}
with indices $a_{i,j}$, which is obtained by symmetrizing/anti-symmetrizing the indices of the tensor basis Eq.~(\ref{eq:tensorrep}) of rank $\sum_i\lambda_i$, such that the indices in the same row are symmetric among each other, and the indices in the same column are antisymmetric among each other. A complete set of the independent basis of this irrep is given by all the possible indices satisfying
\begin{equation}\label{eq:irrep-indices}
1\le a_{i,j}\le a_{i,j+1}\le N\ ,\qquad  1\le a_{i,j}< a_{i+1,j}\le N\ .
\end{equation}
It is then clear that a Young tableau cannot have number of rows larger than $N$.

The number of set of indices $\{a_{i,j}\}$ satisfying Eq.~(\ref{eq:irrep-indices}) gives the dimension of the irrep. Alternatively, the dimension of an SU($N$) irrep represented by a Young tableau $[\lambda_1,\lambda_2,\cdots]_N$ can be conveniently computed by the \emph{Hook Rule}:
\begin{equation}
d_{[\lambda_1,\lambda_2,\cdots]_N}=\prod_{i,j}\frac{N+j-i}{h_{ij}}\ ,
\end{equation}
where $(i,j)$ runs over all the boxes at row $i$ and column $j$ of the Young tableau, and $h_{ij}$ is the \emph{hook number} of box $(i,j)$ defined as follows: one first defines the hook of box $(i,j)$ as the path starting from the rightmost box of row $i$ going leftwards to box $(i,j)$, and then going downwards to the end of row $j$ (see red arrowed dashed line in Fig.~\ref{fig:young}(a)). The hook number $h_{ij}$ is then the number of boxes passed by the hook of box $(i,j)$.

For convenience, if a Young tableau consists of $p$ identical rows of length $\lambda$, we denote the corresponding irrep $[\lambda,\lambda,\cdots]_N$ in short as $[\lambda^p]_N$. Note that by this notation, $[\lambda^0]_N=[0]_N$ is the one-dimensional trivial irrep.

There are two special cases: a Young tableau $[\lambda]_N$ with only one row of $\lambda$ boxes has all the indices of basis Eq.~(\ref{eq:young-basis}) symmetric; while a Young tableau $[1^p]_N$ with only one column ($m$ rows) has all the indices antisymmetric. Their corresponding irrep dimensions are given by
\begin{equation}
d_{[\lambda]_N}=\frac{(N+\lambda-1)!}{\lambda !(N-1)!}\ ,\qquad  d_{[1^p]_N}=\frac{N!}{p !(N-p)!}\ .
\end{equation}
In particular, $[1]_N$ is the fundamental irrep of SU(N), $[1^N]_N$ is an SU($N$) singlet irrep identical to $[0]_N$, and $[1^{N-p}]_N$ is the conjugate irrep of $[1^p]_N$. Lastly, We note that our notation for the fundamental irrep $[1]_N$ and identity irrep $[0]_N$ are simply special cases of the general notation Eq.~(\ref{eq:UNirrep}) for SU($N$) or U($N$) irreps.

\subsection{The U(4)$\times$U(4) Irreps}\label{app:u4u4irreps}

The irreps of the U(4)$\times$U(4) group are simply given by the tensor products of an irrep $[\{\lambda_{1,i}\}]_4=[\lambda_{1,1},\lambda_{1,2},\cdots]_4$ of the first U(4) and an irrep $[\{\lambda_{2,i}\}]_4=[\lambda_{2,1},\lambda_{2,2},\cdots]_4$ of the second U(4), where the notations for U(4) irreps are from Young tableau as we explained in Eq.~(\ref{eq:UNirrep}). We denote such a U(4)$\times$U(4) irrep as 
\begin{equation}
([\{\lambda_{1,i}\}]_4,[\{\lambda_{2,i}\}]_4)\ .
\end{equation}

\section{Exact Ground States in Different Limits}\label{app:exact-GS}

In this appendix, we show that exact ground states can be derived in different limits at appropriate integer fillings $\nu$ (the number of electrons per moir\'e unit cell). All of these exact ground states are of the form of many-body Fock states.

\subsection{Chemical potential shift}\label{app:chempotentialshift}

To identify the exact ground states at certain filling $\nu$, we note that interaction can in general be rewritten as
\begin{equation}
H_{I}=\frac{1}{2\Omega_{\text{tot}}}\sum_{\mathbf{G}\in\mathcal{Q}_{0}}\left[\left(\sum_{\mathbf{q}} (O_{\mathbf{q,G}}-A_{\GG} N_M\delta_{\mathbf{q},0}) (O_{\mathbf{-q,-G}}-A_{-\GG} N_M\delta_{-\mathbf{q},0}) \right) + 2A_{-\GG} N_M O_{\mathbf{0,G}} -A_{-\GG}A_\GG N_M^2 \right]\ ,
\label{eq:shifted-HI}
\end{equation}
where $N_M$ is the total number of moir\'e unit cells, and $A_{\GG}$ is some arbitrarily chosen $\GG$ dependent coefficient satisfying $A_\GG=A_{-\GG}^*$. Note that the first term in Eq.~(\ref{eq:shifted-HI}) is semi-positive definite.

If we further have the following condition - which we call the "flat metric condition" (see Eq.~(\ref{eq-Mq=0}), see also Ref.~\cite{ourpaper5}) that the form factors
\begin{equation}\label{eqn-condition-at-nu}
\boxed{\text{Flat Metric Condition:}\qquad
M_{m,n}^{\left(\eta\right)}\left(\mathbf{k},\mathbf{G}\right)=\xi(\mathbf{G})\delta_{m,n}
}
\end{equation}
is independent of $\mathbf{k}$ (which is always true for $\mathbf{G}=\mathbf{0}$, but generically not true for $\mathbf{G}\neq\mathbf{0}$), one would have the term $O_{\mathbf{0,G}}$ proportional to the total electron number $N$. Accordingly, the second term in Eq.~(\ref{eq:shifted-HI}) term 
\begin{equation}
\frac{1}{2\Omega_{\text{tot}}}\sum_{\GG}2A_{-\GG} N_M O_{\mathbf{0,G}}= \frac{1}{\Omega_M}\left(\sum_{\GG}A_{-\GG} \sqrt{V(\mathbf{G})}\xi(\GG)\right)\sum_{\kk,m,\eta,s} \left(c^\dag_{\mathbf{k},m,\eta,s}c_{\mathbf{k},m,\eta,s} -\frac{1}{2}\right)
\end{equation}
is simply a chemical potential term with chemical potential 
\begin{equation}\label{eq:chemical-potential-shift}
\mu=\frac{1}{N_M\Omega_M}\sum_{\GG}A_{-\GG} \sqrt{V(\mathbf{G})}\sum_{\kk}M_{+1,+1}^{\left(\eta\right)}\left(\mathbf{k},\mathbf{G}\right)=\sum_{\GG}A_{-\GG} \sqrt{V(\mathbf{G})}\xi(\GG)/\Omega_M\ ,
\end{equation}
where $\Omega_M=\Omega_{\text{tot}}/N_M$ is the area of moir\'e unit cell. For a fixed total number of electrons, $N=\sum_{\kk,m,\eta,s} c^\dag_{\mathbf{k},m,\eta,s}c_{\mathbf{k},m,\eta,s} =(\nu+4) N_M$ is a constant, where $\nu$ is the filling fraction (number of doped electrons per moir\'e unit cell) relative to the charge neutrality point (CNP). In this case, the second term and third term in Eq.~(\ref{eq:shifted-HI}) are all constant, and the ground state is solely determined by the first semi-positive definite term. In particular, if a state $|\Psi\rangle$ at certain filling $\nu$ satisfies
\begin{equation}
(O_{\mathbf{q,G}}-A_{\GG} N_M\delta_{\mathbf{q},0}) |\Psi\rangle =0
\end{equation}
for any $\qq$, $\GG$ with some chosen coefficients $A_\GG$, the state $|\Psi\rangle$ is necessarily a ground state of the Hamiltonian $H_I$ at filling $\nu$, if the flat metric condition Eq.~(\ref{eq-Mq=0}) is satisfied. Otherwise, it will be an eigenstate, but not necesarily a ground state.

In the below, we discuss the exact ground states of the (first) chiral-flat U(4)$\times$U(4) limit and the nonchiral-flat U(4) limit, respectively. As we will show, this requires a suitable choice of coefficients $A_\GG$ in Eq.~(\ref{eq:shifted-HI}) depending on the filling fraction.

\subsection{Exact ground states in the (first) chiral-flat U(4)$\times$U(4) limit}\label{app:exactgschiralflat}

We first discuss the (first) chiral-flat limit with exact flat bands and chiral symmetry, where the projected Hamiltonian has the highest U(4)$\times$U(4) symmetry with generators in Eq.~(\ref{eq:U4U4-generator}). We note that in addition to the unitary U(4)$\times$U(4) symmetry, there is also the anti-unitary time-reversal symmetry $T$.

As we have discussed in App.~\ref{app:U(4)U(4)-cf}, the single-particle U(4)$\times$U(4) irreps in the chiral-flat limit are given by the Chern band basis (\ref{eq:Chern-band}). Because of the chiral symmetry, the coefficients $M_{m,n}^{\left(\eta\right)}(\mathbf{k,q})$ satisfies Eq.~(\ref{eq:M-para-chiral}), so the operator $O_{\mathbf{q,G}}$ in Eq.~(\ref{seq-OqG}) can be rewritten under the Chern band basis (\ref{eq:Chern-band}) as Eq.~(\ref{eq-OqG0}), which we reprint here for convenience:
\begin{equation}\label{eq:chiral-OqG}
O_{\mathbf{q,G}}=O_{\mathbf{q,G}}^0=\sum_{\mathbf{k},e_Y,\eta,s} \sqrt{V(\kk+\GG)}M_{e_Y}(\kk,\qq+\GG)\left(d^\dagger_{\kk+\qq,e_Y,\eta,s} d_{\kk,e_Y,\eta,s}-\frac{1}{2}\delta_{\mathbf{q,0}}\right)\ .
\end{equation}
It is diagonal in the valley index $\eta$, spin index $s$ and Chern band index $e_Y$, and we have defined the coefficient
\begin{equation}\label{eq:chiral-OqG0}
M_{e_Y}(\kk,\qq+\GG)=\alpha_0 \left(\mathbf{k},\mathbf{q}+\mathbf{G}\right)+ie_Y\alpha_2 \left(\mathbf{k},\mathbf{q}+\mathbf{G}\right)\ ,
\end{equation}
which satisfies $M_{e_Y}(\kk,\qq+\GG)=M_{e_Y}^*(\kk+\qq,-\qq-\GG)$ due to Eq.~(\ref{eq:alpha-cond1}). 

We now discuss the ground state at integer filling $\nu$. Since we have assumed the flat condition, the projected kinetic term $H_0=0$, and the Hamiltonian is solely the interaction term $H=H_I$.

\subsubsection{Chern insulator eigenstates without the flat metric condition (\ref{eq-Mq=0})}\label{app:c-f-eigenstates}
We first prove that the following Fock state of integer Chern number $\nu_C$ with integer $(\nu+4)$ fully occupied Chern bands is an eigenstate of $H_I$ (without assuming the flat metric condition Eq.~(\ref{eq-Mq=0})):
\begin{equation}\label{eq:U(4)U(4)-GS}
\boxed{
|\Psi_{\nu}^{\nu_+,\nu_-}\rangle =\prod_{\mathbf{k}\in\text{MBZ}} \left(\prod_{j_1=1}^{\nu_+}d^\dag_{\mathbf{k},+1,\eta_{j_1},s_{j_1}} \prod_{j_2=1}^{\nu_-}d^\dag_{\mathbf{k},-1,\eta_{j_2}',s_{j_2}'}\right)|0\rangle
}
\end{equation}
where 
\begin{equation}
\nu_+-\nu_-=\nu_C
\end{equation}
is the total Chern number (integer) of the state, and 
\begin{equation}
\nu_++\nu_-=\nu+4
\end{equation}
is the total number of electrons per moir\'e unit cell in the projected bands, with $0\le \nu_\pm\le 4$, and $\kk$ runs over the entire moir\'e BZ. Here $\nu_\pm$, $\nu$ and $\nu_C$ are all integers. The occupied spin/valley indices $\{\eta_{j_1},s_{j_1}\}$ and $\{\eta_{j_2}',s_{j_2}'\}$ can be arbitrarily chosen. To see it is an eigenstate, we can calculate the following expression:
\begin{equation}\label{eq:U(4)U(4)-OqG-action}
\begin{split}
&O_{\mathbf{q,G}}|\Psi_{\nu}^{\nu_+,\nu_-}\rangle \\
=&\sum_{\mathbf{k},e_Y,\eta,s} \sqrt{V(\mathbf{G}+\mathbf{q})} [\alpha_0(\mathbf{k,q+G})+ie_Y\alpha_2(\mathbf{k,q+G})]\left(d^\dagger_{\kk+\qq,e_Y,\eta,s} d_{\kk,e_Y,\eta,s}-\frac{1}{2}\delta_{\mathbf{q,0}}\right)|\Psi_{\nu}^{\nu_+,\nu_-}\rangle \\
=& \delta_{\mathbf{q,0}} \sum_{\mathbf{k},e_Y,\eta,s} \sqrt{V(\mathbf{G})} [\alpha_0(\mathbf{k,G})+ie_Y\alpha_2(\mathbf{k,G})]\left(d^\dagger_{\kk,e_Y,\eta,s} d_{\kk,e_Y,\eta,s}-\frac{1}{2}\right)|\Psi_{\nu}^{\nu_+,\nu_-}\rangle \\
=& \sqrt{V(\mathbf{G})} \delta_{\mathbf{q,0}}  \left(\sum_\kk \left[(\nu_++\nu_--4)\alpha_0(\mathbf{k,G})+i(\nu_+-\nu_-)\alpha_2(\mathbf{k,G})\right]\right) |\Psi_{\nu}^{\nu_+,\nu_-}\rangle \\
=& \sqrt{V(\mathbf{G})} \delta_{\mathbf{q,0}}  \left(\sum_\kk \left[\nu\alpha_0(\mathbf{k,G})+i\nu_C\alpha_2(\mathbf{k,G})\right]\right)|\Psi_{\nu}^{\nu_+,\nu_-}\rangle \\
=& \sqrt{V(\mathbf{G})} \delta_{\mathbf{q,0}}  \sum_\kk \nu\alpha_0(\mathbf{k,G})|\Psi_{\nu}^{\nu_+,\nu_-}\rangle =\delta_{\mathbf{q,0}} A_\mathbf{G}N_M |\Psi_{\nu}^{\nu_+,\nu_-}\rangle\ ,
\end{split}
\end{equation}
where in the last step we have used the fact that $\alpha_2(\kk,\GG)=-\alpha_2(-\kk,\GG)$ as given in Eq.~(\ref{eq:alpha-cond3}), and we have defined 
\begin{equation}\label{eq:AG-general}
A_\mathbf{G}=\frac{1}{N_M}\sqrt{V(\mathbf{G})}\sum_\kk \nu\alpha_0(\mathbf{k,G})\ .
\end{equation} 
Therefore, we find $|\Psi_{\nu}^{\nu_+,\nu_-}\rangle$ is an eigenstate of operator $O_{\mathbf{q,G}}$, where the eigenvalue is zero if $\mathbf{q\neq 0}$, and is nonzero if $\mathbf{q=0}$. Since the interaction Hamiltonian $H_I$ is a quadratic form of $O_{\mathbf{q,G}}$, we conclude that this state Eq.~(\ref{eq:U(4)U(4)-GS}) is an eigenstate of the Hamiltonian in the chiral-flat limit, and the eigenvalue is given by 
\begin{equation}\label{eq:U(4)U(4)-GSenergy}
H_I|\Psi_{\nu}^{\nu_+,\nu_-}\rangle=\frac{1}{2\Omega_{\text{tot}}}\sum_{\qq,\GG}O_{\mathbf{-q,-G}}O_{\mathbf{q,G}} |\Psi_{\nu}^{\nu_+,\nu_-}\rangle = \frac{\nu^2}{2\Omega_{\text{tot}}}\sum_{\GG} V(\mathbf{G}) \Big(\sum_\kk\alpha_0(\mathbf{k,G})\Big)^2 |\Psi_{\nu}^{\nu_+,\nu_-}\rangle\ ,
\end{equation}
where we have used the fact that $\alpha_0(\mathbf{k,G})=\alpha_0(\mathbf{k,-G})$ is real (Eq.~(\ref{eq:alpha-cond1})). 

Any U(4)$\times$U(4) rotation of this state is also an eigenstate degenerate with this state, and together they form a U(4)$\times$U(4) multiplet. The U(4)$\times$U(4) irrep of this multiplet is given by $\left([N_M^{\nu_+}]_4,[N_M^{\nu_-}]_4\right)$ (the irreps of U(4) and U(4)$\times$U(4) are reviewed via the formalism of Young tableau in App.~\ref{app:U4group}). To see this, we first recall that each Chern number $e_Y=+1$ electron occupies a fundamental irrep $[1]_4$ in the first U(4), which is represented by a Young tableau of one box. In the subspace of Chern number $e_Y=+1$ bands, the wavefunctions of $N_M$ occupied electrons in each fixed valley-spin flavor $\{\eta_j,s_j\}$ are antisymmetric in $\kk$ and symmetric in the valley-spin indices, thus should occupy $N_M$ boxes in the same row in a Young tableau of the first U(4) (recall that U(4) is defined in the valley-spin space). Meanwhile, the wavefunctions of several electrons in the same $\kk$ but different valley-spin favors are symmetric in $\kk$ and antisymmetric in valley-spin indices, thus should occupy boxes in the same column in a Young tableau of the first U(4). This shows the irrep of the multiplet of states $|\Psi_{\nu}^{\nu_+,\nu_-}\rangle$ should occupy an irrep $[N_M^{\nu_+}]_4$ of the first U(4). Similarly, it should occupy an irrep $[N_M^{\nu_-}]_4$ of the second U(4), thus its U(4)$\times$U(4) irrep is given by $\left([N_M^{\nu_+}]_4,[N_M^{\nu_-}]_4\right)$.

In particular, for a fixed filling factor $\nu$, from Eq.~(\ref{eq:U(4)U(4)-GSenergy}) we find that the states with different Chern number $\nu_C$ are all degenerate.

In the special case of charge neutrality $\nu=0$, the U(4)$\times$U(4) multiplet of eigenstate state $|\Psi_0^{\nu_+,\nu_-}\rangle$ with Chern number $\nu_C=\nu_+-\nu_-=0,\pm 2,\pm 4$ has exactly zero energy. Therefore, all these $\nu_C=\nu_+-\nu_-=0,\pm 2,\pm 4$ states are exact degenerate ground states. 

At nonzero fillings $\nu$, generically we cannot guarantee that these eigenstates are ground states (without the flat metric condition Eq.~(\ref{eq-Mq=0})).

\subsubsection{Chern insulator ground states with the flat metric condition (\ref{eq-Mq=0})}

When the the flat metric condition (\ref{eq-Mq=0}) is satisfied, namely, when
\begin{equation}\label{eq:chiral-condition-alpha}
\alpha_0(\mathbf{k,G})=\xi(\GG)\ ,\qquad \alpha_2(\mathbf{k,G})=0 
\end{equation}
for any $\mathbf{k,G}$ in the chiral limit, the eigenstates in Eq.~(\ref{eq:U(4)U(4)-GS}) at nonzero integer fillings $\nu\neq0$ become exact ground states. Then we can rewrite the interaction into the form of Eq.~(\ref{eq:shifted-HI}), and the coefficient $A_\GG$ in Eq.~(\ref{eq:U(4)U(4)-OqG-action}) can be simpliefied as
\begin{equation}\label{eq:condition-AG}
\boxed{
A_\GG=\frac{1}{N_M}\sqrt{V(\mathbf{G})}\sum_\kk \nu\alpha_0(\mathbf{k,G})=\nu\sqrt{V(\mathbf{G})}\xi(\mathbf{G})\ .
}
\end{equation}
By Eq.~(\ref{eq:U(4)U(4)-OqG-action}), we then have
\begin{equation}
(O_{\mathbf{q,G}}-A_{\GG} N_M\delta_{\mathbf{q},0})|\Psi_{\nu}^{\nu_+,\nu_-}\rangle =0\ ,
\end{equation}
for any $\nu_C=\nu_+-\nu_-$. Therefore, we find the first nonnegative term in the rewritten Hamiltonian Eq.~(\ref{eq:shifted-HI}) annihilates the state $|\Psi_{\nu}^{\nu_+,\nu_-}\rangle$, and thus all the eigenstates $|\Psi_{\nu}^{\nu_+,\nu_-}\rangle$ with any Chern number $\nu_C=\nu_+-\nu_-$ are degenerate ground states at filling $\nu$.

These ground states $|\Psi_{\nu}^{\nu_+,\nu_-}\rangle$ will generically be insulators with gapped charge excitations, as we will demonstrate analytically and numerically in Refs.~\cite{ourpaper5,ourpaper6}. This is because there is no remaining symmetry protecting a gapless electron spectrum. At integer fillings $\nu$, the electron spectrum in valley $\eta$ can be gapless only if valley $\eta$ (for a fixed spin $s$) is half-filled and the spinless $C_{2z}T$ symmetry is preserved, so that the $C_{2z}T$ protected fragile topology \cite{po_origin_2018,song_all_2019,po_faithful_2019,ahn_failure_2019,lian2020,po_fragile_2018,cano_fragile_2018,Slager2019WL} of TBG protects the existence of two gapless Dirac points. This is never satisfied by state $|\Psi_{\nu}^{\nu_+,\nu_-}\rangle$, since if a valley $\eta$ (of a given spin $s$) is half-filled, the electrons always fully occupy one Chern band, which breaks the $C_{2z}T$ symmetry because of the Chern number of the Chern band.

Lastly, we note that the eigenstates $|\Psi_{\nu}^{\nu_+,\nu_-}\rangle$ at integer fillings $\nu\neq 0$ would remain the exact ground states of chiral-flat limit if the flat metric condition Eq.~(\ref{eq-Mq=0}) is weakly broken. This is because $|\Psi_{\nu}^{\nu_+,\nu_-}\rangle$ are eigenstates of the Hamiltonian $H=H_I$ regardless of the flat metric condition Eq.~(\ref{eq-Mq=0}), so the ground states will not change unless the flat metric condition Eq.~(\ref{eq-Mq=0}) is largely broken such that other eigenstates (which are above states $|\Psi_{\nu}^{\nu_+,\nu_-}\rangle$ by a finite gap when the flat metric condition Eq.~(\ref{eq-Mq=0}) is satisfied) are brought down to energies lower than that of states $|\Psi_{\nu}^{\nu_+,\nu_-}\rangle$. One could believe that since the spectrum is gapless due to the FM goldstone \cite{ourpaper5}, the state $|\Psi_{\nu}^{\nu_+,\nu_-}\rangle$ could stop being the ground state as soon as leaving the flat metric condition Eq.~(\ref{eq-Mq=0}). However, this is not true, as the "Goldstone" excitation spectrum moves with the state when  the flat metric condition Eq.~(\ref{eq-Mq=0}) is broken \cite{ourpaper5}.

\subsection{Exact ground states in the nonchiral-flat U(4) limit at even fillings}\label{app:exactgsnonchiralflat}

In this subsection we turn to the nonchiral-flat case, which has a U(4) symmetry with generators in Eq.~(\ref{eq:U4-generator}). Without the chiral symmetry, $O_{\mathbf{q,G}}$ is no longer diagonal in the Chern band basis (in the form of Eq.~(\ref{eq:chiral-OqG})) or any certain band basis. Nevertheless, $O_{\mathbf{q,G}}$ is still diagonal in $\eta$ and $s$. More explicitly, under the Chern band basis $d^\dag_{\mathbf{k},e_Y,\eta,s}$ in Eq.~(\ref{eq:Chern-band}), using Eq.~(\ref{eq:M-para}), we can rewrite operator $O_{\mathbf{q,G}}$ as
\begin{equation}\label{seq-OqG-nf}
O_{\mathbf{q,G}}=O_{\mathbf{q,G}}^0+O_{\mathbf{q,G}}^1, 
\end{equation}
where $O_{\mathbf{q,G}}^0$ is defined in Eq.~(\ref{eq:chiral-OqG}), and
\begin{equation}\label{seq-OqG1}
\begin{split}
O_{\mathbf{q,G}}^1=&\sum_{\mathbf{k}\eta s}\sum_{e_Y=\pm1} \eta \sqrt{V(\qq+\GG)}F_{e_Y}(\kk,\qq+\GG)  d^\dag_{\mathbf{k+q},-e_Y,\eta,s}d_{\mathbf{k},e_Y,\eta,s},
\end{split}
\end{equation}
with coefficients defined by
\begin{equation}\label{eq:nonchiral-OqG1}
F_{e_Y}(\kk,\qq+\GG)=\alpha_1 \left(\mathbf{k},\mathbf{q}+\mathbf{G}\right)+ie_Y\alpha_3 \left(\mathbf{k},\mathbf{q}+\mathbf{G}\right)\ .
\end{equation}
The term $O_{\mathbf{q,G}}^1$ therefore is not diagonal in the Chern band basis, and only arises when $w_0>0$.

In this case, we cannot obtain analytical exact ground states (neither eigenstates)
at odd integer fillings $\nu=\pm1,\pm3$. However, for even fillings $\nu=0,\pm2,\pm4$, one can still write down the following Chern number $\nu_C=0$ eigenstate:
\begin{equation}\label{eq:U(4)-GS}
\boxed{
|\Psi_\nu\rangle=\prod_{\mathbf{k}} \left(\prod_{j=1}^{(\nu+4)/2}c^\dag_{\mathbf{k},+,\eta_{j},s_{j}} c^\dag_{\mathbf{k},-,\eta_{j},s_{j}}\right)|0\rangle =\prod_{\mathbf{k}} \left(\prod_{j=1}^{(\nu+4)/2}d^\dag_{\mathbf{k},+1,\eta_{j},s_{j}} d^\dag_{\mathbf{k},-1,\eta_{j},s_{j}}\right)|0\rangle\ ,
}
\end{equation}
where $\{\eta_j,s_j\}$ are distinct valley-spin flavors which are fully occupied. Here we have expressed the state both in the energy band basis $c^\dag_{\mathbf{k},m,\eta,s}$ and in the Chern band basis $d^\dag_{\mathbf{k},e_Y,\eta,s}$. To see it is an eigenstate, we note that each spin-valley flavor is either fully occupied or fully empty, from which we find
\begin{equation}
\begin{split}
&O_{\mathbf{q,G}}|\Psi_\nu\rangle=\sqrt{V(\qq+\GG)}\\
&\ \ \times \sum_{\mathbf{k},e_Y,\eta,s}\left[M_{e_Y}(\kk,\qq+\GG)\left(d^\dagger_{\kk+\qq,e_Y,\eta,s} d_{\kk,e_Y,\eta,s}-\frac{1}{2}\delta_{\mathbf{q,0}}\right) +\eta F_{e_Y}(\kk,\qq+\GG)  d^\dag_{\mathbf{k+q},-e_Y,\eta,s}d_{\mathbf{k},e_Y,\eta,s} \right] |\Psi_\nu\rangle \\
&=\nu \sqrt{V(\mathbf{G})}\delta_{\mathbf{q,0}} \sum_{\kk,m,\eta,s} \alpha_0\left(\mathbf{k},\mathbf{G}\right) |\Psi_\nu\rangle\ ,
\end{split}
\end{equation}
where we have used the properties of $\alpha_j\left(\mathbf{k},\mathbf{G}\right)$ in Eq.~(\ref{eq:alpha-cond3}). 
This shows that the state $|\Psi_\nu\rangle$ is an eigenstate of $O_{\mathbf{q,G}}$, and therefore $|\Psi_\nu\rangle$ is an eigenstate of the interaction Hamiltonian $H_I$. Note that the flat metric condition Eq.~(\ref{eq-Mq=0}) is not needed for $|\Psi_\nu\rangle$ to be an eigenstate. 

Any U(4) rotation of the state $|\Psi_{\nu}\rangle$ is also an eigenstate. The multiplet of $|\Psi_\nu\rangle$ here occupies a nonchiral-flat U(4) irrep $[(2N_M)^{(\nu+4)/2}]_4$. This is because the $2N_M$ electrons occupying the same valley-spin flavor in state $|\Psi_{\nu}\rangle$ of Eq.~(\ref{eq:U(4)-GS}) are symmetric in valley-spin indices, thus occupy the same row of Young tableau of the U(4) irrep; while the electrons at the same $\kk$ and Chern band $e_Y$ but in different valley-spin indices have wavefunctions antisymmetric in valley-spin, thus they occupy the same column in the Young tableau of U(4) irrep.

From the expression in the Chern band basis, it is clear that the U(4) multiplet of ground state $|\Psi_\nu\rangle$ here in Eq.~(\ref{eq:U(4)-GS}) is a subset of the U(4)$\times$U(4) multiplet of state $|\Psi_{\nu}^{\nu_+,\nu_-}\rangle$ with $\nu_+=\nu_-=(\nu+4)/2$ in Eq.~(\ref{eq:U(4)U(4)-GS}). 

If the flat metric condition Eq.~(\ref{eq-Mq=0}) is further satisfied, we can rewrite the interaction Hamiltonian into the form of Eq.~(\ref{eq:shifted-HI}) by choosing coefficient $A_\GG$ as given in Eq.~(\ref{eq:condition-AG}). After this, we find $(O_{\mathbf{q,G}}-A_{\GG} N_M\delta_{\mathbf{q},0})|\Psi_{\nu}\rangle =0$, which indicates $|\Psi_{\nu}\rangle$ is a ground state for even fillings $\nu$. Note that away from the chiral limit, the ground states $|\Psi_{\nu}\rangle$ for even fillings we found here all carry Chern number $0$, as they correspond to filling both Chern-basis bands.

In particular, at the CNP where $\nu=0$, the state Eq.~(\ref{eq:U(4)-GS}) is always a ground state with or without the flat metric condition (Eq.~\ref{eq-Mq=0}).

\section{Nonchiral Perturbation of (first) Chiral-flat Exact Ground States}\label{app:nc-pert-GS}

To understand the low energy states at odd integer fillings and nonzero Chern number states at even integer fillings
in the nonchiral-flat limit, we consider the nonchiral interaction perturbation to the chiral-flat exact ground states in this appendix, while keeping the single-particle bands exactly flat ($H_0=0$). Our treatment is generic for all integer fillings $\nu$ and Chern numbers $\nu_C$. In particular, we note that for Chern number $0$ states at even fillings $\nu=0,\pm2$, the ground states derived from the perturbation theory here become the same as the exact ground states we obtained in Eq.~(\ref{eq:U(4)-GS}).

\subsection{Perturbation energy of Chern insulators}\label{app:perturbation-energy-chern}

In this subsection, we keep the projected bands exactly flat ($H_0=0$), and discuss the nonchiral perturbation (namely, treating $w_0>0$ as a small number) of the Chern insulator states defined in Eq.~(\ref{eq:U(4)U(4)-GS}), which are exact ground states in the U(4)$\times$U(4) (first) chiral limit.

Away from the (first) chiral limit, we have shown in Eq.~(\ref{seq-OqG-nf}) that $O_{\mathbf{q,G}}$ takes the form
\begin{equation}
\begin{split}
&O_{\mathbf{q,G}}=O_{\mathbf{q,G}}^{0}+O_{\mathbf{q,G}}^1\ , \\
\end{split}
\end{equation}
where $O_{\mathbf{q,G}}^0$ and $O_{\mathbf{q,G}}^1$ are defined in Eqs.~(\ref{eq:chiral-OqG}) and~(\ref{seq-OqG1}), respectively. By definition, $O^0_{\mathbf{q,G}}$ and $O^1_{\mathbf{q,G}}$ are diagonal and off-diagonal in the Chern basis $e_Y=\pm1$, respectively. 

The term $O_{\mathbf{q,G}}^1$ only arises when $w_0>0$, namely, when the chiral symmetry is broken. If we treat $w_0>0$ as a perturbation, the term $O_{\mathbf{q,G}}^1$ is proportional to $w_0$ at small $w_0$, and the nonchiral interaction terms containing operators $O_{\mathbf{q,G}}^1$ will yield a perturbation energy to the U(4)$\times$U(4) exact ground states in Eq.~(\ref{eq:U(4)U(4)-GS}). 
To calculate this perturbation energy, we denote the interaction Hamiltonian as $H_I=H_I^{c}+H_I^{nc}$, where the chiral part $H_I^{c}$ and nonchiral part $H_I^{nc}$ are given by
\begin{equation}\label{seq:HI-c-nc}
H_I^{c}=\frac{1}{2\Omega_{\text{tot}}}\sum_{\mathbf{q,G}} O_{\mathbf{-q,-G}}^0O_{\mathbf{q,G}}^0\ ,\quad H_I^{nc}=\frac{1}{2\Omega_{\text{tot}}}\sum_{\mathbf{q,G}} (O_{\mathbf{-q,-G}}^1O_{\mathbf{q,G}}^0+O_{\mathbf{-q,-G}}^0O_{\mathbf{q,G}}^1+O_{\mathbf{-q,-G}}^1O_{\mathbf{q,G}}^1)\ .
\end{equation}
For later convenience, we further divide the nonchiral part into two parts:
\begin{equation}\label{seq:HI-c-nc12}
H_I^{nc}=H_{I}^{nc(1)}+H_I^{nc(2)},\  H_I^{nc(1)}=\frac{1}{2\Omega_{\text{tot}}}\sum_{\mathbf{q,G}}O_{\mathbf{-q,-G}}^1O_{\mathbf{q,G}}^1,\ H_I^{nc(2)}=\frac{1}{2\Omega_{\text{tot}}}\sum_{\mathbf{q,G}}(O_{\mathbf{-q,-G}}^1O_{\mathbf{q,G}}^0+O_{\mathbf{-q,-G}}^0O_{\mathbf{q,G}}^1).
\end{equation}
Note that $H_I^{nc(1)}$ keeps the $e_Y$ index of an electron invariant, while $H_I^{nc(2)}$ flips $e_Y$ of an electron.

For a Chern insulator ground state $|\Psi_\nu^{\nu_+,\nu_-}\rangle$ defined in Eq.~(\ref{eq:U(4)U(4)-GS}), it can be annihilated by $O_{\mathbf{q,G}}^0-A_GN_M\delta_{\mathbf{q,0}}$, and has conserved number of electrons in each Chern number $e_Y$ subspace. As a result, only $H_I^{nc(1)}$ contributes a first order perturbation energy, while $H_I^{nc(2)}$ is non-diagonal in $e_Y$ and only contributes at the second order. We note that since $O_{\mathbf{q,G}}^0$ is of order $1$ and $O_{\mathbf{q,G}}^1$ is of order $w_0$ in the expansion of $w_0$, we have $H_I^{nc(1)}\propto w_0^2$ and $H_I^{nc(2)}\propto w_0$. Therefore, to calculate the energy up to order $w_0^2$, we need to calculate both the first order perturbation of $H_I^{nc(1)}$ and the second order perturbation of $H_I^{nc(2)}$. The first order energy contribution of $H_I^{nc(1)}$ is given by
\begin{equation}\label{seq-E1-nc0}
\begin{split}
&E^{(1)}_{\nu,\nu_C}=\langle \Psi_\nu^{\nu_+,\nu_-}|H_I^{nc}| \Psi_\nu^{\nu_+,\nu_-}\rangle =\langle \Psi_\nu^{\nu_+,\nu_-}|H_I^{nc(1)}| \Psi_\nu^{\nu_+,\nu_-}\rangle \\
&=\frac{1}{2\Omega_{\text{tot}}}\sum_{\mathbf{q,G}}\langle \Psi_\nu^{\nu_+,\nu_-}|O_{\mathbf{-q,-G}}^1O_{\mathbf{q,G}}^1 | \Psi_\nu^{\nu_+,\nu_-}\rangle = \frac{1}{2\Omega_{\text{tot}}} \sum_{j\in \text{half occ}} \sum_{\mathbf{k,q,G}} V(\qq+\GG)\left|F_{+1}(\kk,\qq+\GG)\right|^2\\
&=N_M\sum_{j\in \text{half occ}} U_1\ ,
\end{split}
\end{equation}
where the energy
\begin{equation}\label{seq:U1-def} U_1=\frac{1}{2N_M\Omega_{\text{tot}}} \sum_{\mathbf{k,q,G}} V(\qq+\GG)\left|F_{+1}(\kk,\qq+\GG)\right|^2\ge0\ ,
\end{equation}
and $j$ labels all the spin-valley flavors $\{\eta_{j},s_{j}\}$ where only either the $e_Y=+1$ or $e_Y=-1$ Chern basis are occupied in Eq.~(\ref{eq:U(4)U(4)-GS}), namely, half occupied. This is because if both Chern basis bands of a valley-spin flavor $\{\eta,s\}$ are occupied (empty), $O_{\qq,\GG}^1$ will give zero upon acting on the subspace of flavor $\{\eta,s\}$, and yield zero perturbation energy. Since $O_{\qq,\GG}^1$ is off-diagonal in the Chern basis $e_Y$ index, if a valley-spin flavor $\{\eta,s\}$ only has one of the two Chern basis bands occupied, $O_{\qq,\GG}^1$ will not vanish when acting on $\{\eta,s\}$, and thus contribute a positive perturbation energy to it. Note that the energy $U_1$ we defined here is equivalent to the coupling $\lambda$ in Ref.~\cite{bultinck_ground_2020} (but $U_2$ in Eq.~(\ref{seq:U2-def}) below is not considered in Ref.~\cite{bultinck_ground_2020}). Also, note that this also means for Chern number zero states at even fillings, the perturbation energy (\ref{seq-E1-nc0}) is exactly zero for the states $|\Psi_\nu\rangle$ we defined in Eq.~(\ref{eq:U(4)-GS}), in agreement with our analysis in App.~\ref{app:exactgsnonchiralflat}.

At small $w_0$, the matrix elements $F_{+1}(\kk,\qq+\GG)$ are proportional to $w_0$, so the nonchiral perturbation energy $U_1$ per unit cell is proportional to $w_0^2$. 
Fig. \ref{fig:U1U0}(b) shows the numerically calculated value of $U_1$ as a function of $w_0$ (with $w_1=110$meV fixed) for twist angle $\theta=1.05^\circ$.

We denote $E_{0,\nu,\nu_C}=\langle \Psi_\nu^{\nu_+,\nu_-}|H_I^c+H_I^{nc(1)}| \Psi_\nu^{\nu_+,\nu_-}\rangle$ as the energy of state $|\Psi_\nu^{\nu_+,\nu_-}\rangle$ (with the first-order perturbation energy $E^{(1)}_{\nu,\nu_C}$ in Eq.~(\ref{seq-E1-nc0}) added). The second order contribution of the term $H_I^{nc(2)}$ is then given by
\begin{equation}\label{seq-E2-nc0}
\begin{split}
&E^{(2)}_{\nu,\nu_C}=\langle \Psi_\nu^{\nu_+,\nu_-}|H_I^{nc(2)}(E^c_{0,\nu,\nu_C}-H_I^c)^{-1}H_I^{nc(2)}| \Psi_\nu^{\nu_+,\nu_-}\rangle =-\sum_{j\in \text{half occ}}\sum_\ell \frac{|Y^{nc(2)}_\ell|^2}{E^c_\ell-E^c_{0,\nu,\nu_C}}\ ,
\end{split}
\end{equation}
where $Y_\ell^{nc(2)}=\langle \ell,\Psi_\nu^{\nu_+,\nu_-}|H_I^{nc(2)}|\Psi_\nu^{\nu_+,\nu_-}\rangle$ is the amplitude within a half occupied spin-valley from the state $|\Psi_\nu^{\nu_+,\nu_-}\rangle$ to excited eigenstates $|\ell,\Psi_\nu^{\nu_+,\nu_-}\rangle$ at energies $E^c_\ell=\langle \ell,\Psi_\nu^{\nu_+,\nu_-}|H_I^c|\ell,\Psi_\nu^{\nu_+,\nu_-}\rangle$. Note that since $H_I^{nc(2)}\propto O^{1}_{\qq,\GG}\propto w_0$, this perturbation energy $E^{(2)}_{\nu,\nu_C}$ is also proportional to $w_0^2$. The action of $H_I^{nc(2)}$ on the state $|\Psi_\nu^{\nu_+,\nu_-}\rangle$ can be simplified by noting that 
\begin{equation}
\begin{split}
&\frac{1}{2\Omega_{\text{tot}}}\sum_{\qq,\GG}[O^0_{-\qq,-\GG},O^1_{\qq,\GG}] =\frac{1}{2\Omega_{\text{tot}}}\sum_{\qq,\GG,\kk,e_Y}\eta  V(\qq+\GG)[M_{-e_Y}(\kk+\qq,-\qq-\GG)F_{e_Y}(\kk,\qq+\GG) d^\dag_{\kk,-e_Y,\eta,s}d_{\kk,e_Y,\eta,s} \\
&\qquad -M_{e_Y}(\kk+\qq,-\qq-\GG)F_{e_Y}(\kk,\qq+\GG)d^\dag_{\kk+\qq,-e_Y,\eta,s}d_{\kk+\qq,e_Y,\eta,s}]\\
&=\frac{1}{2\Omega_{\text{tot}}}\sum_{\qq,\GG,\kk,e_Y}\eta V(\qq+\GG)  [M_{-e_Y}^*(\kk,\qq+\GG)F_{e_Y}(\kk,\qq+\GG)-M_{e_Y}(\kk,\qq+\GG)F_{-e_Y}^*(\kk,\qq+\GG)]d^\dag_{\kk,-e_Y,\eta,s}d_{\kk,e_Y,\eta,s}\\
&=0\ ,
\end{split}
\end{equation}
where we have used the fact that $M_{e_Y}^*(\kk,\qq+\GG)=M_{-e_Y}(\kk,\qq+\GG)=M_{-e_Y}^*(\kk+\qq,-\qq-\GG)$, and $F_{-e_Y}^*(\kk,\qq+\GG)=F_{e_Y}(\kk,\qq+\GG)=F_{-e_Y}^*(\kk+\qq,-\qq-\GG)$. Therefore, using $O_{\qq,\GG}^0|\Psi_\nu^{\nu_+,\nu_-}\rangle=A_\GG N_M\delta_{\qq,\mathbf{0}}|\Psi_\nu^{\nu_+,\nu_-}\rangle$ with $A_\GG=\frac{\nu }{N_M} \sqrt{V(\mathbf{G})} \sum_\mathbf{k} \alpha_0(\mathbf{k,G})$ (see Eq.~(\ref{eq-AG})), we have 
\begin{equation}\label{seq:nonzero-pert-Hnc2}
\begin{split}
&H_I^{nc(2)}|\Psi_\nu^{\nu_+,\nu_-}\rangle=\frac{1}{2\Omega_{\text{tot}}}\sum_{\qq,\GG}V(\qq+\GG)\Big([O^0_{-\qq,-\GG},O^1_{\qq,\GG}]+2O^1_{\qq,\GG}O^0_{-\qq,-\GG}\Big)|\Psi_\nu^{\nu_+,\nu_-}\rangle =\frac{1}{\Omega_{\text{tot}}}\sum_{\qq,\GG}O_{-\qq,-\GG}^1O_{\qq,\GG}^0|\Psi_\nu^{\nu_+,\nu_-}\rangle\\
&=\frac{\nu}{\Omega_{\text{tot}}} \sum_{\GG,e_Y,\kk,\eta,s} V(\GG)F_{e_Y}(\kk,-\GG)\left(\sum_\mathbf{k'} \alpha_0(\mathbf{k',G})\right) d^\dag_{\kk,-e_Y,\eta,s}d_{\kk,e_Y,\eta,s}|\Psi_\nu^{\nu_+,\nu_-}\rangle\ .
\end{split}
\end{equation}
This shows that the amplitude $Y_\ell^{nc(2)}$ is proportional to the filling $\nu$, and thus generically we have the second order perturbation energy
\begin{equation}\label{seq:U2-def}
E^{(2)}_{\nu,\nu_C}=-N_M\sum_{j\in \text{half occ}}\nu^2U_2\ ,\qquad U_2=\frac{1}{N_M}\sum_\ell \frac{|Y^{nc(2)}_\ell|^2}{E_\ell-E_{0,\nu,\nu_C}}\Big|_{|\nu|=1}\ .
\end{equation}
where $U_2\ge0$.

In particular, if we impose the FMC in Eq.~(\ref{eq-Mq=0}) which implies $F_{e_Y}(\kk,\GG)=-F_{e_Y}(-\kk,\GG)=0$, we find that 
\begin{equation}\label{seq:zero-pert-Hnc2-0}
H_I^{nc(2)}|\Psi_\nu^{\nu_+,\nu_-}\rangle=0\ ,
\end{equation}
and thus
\begin{equation}\label{seq:zero-pert-Hnc2}
Y_\ell^{nc(2)}=0\ ,\qquad U_2=0\ ,\qquad E^{(2)}_{\nu,\nu_C}=0\ ,
\end{equation}
namely, the second order perturbation energy of $H_I^{nc(2)}$ is zero. Note that Eq.~(\ref{seq:zero-pert-Hnc2-0}) also indicates that in the presence of FMC, the eigenstate wavefunction $|\Psi_\nu^{\nu_+,\nu_-}\rangle$ is not only accurate to the zeroth order, but also accurate to the first order of the perturbation $H_I^{nc(2)}$. Since $H_I^{nc(2)}\propto w_0$ and $H_I^{nc(1)}\propto w_0^2$, this means the wavefunction $|\Psi_\nu^{\nu_+,\nu_-}\rangle$ is accurate to the linear order of $w_0$ with the FMC.

Without the FMC, the numerical values of $U_2$ with respect to $w_0/w_1$ (with $w_1=110$meV fixed) at $\theta=1.05^\circ$ is shown in Fig. \ref{fig:U1U0}(b). The numerical calculations are done by summing over all the one electron-hole pair charge neutral excitations which are exactly solvable in the chiral-flat limit by diagonalizing a one-body Hamiltonian, as derived in Ref.~\cite{ourpaper5}. This is because $H_I^{nc(2)}$ only excites to states one electron-hole pair away from the state $|\Psi_\nu^{\nu_+,\nu_-}\rangle$, as we demonstrated in Eq.~(\ref{seq:nonzero-pert-Hnc2}). We take a $12\times12$ momentum lattice discretization of the MBZ, which is sufficiently close to the thermodynamic limit.

Therefore, we find the total perturbation energy proportional to $w_0^2$ is given by
\begin{equation}\label{seq-E1-nc12}
E^{(nc)}_{\nu,\nu_C}=E^{(1)}_{\nu,\nu_C}+E^{(2)}_{\nu,\nu_C}=N_M\sum_{j\in \text{half occ}} (U_1-\nu^2U_2)\ .
\end{equation}

We note that the perturbation due to $O_{\mathbf{q,G}}^1$ is a nondegenerate perturbation. This is because the nonchiral interaction terms $O_{\mathbf{-q,-G}}^1O_{\mathbf{q,G}}^1$ (or $O_{\mathbf{-q,-G}}^0O_{\mathbf{q,G}}^1$, $O_{\mathbf{-q,-G}}^1O_{\mathbf{q,G}}^0$) either leave the state $| \Psi_\nu^{\nu_+,\nu_-}\rangle$ invariant, or flip the index $e_Y$ of 2 (or $1$) electrons. However, an operation flipping $e_Y$ does not belong to the chiral-flat U(4)$\times$U(4) symmetry group (which can be seen from the generators in Eq.~(\ref{seq-Sabpm})), thus the resulting state cannot be a U(4)$\times$U(4) rotation of the state $| \Psi_\nu^{\nu_+,\nu_-}\rangle$ (or in other words, a Goldstone mode of zero momentum), but necessarily has an energy cost. Therefore, the nonchiral perturbation is nondegenerate, and our above nondegenerate perturbation calculation is valid.

In particular, our numerical calculation shows that $U_1-\nu^2U_2>0$ for any $0<w_0/w_1\le 1$ and any $|\nu|\le 3$. It is therefore clear that the nonchiral interaction perturbation favors the lower Chern number $|\nu_C|$ insulator states, with as many spin-valley flavors fully occupied or fully empty as possible. Such a state can be generically written as
\begin{equation}\label{seq-Psi-nu-nuC}
|\Psi_{\nu,\nu_C}\rangle =\prod_{\mathbf{k}} \prod_{j=1}^{\nu_+}d^\dag_{\mathbf{k},+1,\eta_{j},s_{j}} \prod_{j=1}^{\nu_-}d^\dag_{\mathbf{k},-1,\eta_{j},s_{j}}|0\rangle\ ,
\end{equation}
which fully occupies valley-spin flavors $\{\eta_j,s_j\}$ with $1\le j\le \text{min}(\nu_+,\nu_-)$. Any chiral-nonflat U(4) rotations of the state $|\Psi_{\nu,\nu_C}\rangle$ are also degenerate, and all the degenerate states form a chiral-nonflat U(4) irrep $[(2N_M)^{(\nu-|\nu_C|+4)/2},N_M^{|\nu_C|}]_4$. This notation means the irrep of a Young tableau with $2N_M$ boxes in each of the first $(\nu-|\nu_C|+4)/2$ rows, and $N_M$ boxes in each of the next $|\nu_C|$ rows. This is because there are $(\nu-|\nu_C|+4)/2$ valley-spin flavors where both Chern bands are fully occupied, and the $2N_M$ electrons in each of these flavors are antisymmetric in $\kk$ and symmetric in the valley-spin U(4) space, thus lying in the same row of Young tableau. Then there are $|\nu_C|$ flavors where only one Chern band is occupied, and the $N_M$ electrons in such a Chern band are antisymmetric in $\kk$ and symmetric in valley-spin space, thus lying in the same row of Young tableau. Note that the U(4) multiplet of state $|\Psi_{\nu,\nu_C}\rangle$ is a subset of the U(4)$\times$U(4) multiplet of state $| \Psi_\nu^{\nu_+,\nu_-}\rangle$ in Eq.~(\ref{eq:U(4)U(4)-GS}). By Eq.~(\ref{seq-E1-nc12}), we find the perturbation energy of state $|\Psi_{\nu,\nu_C}\rangle$ in Eq.~(\ref{seq-Psi-nu-nuC}) up to order $w_0^2$ is
\begin{equation}\label{seq-E1-nc}
E^{(nc)}_{\nu,\nu_C}=|\nu_C|N_M (U_1-\nu^2U_2)\ ,
\end{equation}
with $U_1-\nu^2U_2>0$ for any $|\nu|\le 3$. For $\nu=\pm4$, the only available state has Chern number $\nu_C=0$, thus the perturbation energy is zero. 
This clearly indicates that the states with a lower Chern number $|\nu_C|$ have a lower energy in the nonchiral-flat limit.

In particular, we note that for even fillings $\nu=0,\pm2,\pm4$ and $\nu_C=0$, the states $|\Psi_{\nu,0}\rangle$ in Eq.~(\ref{seq-Psi-nu-nuC}) become the exact ground states $|\Psi_{\nu}\rangle$ in Eq.~(\ref{eq:U(4)-GS}).

\subsection{Summary of U(4) irreps of various ground states}\label{app:u4irreps-GS}

This appendix summarizes the ground states we discussed in Apps.~\ref{app:exact-GS} and~\ref{app:nc-pert-GS} above, which are listed in Table~\ref{Tab-QN}. Since the system is PH symmetric about $\nu=0$, here we only list those ground states with integer filling fraction $\nu\le 0$, and we have labeled under which limit they are exact. A ground state exact under the nonchiral-flat U(4) limit is also exact under the (first) chiral-flat U(4)$\times$U(4) limit.

\begin{table}[htbp]
  \centering
  \begin{tabular}{c|c|c|c|c|c}
  \hline
 filling $\nu$ & Chern number $\nu_C$ & nonchiral-flat U(4) irrep & chiral-flat U(4)$\times$U(4) irrep & exact under &  if nonchiral-flat GS  \\
  \hline
  $-3$ & $\pm1$ & $[N_M]_4$ & $([N_M]_4,[0]_4)$ & U(4)$\times$U(4) & yes (perturbative) \\
  \hline
  $-2$ & 0 & $[2N_M]_4$ & $([N_M]_4,[N_M]_4)$ & U(4) & yes (exact) \\
  \hline
  $-2$ & $\pm2$ & $[N_M^2]_4$ & $([N_M^2]_4,[0]_4)$ & U(4)$\times$U(4)  & no\\
  \hline
  $-1$  & $\pm1$ & $[2N_M,N_M]_4$ & $([N_M^2]_4,[N_M]_4)$ & U(4)$\times$U(4) & yes (perturbative) \\
  \hline
  $-1$ & $\pm3$ & $[N_M^3]_4$ & $([N_M^3]_4,[0]_4)$ & U(4)$\times$U(4) & no\\
  \hline
  $0$ & 0 &  $[(2N_M)^2]_4$ & $([N_M^2]_4,[N_M^2]_4)$ & U(4)  & yes (exact) \\
  \hline
  $0$ & $\pm2$ & $[2N_M,N_M,N_M]_4$ & $([N_M^3]_4,[N_M]_4)$ & U(4)$\times$U(4) & no\\
  \hline
  $0$ & $\pm4$ & $[0]_4$ & $([0]_4,[0]_4)$ & U(4)$\times$U(4) & no\\
  \hline
  \end{tabular}
  \caption{Representations of various TBG insulating Phases we derived. $N_M$ is the number of moir\'e unit cells, and the number of electrons is $N=(\nu+4)N_M$. Here U(4)$\times$U(4) refers to the first chiral limit. In the table, we only list the U(4)$\times$U(4) irrep $([\{\lambda_1\}]_4,[\{\lambda_2\}]_4)$ of the state with Chern number $\nu_C\ge0$, while the state with Chern number $-\nu_C$ in the same line has irrep $([\{\lambda_2\}]_4,[\{\lambda_1\}]_4)$. In the nonchiral-flat U(4) limit, the states with smaller Chern number $|\nu_C|$ would have a lower energy. The last column denotes whether each state is the ground state (GS) in the nonchiral-flat limit, either exactly or in the perturbation theory.}\label{Tab-QN}
\end{table}

\section{Kinetic Term Perturbation to the Ground States}\label{app:kineticperturbartion}
In this appendix, we consider the perturbation effects of the kinetic term on the exact ground states in App.~\ref{app:exact-GS} and the nonchiral-flat states in App.~\ref{app:nc-pert-GS}. Throughout this appendix, we shall assume the flat metric condition Eq.~(\ref{eq-Mq=0}) holds exactly or approximately, so that the eigenstates $|\Psi_\nu^{\nu_+,\nu_-}\rangle$ and $|\Psi_\nu\rangle$ we discussed in App.~\ref{app:exact-GS} are the ground states in the absence of the kinetic energy. We discuss the U(4) case and the chiral limit U(4)$\times$U(4) case separately.

\subsection{The kinetic Hamiltonian}\label{app:perturb-kinetic-term}

Before we proceed to perturbations, we first recall and rewrite the projected kinetic Hamiltonian of TBG. By Eq.~(\ref{eq:H0-proj}), the TBG kinetic energy term has the form
\begin{equation}
H_0=\sum_{\mathbf{k},m,\eta,s} \epsilon_{m,\eta}(\mathbf{k})c^\dag_{\mathbf{k},m,\eta,s}c_{\mathbf{k},m,\eta,s}\ ,
\end{equation}
where the single-particle energy satisfies $\epsilon_{m,\eta}(\mathbf{k})=\epsilon_{m,-\eta}(-\mathbf{k})=-\epsilon_{-m,\eta}(-\mathbf{k})=-\epsilon_{-m,-\eta}(\mathbf{k})$ due to $C_{2z}$ symmetry and the anti-commuting unitary symmetry $P$. The kinetic term breaks the $C_{2z}P$ in the nonchiral-flat U(4). We can rewrite the kinetic energy term into two parts:
\begin{equation}\label{seq:H0pm}
H_0=H_0^+ +H_0^-\ ,\quad H_0^+=\sum_{\mathbf{k}} \epsilon_+(\mathbf{k}) c^\dag_{\mathbf{k}}(\zeta^z\tau^0s^0)c_{\mathbf{k}}\ ,\quad H_0^-=\sum_{\mathbf{k}} \epsilon_-(\mathbf{k}) c^\dag_{\mathbf{k}} (\zeta^0 \tau^z s^0)c_{\mathbf{k}}\ ,
\end{equation}
where band, valley and spin indices are written into the matrix form, $c_\kk$ is the column vector of the eight fermion operators $c_{\kk,m,\eta,s}$ of all band, valley and spin indices, and $\epsilon_{\pm}(\mathbf{k})=[ \epsilon_{+1,+}(\mathbf{k})\mp \epsilon_{-1,+}(\mathbf{k})]/2 $. These two functions satisfy 
\begin{equation}
\epsilon_{\pm}(\mathbf{k})=\pm\epsilon_{\pm}(-\mathbf{k})\ .
\end{equation}

In the chiral limit, we have additionally $\epsilon_{m,\eta}(\mathbf{k})=-\epsilon_{-m,\eta}(\mathbf{k})$. This ensures that $H_0^-=0$ in the chiral limit, so the kinetic term is solely $H_0=H_0^+$. Accordingly, in the chiral-nonflat limit there is a remaining U(4) symmetry unbroken by $H_0$, with generators given by $\zeta^0\tau^as^b$ (see Eq.~(\ref{eq:U4-generator-chiral-nonflat}) and details in Ref.~\cite{ourpaper3}).

\subsection{Kinetic perturbation in the (first) chiral-flat U(4)$\times$U(4) limit}\label{app:kineticpert-u4u4}

In this subsection, we study the perturbation of the kinetic term up to the second order on the exact ground states $|\Psi_\nu^{\nu_+,\nu_-}\rangle$ in the (first) chiral-flat U(4)$\times$U(4) limit given in Eq.~(\ref{eq:U(4)U(4)-GS}). We note that here we do not require the FMC in Eq.~(\ref{eq-Mq=0}) to hold, we only require that $|\Psi_\nu^{\nu_+,\nu_-}\rangle$ are ground states in the chiral-flat limit without the FMC (which is verified in our exact digonalization in Ref.~\cite{ourpaper6}).

In the chiral limit, the kinetic term is simply $H_0=H_0^+$, since the term $H_0^-=0$ as discussed in App.~\ref{app:perturb-kinetic-term}. After adding the kinetic term $H_0^+$, there is still a remaining U(4) symmetry with generators proportional to $\zeta^0\tau^a s^b$ ($a,b=0,x,y,z$, see Eq.~(\ref{eq:U4-generator-chiral-nonflat})), so one is free (without energy cost) to rotate valley $\eta$ and spin $s$ without affecting the space of band indices.

When the perturbation $H_0^+$ is added, we want to find the lowest state among the U(4)$\times$U(4) multiplet of the Chern insulator state $|\Psi_{\nu}^{\nu_+,\nu_-}\rangle$ defined in Eq.~(\ref{eq:U(4)U(4)-GS}).
For a fixed filling $\nu=\nu_++\nu_--4$ and Chern number $\nu_C=\nu_+-\nu_-$, we examine which choice of occupied valley-spin flavors for the Chern number $e_Y=\pm1$ bands gives the lowest energy. The kinetic Hamiltonian can be rewritten in the Chern band basis as
\begin{equation}
H_0=H_0^+ = \sum_{\mathbf{k},\eta,s} \epsilon_+(\mathbf{k}) \left(d^\dag_{\mathbf{k},+,\eta,s}d_{\mathbf{k},-,\eta,s}+d^\dag_{\mathbf{k},-,\eta,s}d_{\mathbf{k},+,\eta,s}\right) \ ,
\end{equation}
which flips the Chern band index $e_Y$ of an electron within the same valley and spin. Such flipping of Chern band index $e_Y$ in a valley-spin flavor $\{\eta,s\}$ is possible only if the valley-spin flavor $\{\eta,s\}$ is half-occupied, namely, only one of the $e_Y=\pm1$ Chern band basis is fully occupied.

\emph{Second order perturbation.} Since $H_0^+$ is off-diagonal in the Chern band basis, it does not yield a first order perturbation energy to the state $|\Psi_{\nu}^{\nu_+,\nu_-}\rangle$. It does give a 2nd order perturbation energy by exciting the ground state into some high energy states.

We now consider the Hilbert space of such reachable excited states by acting $H_0$ once on the state $|\Psi_{\nu}^{\nu_+,\nu_-}\rangle$. Assume the valley-spin flavor $\{\eta,s\}$ has its Chern band basis $-e_Y$ fully occupied and its Chern band basis $e_Y$ fully empty. We consider the following sets of states
\begin{equation}
|\mathbf{k},e_Y,\eta,s,\Psi_{\nu}^{\nu_+,\nu_-}\rangle=d^\dag_{\mathbf{k},e_Y,\eta,s} d_{\mathbf{k},-e_Y,\eta,s}|\Psi_{\nu}^{\nu_+,\nu_-}\rangle\ ,
\end{equation}
which can be reached by acting $H_0$ onto the ground state $|\Psi_{\nu}^{\nu_+,\nu_-}\rangle$ (restricted within the valley-spin flavor $\{\eta,s\}$), where the amplitude is $\epsilon_+(\mathbf{k})$. These states are not eigenstates of the interaction Hamiltonian $H_I$. The eigenstates will be found and discussed in Ref.~\cite{ourpaper5}. However, we here show, for self-consistencey, that these states of different momentum $\kk$ with fixed $e_Y,\eta,s$ form a closed subspace under the interaction $H_I$. 
To see this, we note that in the chiral limit, the operator $O_{\qq,\GG}$ in $H_I$ is given by Eq.~(\ref{eq:chiral-OqG}).
Note the fact that
\begin{equation}\label{seq:action-OqG-neutral-ex}
\begin{split}
&[O_{\mathbf{q,G}},d^\dag_{\mathbf{k},e_Y,\eta,s} d_{\mathbf{k},-e_Y,\eta,s}]\\
&=\sqrt{V(\qq+\GG)} (M_{e_Y}(\kk,\qq+\GG) d^\dag_{\mathbf{k+q},e_Y,\eta,s} d_{\mathbf{k},-e_Y,\eta,s}- M_{-e_Y}(\kk-\qq,\qq+\GG) d^\dag_{\mathbf{k},e_Y,\eta,s} d_{\mathbf{k-q},-e_Y,\eta,s})\ ,
\end{split}
\end{equation}
and the equality $M_{e_Y}(\kk,\qq+\GG)=M_{e_Y}^*(\kk+\qq,-\qq-\GG)$ (due to Eq.~(\ref{eq:alpha-cond1})), 
we find the interaction Hamiltonian $H_I$ satisfies (see the calculation of charge neutral excitations in Ref.~\cite{ourpaper5} for details):
\begin{equation}
\left[H_I-\mu N,d^\dag_{\mathbf{k},e_{Y},\eta,s} d_{\mathbf{k},-e_{Y},\eta,s}\right] =\frac{1}{2\Omega_{\text{tot}}}\sum_\qq S _{e_{Y};-e_{Y}}(\mathbf{k},\mathbf{0};\qq)d^\dag_{\mathbf{k+q},e_{Y},\eta,s} d_{\mathbf{k+q},e_{Y},\eta,s}\ ,
\end{equation}
where (see Ref.~\cite{ourpaper5} for notations)
\begin{equation}\label{seq:Seyney}
S _{e_{Y};-e_{Y}}(\mathbf{k},\mathbf{0};\qq) =  2\sum_{\GG}\Big\{-V(\qq+\GG)M_{e_Y}(\mathbf{k,q+G})^2 +\delta_{\mathbf{q,0}}\sum_{\qq'}V(\qq'+\GG)|M_{e_Y}(\mathbf{k,q'+G})|^2 \Big\}\ .
\end{equation}
This expression in the chiral limit is independent of filling $\nu$, and does not require the FMC in Eq.~(\ref{eq-Mq=0}) (see derivation in Ref.~\cite{ourpaper5}). 
Moreover, by Eq.~(\ref{seq:action-OqG-neutral-ex}), and using the fact that $\sum_\kk\alpha_2(\kk,\GG)=0$ at $\qq=\mathbf{0}$, we have
\begin{equation}
O_{\mathbf{0,G}}|\mathbf{k},e_Y,\eta,s,\Psi_{\nu}^{\nu_+,\nu_-}\rangle = N_M A_\GG|\mathbf{k},e_Y,\eta,s,\Psi_{\nu}^{\nu_+,\nu_-}\rangle\ ,
\end{equation}
where $A_\GG$ depends on filling $\nu$ as given by Eq.~(\ref{eq-AG}). 
As a result, we find $H_I$ is closed within the Hilbert subspace of states $|\mathbf{k},e_Y,\eta,s,\Psi_{\nu}^{\nu_+,\nu_-}\rangle$, and satisfy
\begin{equation}
(H_I-E_{0,\nu})|\mathbf{k},e_Y,\eta,s,\Psi_{\nu}^{\nu_+,\nu_-}\rangle=\sum_{\qq} \mathcal{H}^{e_Y,\eta,s}_{\kk+\qq,\kk}|\mathbf{k+q},e_Y,\eta,s,\Psi_{\nu}^{\nu_+,\nu_-}\rangle\ ,
\end{equation}
where $E_{0,\nu}$ is the unperturbed energy of the ground state $|\Psi_{\nu}^{\nu_+,\nu_-}\rangle$ (which only depends on $\nu$ but not $\nu_C=\nu_+-\nu_-$), and the sub-Hamiltonian
\begin{equation}
\mathcal{H}^{e_Y,\eta,s}_{\kk+\qq,\kk} =\frac{1}{2\Omega_{\text{tot}}}S _{e_{Y};-e_{Y}}(\mathbf{k},\mathbf{0};\qq)\ .
\end{equation}
Note that we do not need to assume the flat metric condition Eq.~(\ref{eq-Mq=0}) here. In particular, one finds the sub-Hamiltonian $\mathcal{H}^{e_Y,\eta,s}=(\mathcal{H}^{-e_Y,\eta,s})^*$ and is independent of $\eta$ and $s$. Therefore, we conclude that in each sector of $\{e_Y,\eta,s\}$, the eigenstates in this subspace of excited states have identical spectrum. We denote these eigenstates as $|\ell,e_Y,\eta,s,\Psi_{\nu}^{\nu_+,\nu_-}\rangle$, which has energy $E_\ell$ under $H_I$ (or equivalently, the sub-Hamiltonian $\mathcal{H}^{e_Y,\eta,s}+E_{0,\nu}$) independent of $\{e_Y,\eta,s\}$. These eigenstates can be interpreted as exciton states on top of the occupied Chern band $e_Y$ within valley $\eta$ and spin $s$ \cite{ourpaper5}. We note that these excitations flipping $e_Y$ are all gapped, since the operation of flipping $e_Y$ is not within the U(4)$\times$U(4) symmetry group (which can be seen from the generators in Eq.~(\ref{seq-Sabpm})), so the perturbation is non-degenerate. Instead, it can be proved that the excitations not flipping $e_Y$ (but change valley/spin, discussed in Ref.~\cite{ourpaper5}) contain the Goldstone modes \cite{ourpaper5}.

This leads to the following 2nd order perturbation energy:
\begin{equation}\label{eq:chern-energy-H0-2nd-order}
\widetilde{E}_{\nu,\nu_C}^{(2)}=-\sum_{\{{e_Y,\eta,s}\}}\sum_{\ell} \frac{|Y_\ell^{e_Y,\eta,s}|^2}{E_{\ell}-E_{0,\nu}}=-N_M\sum_{\{{\eta,s}\}\in\text{half occ}}J_0\ ,\qquad J_0=\frac{1}{N_M}\sum_{\ell} \frac{|Y_\ell|^2}{E_{\ell}-E_{0,\nu}}\ ,
\end{equation}
where the summation is over all valley-spin flavors $\{\eta,s\}$ which are half-occupied, and $E_{0,\nu}$ denotes the unperturbed ground state energy of the chiral-flat state $|\Psi_{\nu}^{\nu_+,\nu_-}\rangle$. We have defined the amplitude  $Y_\ell^{e_Y,\eta,s}=\langle \ell,e_Y,\eta,s,\Psi_{\nu}^{\nu_+,\nu_-}|H_0|\Psi_{\nu}^{\nu_+,\nu_-}\rangle$, which is nonzero and has a norm independent of $e_Y,\eta,s$ if the Chern basis of $\{e_Y,\eta,s\}$ is empty and $\{-e_Y,\eta,s\}$ is fully occupied. For short, we denote the norm of these nonzero amplitude $Y_\ell^{e_Y,\eta,s}$ independent of $e_Y,\eta,s$ as $|Y_\ell|$. We note that this perturbation energy per moir\'e unit cell $J_0$ is equivalent to the coupling $J$ defined in Ref.~\cite{bultinck_ground_2020}.

Therefore, the second order perturbation energy is the lowest if the state $|\Psi_{\nu}^{\nu_+,\nu_-}\rangle$ has as many half-occupied valley-spin flavors $\{\eta,s\}$ as possible. However, all the states $|\Psi_{\nu}^{\nu_+,\nu_-}\rangle$ with equal number of half-occupied valley-spin flavors $\{\eta,s\}$ have the same the second order perturbation energy, and are still degenerate. This selects the following subset of the multiplet $|\Psi_{\nu}^{\nu_+,\nu_-}\rangle$ at filling $\nu=\nu_++\nu_--4$ and Chern number $\nu_C=\nu_+-\nu_-$ as the lowest states:
\begin{equation}\label{seq-Psi-nu-nuC2}
|\widetilde{\Psi}_{\nu,\nu_C}\rangle =\prod_{\mathbf{k}} \prod_{j=1}^{\nu_+}d^\dag_{\mathbf{k},+1,\eta_{j},s_{j}} \prod_{j=5-\nu_-}^{4}d^\dag_{\mathbf{k},-1,\eta_{j},s_{j}}|0\rangle\ ,
\end{equation}
where $\{\eta_j,s_j\}$ are the 4 valley-spin flavors arbitrarily sorted in $j$ ($1\le j\le 4$). This state has $4-|\nu|$ valley-spin flavors half-occupied, and has a second order perturbation energy
\begin{equation}\label{seq-E2-nf}
\widetilde{E}_{\nu,\nu_C}^{(2)}=-(4-|\nu|)N_M J_0\ .
\end{equation}
Note that $\widetilde{E}_{\nu,\nu_C}^{(2)}$ is independent of $\nu_C$. Therefore, the states $|\widetilde{\Psi}_{\nu,\nu_C}\rangle$ for fixed $\nu$ with different Chern numbers $\nu_C$ are degenerate up to the second order perturbation of $H_0$. This degeneracy between different absolute values of the Chern numbers $|\nu_C|$ is expected to be broken at higher order perturbations, since there is no symmetry protecting this degeneracy (the $\nu_C$ and $-\nu_C$ states have to be degenerate with each other due to time-reversal symmetry).

If $\nu_-=0,4$ or $\nu_+=0,4$, in which case $\nu_C=\pm(4-|\nu|)$, the state $|\widetilde{\Psi}_{\nu,\nu_C}\rangle$ in Eq.~(\ref{seq-Psi-nu-nuC2}) falls into a U(4) irrep $[N_M^{4-|\nu|}]_4$. However, if both $\nu_+$ and $\nu_-$ are nonzero and not equal to $4$, the state $|\widetilde{\Psi}_{\nu,\nu_C}\rangle$ in Eq.~(\ref{seq-Psi-nu-nuC2}) is not in a U(4) irrep that is easy to write down, but resembles a U(4) version of the Neel-ordered antiferromagnetic state (if $\nu_+=\nu_-$) or a ferrimagnetic state (if $\nu_+\neq\nu_-$). This is because, in the expression of the wavefunction $|\widetilde{\Psi}_{\nu,\nu_C}\rangle$ in Eq.~(\ref{seq-Psi-nu-nuC2}) (which is not exact), it is not clear whether the occupied electron at momentum $\kk$ in Chern basis $e_Y=+1$ in valley-spin flavor $j=1$ and the occupied electron at momentum $\kk$ in Chern basis $e_Y=-1$ in valley-spin flavor $j=4$ are symmetric or antisymmetric in the U(4) valley-spin indices (because they occupy different $e_Y$ basis). If at all momenta $\kk$ such two electrons are considered antisymmetric in valley-spin flavors (as suggested by the fact that they occupy different valley-spin flavors), they should occupy the same column in a Young tableau, and the U(4) irrep of state $|\widetilde{\Psi}_{\nu,\nu_C}\rangle$ will be given by $[N_M^{4-|\nu|}]_4$. However, this cannot be exact (similar to the Neel order state in a SU(2) spin system, where the ground state cannot be a total antisymmetric singlet due to symmetry breaking), so we would expect the irrep of state $|\widetilde{\Psi}_{\nu,\nu_C}\rangle$ to be close to $[N_M^{4-|\nu|}]_4$, differing from it only by a few boxes ($\ll N_M$) moved among different rows in the Young tableau.

Lastly, we note that both the operator $O_{\qq,\GG}^1$ in the nonchiral-flat limit and the kinetic term $H_0^+$ flip the Chern basis $e_Y$. Thus the criteria for their energy perturbations both rely on half occupied spin-valley flavors. However, and as opposed to the nonchiral-flat limit (where the perturbation is in fact first order given by a positive operator $O_{-\qq,-\GG}^1O_{\qq,\GG}^1$), here in the chiral-nonflat limit the kinetic term $H_0^+$ is one-body. As a consequence, $H_0^+$ only contributes in the second order perturbation, involving higher states and thus lowering the energy.

\subsection{Kinetic perturbation in the nonchiral-flat U(4) limit}\label{app:kineticpert-nonchiralflat}

We now consider the kinetic term perturbation to the nonchiral-flat ground/low-energy states, which makes the system nonchiral-nonflat, and corresponds to the experimental situation. In the case where the wavefunctions in Eq. (\ref{seq-Psi-nu-nuC}) (see also Eq. (\ref{seq-Psi-nu-nuC-rotate})) are not exact eigenstates in the nonchiral-flat limit (when the Chern number $\nu_C\neq0$), we assume the single-particle bandwidth $t<w_0$, such that we can regard the kinetic term $H_0$ as a small perturbation to the nonchiral-flat limit (exact or approximate) eigenstates.

In the nonchiral-flat U(4) limit, we have a set of approximate/exact insulator states $|\Psi_{\nu,\nu_C}\rangle$ in Eq.~(\ref{seq-Psi-nu-nuC}) with Chern number $\nu_C$ at integer fillings $\nu$, where $\nu_C=4-|\nu|,2-|\nu|,\cdots,-4+|\nu|$. The U(4) rotations of $|\Psi_{\nu,\nu_C}\rangle$ form a degenerate U(4) multiplet of states. In particular, for even fillings $\nu=0,\pm2,\pm4$, the insulator states $|\Psi_{\nu,\nu_C}\rangle$ with Chern number $\nu_C=0$ become the exact ground states $|\Psi_\nu\rangle$ defined in Eq.~(\ref{eq:U(4)-GS}) for even fillings $\nu$. We shall assume the nonchiral interaction terms due to nonzero $w_0$ are larger than the energy scale of the kinetic term $H_0$ (which is the case near magic angle with a realistic $w_0\approx0.8w_1$), so that $H_0$ can be treated as perturbation on top of the nonchiral-flat exact/approximate insulator states $|\Psi_{\nu,\nu_C}\rangle$.

We shall examine the perturbation of kinetic term $H_0$ to the state $|\Psi_{\nu,\nu_C}\rangle$. Since $H_0$ breaks the U(4) symmetry down to the U(2)$\times$U(2) spin-charge rotational symmetry of two valleys, one expects it to select a subset of the U(4) multiplet $|\Psi_{\nu,\nu_C}\rangle$ as the lowest states. To examine which states in the U(4) multiplet are preferred, we start with the state $|\Psi_{\nu,\nu_C}\rangle$ in Eq.~(\ref{seq-Psi-nu-nuC}) with the valley-spin flavors $\{\eta_j,s_j\}$ ($1\le j\le 4$) therein sorted in a certain order (which will be specified below). We then consider the following U(4) rotated state relative to the expression of $|\Psi_{\nu,\nu_C}\rangle$ in Eq.~(\ref{seq-Psi-nu-nuC}):
\begin{equation}\label{seq-Psi-nu-nuC-rotate}
|\Psi_{\nu,\nu_C}(\varphi_s)\rangle =U(\varphi_s)|\Psi_{\nu,\nu_C}\rangle = U(\varphi_s) \prod_{\mathbf{k}} \prod_{j=1}^{\nu_+}d^\dag_{\mathbf{k},+1,\eta_{j},s_{j}} \prod_{j=1}^{\nu_-}d^\dag_{\mathbf{k},-1,\eta_{j},s_{j}}|0\rangle\ ,
\end{equation}
where $s=\uparrow,\downarrow$ stands for spin, and we have defined the following generic U(4) rotation in terms of the angles $\varphi_s$:
\begin{equation}\label{seq:U4-rotator}
U(\varphi_s)=e^{i\varphi S^{y0}/2} e^{i\phi S^{yz}/2}\ ,\qquad \varphi=\frac{\varphi_\uparrow+\varphi_\downarrow}{2}\ ,\quad \phi=\frac{\varphi_\uparrow-\varphi_\downarrow}{2}\ .
\end{equation}
Here $S^{ab}$ are the generators given in Eq.~(\ref{eq:U4-generator0}), which have representation matrices $s^{ab}(e_Y)=\{ \tau^0s^b,\ e_Y\tau^x s^b,\ e_Y\tau^y s^b,\ \tau^zs^b \}$ under the Chern basis $e_Y$ (see the derivation in Ref.~\cite{ourpaper3}). 
Note that the first generator 
$S^{yz}$ rotates the valley polarization in the spin $\uparrow$ and $\downarrow$ subspaces oppositely, which changes the spin configuration; while the second generator $S^{y0}$ rotates the valley polarization as a whole without affecting the total spin. Therefore, the valley polarization of the spin $\uparrow$ and $\downarrow$ electrons are rotated differently by angles $\varphi_\uparrow$ and $\varphi_\downarrow$, respectively. Also, note that the state $|\Psi_{\nu,\nu_C}\rangle$ in Eq.~(\ref{seq-Psi-nu-nuC-rotate}) has as many fully occupied (or fully empty) valley-spin flavors as possible, due to the nonchiral perturbation energy in Eq. (\ref{seq-E1-nc12}).

The energy of state $|\Psi_{\nu,\nu_C}(\varphi_s)\rangle$ in Eq.~(\ref{seq-Psi-nu-nuC-rotate}) depends on the order of the valley-spin flavors $\{\eta_j,s_j\}$ chosen, and the angles $\varphi_s$ ($s=\uparrow,\downarrow$). One is further allowed to do arbitrary spin-charge rotations in each valley to the state $|\Psi_{\nu,\nu_C}(\varphi_s)\rangle$ in Eq.~(\ref{seq-Psi-nu-nuC-rotate}), which would not affect the energy due to the remaining U(2)$\times$U(2) symmetry. Therefore, for the purpose of examining energies, it is sufficient to consider states of the form in Eq.~(\ref{seq-Psi-nu-nuC-rotate}).

While the calculations below are generic for any order of valley-spin flavors $\{\eta_j,s_j\}$ in Eq.~(\ref{seq-Psi-nu-nuC-rotate}), for concreteness we will dominantly present the case where the 4 valley-spin flavors $\{\eta_j,s_j\}$ ($1\le j\le 4$) in Eq. (\ref{seq:U4-rotator}) are sorted in the order of
\begin{equation}\label{seq-flavor-order1}
\{\eta_j,s_j\}=\{+,\uparrow\},\{+,\downarrow\},\{-,\downarrow\},\{-,\uparrow\}\ ,\qquad (1\le j\le 4),
\end{equation}
which turns out to give the lowest nonchiral-nonflat states at all integer fillings $\nu$. This is because such an order of $\{\eta_j,s_j\}$ makes the state $|\Psi_{\nu,\nu_C}\rangle$ in Eq.~(\ref{seq-Psi-nu-nuC}) maximally valley polarized (always filling the valley $\eta=+$ first), thus it gains as much perturbation energies (with proper rotation angles $\varphi_s$) from the kinetic term which breaks the SU(2) valley rotational symmetry (a subgroup of the U(4)). Another order of $\{\eta_j,s_j\}$ that maximizes the valley polarization of state $|\Psi_{\nu,\nu_C}\rangle$ is 
\begin{equation}\label{seq-flavor-order2}
\{\eta_j,s_j\}=\{+,\uparrow\},\{+,\downarrow\},\{-,\uparrow\},\{-,\downarrow\}\ ,\qquad (1\le j\le 4),
\end{equation}
which is also compared in Tab. \ref{Tab-nu123}, 
but always yields an equal or higher energy state than the order of Eq. (\ref{seq-flavor-order2}). All the other flavor orders which do not maximize the valley polarization yield even higher perturbation energies, and will not be presented here.

\subsubsection{Action of the kinetic term}

In the absence of the chiral symmetry, the kinetic term $H_0=H_0^++H_0^-$ is given by Eq.~(\ref{seq:H0pm}), which can be rewritten into the Chern band basis $d^\dag_{\kk,e_Y}$ (defined as the row vector of 4 spin-valley components $d^\dag_{\kk,e_Y,\eta,s}$) as 
\begin{equation}\label{seq:H0pm-chern}
H_0=H_0^+ +H_0^-\ ,\quad H_0^+=\sum_{\mathbf{k},e_Y} \epsilon_+(\mathbf{k}) d^\dag_{\mathbf{k},-e_Y}(\tau^0s^0)d_{\mathbf{k},e_Y}\ ,\quad H_0^-=\sum_{\mathbf{k},e_Y} \epsilon_-(\mathbf{k}) d^\dag_{\mathbf{k},e_Y} (\tau^z s^0)d_{\mathbf{k},e_Y}\ ,
\end{equation}
where $\epsilon_{\pm}(\kk)=\pm\epsilon_{\pm}(-\kk)$. For later use, here we examine the action of $H_0$ on the wavefunction $|\Psi_{\nu,\nu_C}(\varphi_s)\rangle$ in Eq.~(\ref{seq-Psi-nu-nuC-rotate}) (not the more accurate wavefunction in Eq.~(\ref{seq-Psi-nu-nuC-rotate-2nd})). Recall the representation matrices of $S^{ab}$ under the Chern band basis are given by Eq.~(\ref{seq-SabeY}). Therefore, the action of $H_0$ on the rotated state Eq.~(\ref{seq-Psi-nu-nuC-rotate}) is then
\begin{equation}\label{eq:H0Psi-0}
\begin{split}
&H_0|\Psi_{\nu,\nu_C} (\varphi_s)\rangle=(H_0^++H_0^-)U(\varphi_s)|\Psi_{\nu,\nu_C}\rangle \\
=&U(\varphi_s)\sum_{\mathbf{k},e_Y} \{\epsilon_+(\mathbf{k})d_{\mathbf{k},-e_Y}^\dag [\tau^0(s^0\cos\varphi\cos\phi-s^z\sin\varphi\sin\phi)+ie_Y\tau^y(s^0\sin\varphi\cos\phi+s^z\cos\varphi\sin\phi)]d_{\mathbf{k},e_Y} \\
&\quad +\epsilon_-(\mathbf{k})d^\dag_{\kk,e_Y}[\tau^z(s^0\cos\varphi\cos\phi-s^z\sin\varphi\sin\phi)+e_Y\tau^x(s^0\sin\varphi\cos\phi+s^z\cos\varphi\sin\phi)]d_{\mathbf{k},e_Y}\} |\Psi_{\nu,\nu_C}\rangle \\
=&U(\varphi_s)\sum_s\Big[W_{1,s}\cos\varphi_s-(W_{2,s}-W_{3,s})\sin\varphi_s\Big]|\Psi_{\nu,\nu_C}\rangle\ ,
\end{split}
\end{equation}
where 
\begin{equation}\label{eq:H0Psi-0-W}
\begin{split}
&W_{1,s}=\sum_{\mathbf{k},e_Y,\eta}\epsilon_+(\mathbf{k}) d^\dag_{\mathbf{k},-e_Y,\eta,s}d_{\mathbf{k},e_Y,\eta,s}\ ,\qquad W_{2,s}=\sum_{\mathbf{k},e_Y,\eta}e_Y\eta\epsilon_+(\mathbf{k}) d^\dag_{\mathbf{k},-e_Y,-\eta,s}d_{\mathbf{k},e_Y,\eta,s}\ ,\\
& W_{3,s}=\sum_{\mathbf{k},e_Y,\eta}e_Y\epsilon_-(\mathbf{k}) d^\dag_{\mathbf{k},e_Y,-\eta,s}d_{\mathbf{k},e_Y,\eta,s}\ ,
\end{split}
\end{equation}
and we have used the relation $\varphi_s=\varphi+s\phi$ for $s=\uparrow,\downarrow$, and have used the fact that $\sum_{\kk}\epsilon_-(\kk)=0$. It is easy to verify that $\langle\Psi_{\nu,\nu_C}(\varphi_s)|H_0|\Psi_{\nu,\nu_C}(\varphi_s)\rangle=0$, since the unrotated state $|\Psi_{\nu,\nu_C}\rangle$ is diagonal in $e_Y,\eta$ and $s$.

\subsubsection{Perturbation to the exact nonchiral-flat ground states at even fillings}

For the exact ground states $|\Psi_{\nu,0}\rangle=|\Psi_\nu\rangle$ at even fillings $\nu=0,\pm2$, the first term $W_{1,s}$ in Eq.~(\ref{eq:H0Psi-0}) with coefficient $\cos\varphi_s$ gives $0$ when acting on $|\Psi_{\nu,0}\rangle$. Besides, all the other terms only excite the state $|\Psi_{\nu,0}\rangle$ into gapped excitations (instead of gapless Goldstone modes): (1) any operators flipping $e_Y$ (regardless of whether flipping $\eta,s$ or not) will not be generators of the nonchiral-flat U(4) symmetry group (see the U(4) generators in Eq.~(\ref{eq-SabeY})). Therefore, the term $W_{2,s}$ in Eq.~(\ref{eq:H0Psi-0}) will map the state $|\Psi_{\nu,0}\rangle$ to gapped excitations rather than Goldstone modes (the Goldstone modes are obtained by acting U(4) generators on the state $|\Psi_{\nu,0}\rangle$). (2) The operator $W_{3,s}$ in Eq.~(\ref{eq:H0Psi-0}) does not flip $e_Y$. But the U(4) generators that flip $\eta$ and yields a gapless Goldstone mode (at zero momentum), e.g., $S^{x0}=\sum_{\kk,\eta}d^\dag_{\mathbf{k},e_Y,-\eta,s} d_{\mathbf{k},e_Y,\eta,s}$ (which generate a Goldstone mode $S^{x0}|\Psi_{\nu,0}\rangle$), satisfy
\begin{equation}\label{seq:kinetic-goldstone}
\langle \Psi_{\nu,0} |S^{x0} W_{3,s}| \Psi_{\nu,0}\rangle=\frac{4-|\nu|}{2}\sum_{\kk}\epsilon_-(\kk) =0\ .
\end{equation}
Thus, we conclude that the resulting states of Eq.~(\ref{eq:H0Psi-0}) are orthogonal to the gapless Goldstone modes. This allows us to use nondegenerate perturbation theory. 

From Eq.~(\ref{eq:H0Psi-0}), we can see that the first order perturbation energy is $\langle\Psi_{\nu,0}(\varphi_s)|H_0|\Psi_{\nu,0}(\varphi_s)\rangle=0$.

Now we examine the (nondegenerate) 2nd order perturbation energy. Since $W_{1,s}$ in Eq.~(\ref{eq:H0Psi-0}) yields zero, we find the 2nd order perturbation energy is given by
\begin{equation}\label{seq:approx-E-H0-ncnf-exact}
E^{(2)'}_{\nu,0}=\langle \Psi_{\nu,0}(\varphi_s)|H_0(E_{0,\nu,0}-H_I)^{-1}H_0|\Psi_{\nu,0}(\varphi_s)\rangle=-N_M J'(w_0) \sum_{s=\uparrow,\downarrow}\nu_{s}\sin^2\varphi_s\ ,
\end{equation}
where $\nu_s=1$ if only one valley of the spin $s$ sector in the state $|\Psi_{\nu,0}\rangle$ is fully occupied, and $\nu_s=0$ if both valleys of the spin $s$ sector in the state $|\Psi_{\nu,0}\rangle$ are fully occupied or fully empty. The perturbation energy per moir\'e unit cell $J'(w_0)$ (which depends on $w_0$, and here we assume $w_1=110$meV is fixed) is defined by
\begin{equation}
J'(w_0)=\frac{1}{N_M}\sum_{\ell} \frac{|Y_{\ell}|^2}{E_{\ell}-E_{0,\nu,0}}\ ,
\end{equation}
where $E_{0,\nu,0}$ stands for the unperturbed energy of state $|\Psi_{\nu,0}\rangle$ in the nonchiral-flat limit, while $\ell$ runs over all the excited states $|\ell,\Psi_{\nu,0}\rangle$ reachable by the operator $W_{2,s}-W_{3,s}$ from a state $|\Psi_{\nu,0}\rangle$ with both bands in valley-spin flavor $\{+,s\}$ are fully occupied and both bands in valley-spin flavor $\{-,s\}$ are fully empty. The corresponding amplitude is defined as $Y_{\ell}=\langle \ell,\Psi_{\nu,0}\rangle|W_{2,s}-W_{3,s} |\Psi_{\nu,0}\rangle$. 
The unperturbed energies of the excited states in the nonchiral-flat limit are denoted by $E_{\ell}$. From the more generic expression in Eq.~(\ref{seq:approx-E-H0v2}), one can see that $J'(w_0)=J_2(w_0)+J_3(w_0)$, where $J_i(w_0)$ is defined in Eq.~(\ref{seq:approx-E-H0-J}). 
Generically, since $E_\ell>E_{0,\nu,0}$, we have $J'(w_0)>0$. Therefore, it is clear that the lowest energy is reached when $\varphi_s$ for all $s$ with $\nu_s=1$ are at 
\begin{equation}\label{seq:valley-pol1}
\varphi_s=\pi/2\ ,
\end{equation}
namely, when the valley polarizations of both spins $s$ are maximally polarized in the valley $x$-$y$ plane. This corresponds to $\varphi=\pi/2$ and $\phi=0$ in Eq.~(\ref{seq:U4-rotator}).

\subsubsection{Perturbation to the generic nonchiral-flat Chern states (part I): 1st order}

For generic approximate (Chern) insulator states with Chern number $\nu_C\neq 0$, the state $|\Psi_{\nu,0}\rangle$ in Eq.~(\ref{seq-Psi-nu-nuC-rotate}) is only an approximate eigenstate in the nonchiral-flat limit, which is obtained by adding nonchiral perturbations to the chiral-flat limit. Since we are going to further add the kinetic perturbation $H_0$ to the nonchiral-flat limit, we want to start from a more accurate eigenstate of the nonchiral-flat limit. Recall that as defined in Eq.~(\ref{seq:HI-c-nc12}), the nonchiral interaction terms $H_I^{nc(1)}\propto O^1_{-\qq,-\GG}O^1_{\qq,\GG}\propto w_0^2$, and $H_I^{nc(2)}\propto O^0_{-\qq,-\GG}O^1_{\qq,\GG}\propto w_0$. Therefore, we define the wavefunction accurate up to order $w_0^2$ in the nonchiral-flat limit, which includes the first order perturbation of the nonchiral term $H_I^{nc(1)}$ and second order perturbation of $H_I^{nc(2)}$ to the chiral-flat limit (see Ref.~\cite{winklerbook} for the generic expression of the second order perturbed wavefunction in quantum mechanics):
\begin{equation}\label{seq-Psi-nu-nuC-rotate-2nd}
\begin{split}
|\Psi_{\nu,\nu_C}^{(acc)}(\varphi_s)\rangle&=\mathcal{N}^{-1/2}U(\varphi_s)\Big[1+(E_{0,\nu,\nu_C}^{c}-H_I^{c})^{-1}(\overline{H}_I^{nc(1)}+H_I^{nc(2)})\\
& \qquad + (E_{0,\nu,\nu_C}^{c}-H_I^{c})^{-1}\overline{H_I^{nc(2)}(E_{0,\nu,\nu_C}^{c}-H_I^{c})^{-1}H_I^{nc(2)}} \Big]|\Psi_{\nu,\nu_C}\rangle\\
=&\mathcal{N}^{-1/2}\Big[1+(E_{0,\nu,\nu_C}^{c}-H_I^{c})^{-1}(\overline{H}_I^{nc(1)}+H_I^{nc(2)})\\
&\qquad + (E_{0,\nu,\nu_C}^{c}-H_I^{c})^{-1}\overline{H_I^{nc(2)}(E_{0,\nu,\nu_C}^{c}-H_I^{c})^{-1}H_I^{nc(2)}} \Big]|\Psi_{\nu,\nu_C}(\varphi_s)\rangle\ .
\end{split}
\end{equation}
Here the overlined operator $\overline{Q}=Q-\langle \Psi_{\nu,\nu_C}|Q|\Psi_{\nu,\nu_C}\rangle$ stands for the part of operator $Q$ not mapping back to state $|\Psi_{\nu,\nu_C}\rangle$, and $E_{0,\nu,\nu_C}^{c}$ is the chiral-flat part of the energy of state $|\Psi_{\nu,\nu_C}\rangle$ defined by $H_I^{c}|\Psi_{\nu,\nu_C}\rangle=E_{0,\nu,\nu_C}^{c}|\Psi_{\nu,\nu_C}\rangle$. $\mathcal{N}$ is the normalization factor. 

We shall now regard the state $|\Psi_{\nu,\nu_C}^{(acc)}(\varphi_s)\rangle$ in Eq.~(\ref{seq-Psi-nu-nuC-rotate-2nd}) as an accurate eigenstate of the nonchiral-flat interaction Hamiltonian $H_I=H_I^{c}+H_I^{nc}$, on top of which we add the kinetic perturbation $H_0=H_0^++H_0^-$. Generically, if we define $t$ as the energy scale of the kinetic term $H_0^+$, the other kinetic term $H_0^-$ will be of order $w_0t$, since $H_0^-$ vanishes when $w_0=0$. 

The first order perturbation energy of $H_0$ to the state $|\Psi_{\nu,\nu_C}^{(acc)}(\varphi_s)\rangle$ in Eq.~(\ref{seq-Psi-nu-nuC-rotate-2nd}) is then given by
\begin{equation}\label{seq:H0pert-1st-nc-f}
\begin{split}
&E_{\nu,\nu_C}^{(1)'}(\varphi_s)=\langle \Psi_{\nu,\nu_C}^{(acc)}(\varphi_s)|H_0 |\Psi_{\nu,\nu_C}^{(acc)}(\varphi_s)\rangle \\
&=\mathcal{N}\langle \Psi_{\nu,\nu_C}(\varphi_s)| \left[1+(\overline{H_I^{nc(2)}(E_{0,\nu,\nu_C}^{c}-H_I^{c})^{-1}H_I^{nc(2)}}+H_I^{nc(2)}+\overline{H}_I^{nc(1)}) (E_{0,\nu,\nu_C}^{c}-H_I^{c})^{-1}\right]  (H_0^++H_0^-) \\
&\quad \times \left[1+(E_{0,\nu,\nu_C}^{c}-H_I^{c})^{-1}(\overline{H}_I^{nc(1)}+H_I^{nc(2)}+\overline{H_I^{nc(2)}(E_{0,\nu,\nu_C}^{c}-H_I^{c})^{-1}H_I^{nc(2)}})\right]|\Psi_{\nu,\nu_C}(\varphi_s)\rangle\ .
\end{split}
\end{equation}
We now show this first order perturbation energy is zero. To see this, we first define a unitary $\kk$ flipping operation $I$ which acts as 
\begin{equation}\label{seq-operator-I}
Id^\dag_{\kk,e_Y,\eta,s} I^{-1}=d^\dag_{-\kk,e_Y,\eta,s}\ ,\qquad Id_{\kk,e_Y,\eta,s} I^{-1}=d_{-\kk,e_Y,\eta,s}\ .
\end{equation}
Note that because of the properties of coefficients $\alpha_j(\kk,\qq+\GG)$ in Eq.~(\ref{eq:alpha-cond2}), we have $I O_{\qq,\GG}^0 I^{-1}= O_{-\qq,-\GG}^0$ and $I O_{\qq,\GG}^1I^{-1} = -O_{-\qq,-\GG}^1$. Besides, the kinetic terms satisfy $I H_0^{\pm} I^{-1}I^{-1}=\pm H_0^{\pm}$, because $\epsilon_\pm(\kk)=\pm \epsilon_\pm(-\kk)$. Since the chiral term $H_I^{c}\propto \sum_{\qq,\GG}O_{-\qq,-\GG}^0O_{\qq,\GG}^0$, and the nonchiral terms $H_I^{nc(1)}\propto \sum_{\qq,\GG}O_{-\qq,-\GG}^1O_{\qq,\GG}^1$, $H_I^{nc(2)}\propto \sum_{\qq,\GG}O_{-\qq,-\GG}^0O_{\qq,\GG}^1$, we find that
\begin{equation}\label{seq:nc-f-1st-pert-energy1}
IH_I^{c}I^{-1}= H_I^{c}\ ,\quad IH_I^{nc(1)}I^{-1}= H_I^{nc(1)}\ ,\quad IH_I^{nc(2)}I^{-1}= -H_I^{nc(2)}\ , \quad IH_0^\pm I^{-1}= \pm H_0^\pm\ .
\end{equation}
Furthermore, one can see that the operators
\begin{equation}\label{seq:nc-f-1st-pert-energy2}
H_I^{c}\ ,\ H_I^{nc(1)}\ , H_0^-\qquad \text{do not flip the index}\ e_Y\ \text{of an electron}\ ,
\end{equation}
and
\begin{equation}\label{seq:nc-f-1st-pert-energy3}
H_I^{nc(2)}\ , H_0^+\qquad \text{flip the index}\ e_Y\ \text{of an electron}\ .
\end{equation}

Then, note that the state $|\Psi_{\nu,\nu_C}\rangle $ has conserved number of electrons in each $e_Y$ sector, and satisfies $I|\Psi_{\nu,\nu_C}\rangle= e^{i\alpha_{\nu,\nu_C}} |\Psi_{\nu,\nu_C}\rangle$ under the $\kk$ flipping operation $I$, where $\alpha_{\nu,\nu_C}$ being some phase. Therefore, an operator $Q$ can have nonzero expectation $\langle\Psi_{\nu,\nu_C}|Q|\Psi_{\nu,\nu_C}\rangle$ only if it contains a part that is even under $I$ and does not flip $e_Y$. However, we find every term in Eq.~(\ref{seq:H0pert-1st-nc-f}) either is odd under $I$, or flips the $e_Y$ index of at least one electron. For instance, the term $\overline{H_I^{nc(2)}(E_{0,\nu,\nu_C}^{c}-H_I^{c})^{-1}H_I^{nc(2)}} (E_{0,\nu,\nu_C}^{c}-H_I^{c})^{-1}H_0^+ \overline{H}_I^{nc(1)}$ will flip $e_Y$, while the term $\overline{H_I^{nc(2)}(E_{0,\nu,\nu_C}^{c}-H_I^{c})^{-1}H_I^{nc(2)}} (E_{0,\nu,\nu_C}^{c}-H_I^{c})^{-1}H_0^- \overline{H}_I^{nc(1)}$ is odd under the $\kk$ flipping operation $I$. Therefore, we conclude that the first order perturbation of $H_0$ is zero:
\begin{equation}\label{seq:nc-f-1st-pert-energy}
E_{\nu,\nu_C}^{(1)'}(\varphi_s)=0\ .
\end{equation}

\subsubsection{Perturbation to the generic nonchiral-flat Chern states (part II): 2nd order}\label{app:kineticpert-nonchiralflat2}

For the second order perturbation of $H_0$, since the energy scale of the second-order perturbation of $H_0$ is smaller (than the first-order perturbation of $H_0$, if nonzero), it is sufficient to approximate the nonchiral-flat state $|\Psi_{\nu,\nu_C}^{(acc)} (\varphi_s)\rangle$ in Eq.~(\ref{seq-Psi-nu-nuC-rotate-2nd}) as the less accurate expression $|\Psi_{\nu,\nu_C} (\varphi_s)\rangle$ in (\ref{seq-Psi-nu-nuC-rotate}) for the purpose here. Similar to the argument in Eq.~(\ref{seq:kinetic-goldstone}), all the resulting states of Eq.~(\ref{eq:H0Psi-0}) are orthogonal to the gapless Goldstone modes and thus are gapped, so we can employ the nondegenerate perturbation theory. The second order perturbation energy is then given by
\begin{equation}
\begin{split}
&E^{(2)'}_{\nu,\nu_C}(\varphi_s)
= \langle \Psi_{\nu,\nu_C} (\varphi_s)|H_0(E_{0,\nu,\nu_C}-H_I)^{-1}H_0|\Psi_{\nu,\nu_C} (\varphi_s)\rangle\ ,
\end{split}
\end{equation}
where $H_I$ is the nonchiral-flat interaction Hamiltonian, and $E_{0,\nu,\nu_C}=\langle \Psi_{\nu,\nu_C}|H_I|\Psi_{\nu,\nu_C}\rangle $ is the energy of the state $|\Psi_{\nu,\nu_C}\rangle$ in the nonchiral-flat limit. 
All the terms $W_{j,s}$ ($1\le j\le3$) in Eq.~(\ref{eq:H0Psi-0}) contribute here, and yield a 2nd order perturbation energy expression:
\begin{equation}\label{seq:approx-E-H0}
\begin{split}
E^{(2)'}_{\nu,\nu_C}(\varphi_s)=-N_M\sum_{s=\uparrow,\downarrow} \left[  \nu_{s}^{(1)}J_1(w_0)\cos^2\varphi_s + \sin^2\varphi_s \Big(\nu_{s}^{(2)}  J_2(w_0)+\nu_{s}^{(3)}  J_3(w_0) \Big)\right],
\end{split}
\end{equation}
where the three different coefficients are given by
\begin{equation}\label{seq:approx-E-H0-J}
J_1(w_0)=\frac{1}{N_M}\sum_{\ell} \frac{|Y_{1,\ell}|^2}{E_{\ell}-E_{0,\nu,\nu_C}} ,\ J_2(w_0)=\frac{1}{N_M}\sum_{\ell} \frac{|Y_{2,\ell}|^2}{E_{\ell}-E_{0,\nu,\nu_C}} , \  J_3(w_0)=\frac{1}{N_M}\sum_{\ell} \frac{|Y_{3,\ell}|^2+Y_{2,\ell}^*Y_{3,\ell}+Y_{2,\ell}Y_{3,\ell}^*}{E_{\ell}-E_{0,\nu,\nu_C}} .
\end{equation}
In the main text Eq.~(\ref{eq-E2-nc-nf}), we simply write $J_i=J_i(w_0)$ for short, while here we keep the variable $w_0$ to remind the readers that they depend on $w_0$. The three terms in Eq.~(\ref{seq:approx-E-H0}) are from the following second order perturbations:

(1) By the amplitudes $Y_{1,\ell}=\langle \ell,\Psi_{\nu,\nu_C}|W_{1,s}|\Psi_{\nu,\nu_C}\rangle$ of the term $W_{1,s}= \sum_{\kk} \epsilon_+(\mathbf{k}) d^\dag_{\mathbf{k},-e_Y,\eta,s}d_{\mathbf{k},e_Y,\eta,s}$ (see Eq.~(\ref{eq:H0Psi-0-W})) from the insulating state $|\Psi_{\nu,\nu_C}\rangle$ to all the excited states $|\ell,\Psi_{\nu,\nu_C}\rangle$, where the band $\{e_Y,\eta,s\}$ is occupied while the band $\{-e_Y,\eta,s\}$ is empty. The energies of the excited states $|\ell,\Psi_{\nu,\nu_C}\rangle$ relative to the energy of state $|\Psi_{\nu,\nu_C}\rangle$ are denoted by $E_\ell-E_{0,\nu,\nu_C}$. The number of such indices $\{e_Y,\eta,s\}$ for a fixed $s$ is denoted by $\nu_s^{(1)}\ge0$.

(2) By the amplitudes $Y_{2,\ell}=\langle \ell,\Psi_{\nu,\nu_C}|W_{2,s}|\Psi_{\nu,\nu_C}\rangle$ induced by the term $W_{2,s}= \sum_{\kk} \epsilon_+(\mathbf{k}) d^\dag_{\mathbf{k},-e_Y,-\eta,s}d_{\mathbf{k},e_Y,\eta,s}$ (see Eq.~(\ref{eq:H0Psi-0-W})) from the insulating state $|\Psi_{\nu,\nu_C}\rangle$ to all the excited states $|\ell,\Psi_{\nu,\nu_C}\rangle$, where the band $\{e_Y,\eta,s\}$ is occupied while the band $\{-e_Y,-\eta,s\}$ is empty. The number of such indices $\{e_Y,\eta,s\}$ for a fixed $s$ is denoted by $\nu_s^{(2)}\ge0$.

(3) By the amplitudes $Y_{3,\ell}=\langle \ell,\Psi_{\nu,\nu_C}|W_{3,s}|\Psi_{\nu,\nu_C}\rangle$ of the term $W_{3,s}= \sum_{\kk} \epsilon_-(\mathbf{k}) d^\dag_{\mathbf{k},e_Y,-\eta,s}d_{\mathbf{k},e_Y,\eta,s}$ (see Eq.~(\ref{eq:H0Psi-0-W})) from the insulating state $|\Psi_{\nu,\nu_C}\rangle$ to all the excited states $|\ell,\Psi_{\nu,\nu_C}\rangle$, where the band $\{e_Y,\eta,s\}$ is occupied while the band $\{e_Y,-\eta,s\}$ is empty. Both $|Y_{3,\ell}|^2$ and the cross terms between $Y_{2,\ell}$ and $Y_{3,\ell}$ contribute. The number of such indices $\{e_Y,\eta,s\}$ for a fixed $s$ is denoted by $\nu_s^{(3)}\ge0$.

For numerical calculations of the coefficients $J_{i}(w_0)$ (see App. \ref{app:kineticpert-nonchiralflat3} for results), we only restrict the excitation levels $|\ell,\Psi_{\nu,\nu_C}\rangle$ within the exactly solvable excitations found in Ref.~\cite{ourpaper5} in the sub-Hilbert space of basis $d^\dag_{\kk,e_Y,\eta,s}d_{\kk,e_Y',\eta',s'}|\Psi_{\nu,\nu_C}\rangle$ (of any $\kk,e_Y,\eta,s,e_Y',\eta',s'$). 
These excitations can be obtained by diagonalizing a one-body Hamiltonian matrix in this sub-Hilbert space \cite{ourpaper5}. Such excitations are the ones $H_0|\Psi_{\nu,\nu_C}\rangle$ dominantly overlap with, since $H_0$ excites only one electron-hole pair with zero total momentum.

In particular, when the interaction $H_I$ is in the chiral limit ($w_0=0$), we would have the contribution of the $|Y_{1,\ell}|$ term being exactly the same as that of the $|Y_{2,\ell}|$ term, which is no different from the amplitude $|Y_\ell|$ in the chiral limit in Eq.~(\ref{eq:chern-energy-H0-2nd-order}). Indeed these two terms are related by a chiral-flat U(4)$\times$U(4) rotation in the Chern number $-e_Y$ basis. Besides, $H_0^-=0$ when $w_0=0$, so $|Y_{3,\ell}|=0$ as well. Therefore, in the chiral limit $w_0=0$, we have
\begin{equation}
J_1(w_0=0)=J_2(w_0=0)=J_0\ ,\qquad  J_3(w_0=0)=0\ ,
\end{equation}
where $J_0$ is the coupling in the chiral limit defined in Eq.~(\ref{eq:chern-energy-H0-2nd-order}). When $w_0>0$ is small, we have $J_1(w_0)-J_2(w_0)\propto w_0^2t^2$, where $t$ is the bandwidth. This is because the differences between the excitation energies $E_{1,\ell}$ and $E_{2,\ell}$ are proportional to the coefficients $F_{e_Y}^2(\kk,\qq+\GG)\propto w_0^2$. Besides, the third coupling $J_3(w_0)\propto w_0^2 t^2$, which can be seen by noting that the kinetic term $H_0^-\propto \epsilon_-(\kk)\propto w_0 t$, and the fact that the cross terms are proportional to coefficients $\epsilon_-\epsilon_+F_{e_Y}$. We also note that by definition, the coupling $J'(w_0)$ in Eq.~(\ref{seq:approx-E-H0-ncnf-exact}) is equal to $J'(w_0)=J_2(w_0)+J_3(w_0)$.

\begin{table}[tbp]
\centering
\begin{tabular}{c|c|c|c|c|c|c|c|c|c}
\hline
\multicolumn{10}{c}{flavor order $\{+,\uparrow\}$, $\{+,\downarrow\}$, $\{-,\downarrow\}$, $\{-,\uparrow\}$ (giving the lowest state)} \\
\hline
 filling $\nu$ & Chern number $|\nu_C|$ & $\nu_\uparrow^{(1)}$ & $\nu_\uparrow^{(2)}$ & $\nu_\uparrow^{(3)}$ & $\nu_\downarrow^{(1)}$ & $\nu_\downarrow^{(2)}$ & $\nu_\downarrow^{(3)}$ & $\varphi_\uparrow$ & $\varphi_\downarrow$ \\
\hline
0 & 0 & 0 & 2 & 2 & 0 & 2 & 2 & $\pi/2$ & $\pi/2$ \\
0 & 2 & 0 & 2 & 2 & 2 & 2 & 0 & $\pi/2$ & $0$ \\
0 & 4 & 2 & 2 & 0 & 2 & 2 & 0 & $0$ & $0$ \\
-1 & 1 & 0 & 2 & 2 & 1 & 1 & 1 & $\pi/2$ & $0$ \\
-1 & 3 & 1 & 1 & 1 & 2 & 2 & 0 & $0$ & $0$ \\
-2 & 0 & 0 & 2 & 2 & 0 & 0 & 0 & $\pi/2$ & - \\
-2 & 2 & 1 & 1 & 1 & 1 & 1 & 1 & $0$ & $0$ \\
-3 & 1 & 1 & 1 & 1 & 0 & 0 & 0 & $0$ & - \\
\hline
\end{tabular}
\qquad \quad
\begin{tabular}{c|c|c|c|c|c|c|c|c|c}
\hline
\multicolumn{10}{c}{flavor order $\{+,\uparrow\}$, $\{+,\downarrow\}$, $\{-,\uparrow\}$, $\{-,\downarrow\}$} \\
\hline
 $\nu$ & $|\nu_C|$ & $\nu_\uparrow^{(1)}$ & $\nu_\uparrow^{(2)}$ & $\nu_\uparrow^{(3)}$ & $\nu_\downarrow^{(1)}$ & $\nu_\downarrow^{(2)}$ & $\nu_\downarrow^{(3)}$ & $\varphi_\uparrow$ & $\varphi_\downarrow$ \\
\hline
0 & 0 & 0 & 2 & 2 & 0 & 2 & 2 & $\pi/2$ & $\pi/2$ \\
0 & 2 & 1 & 1 & 1 & 1 & 1 & 1 & $0$ & $0$ \\
0 & 4 & 2 & 2 & 0 & 2 & 2 & 0 & $0$ & $0$ \\
-1 & 1 & 0 & 2 & 2 & 1 & 1 & 1 & $\pi/2$ & $0$ \\
-1 & 3 & 2 & 2 & 0 & 1 & 1 & 1 & $0$ & $0$ \\
-2 & 0 & 0 & 2 & 2 & 0 & 0 & 0 & $\pi/2$ & - \\
-2 & 2 & 1 & 1 & 1 & 1 & 1 & 1 & $0$ & $0$ \\
-3 & 1 & 1 & 1 & 1 & 0 & 0 & 0 & $0$ & - \\
\hline
\end{tabular}
\caption{(a) \emph{Left table}: the coefficients $\nu_s^{(j)}$ of state (\ref{seq-Psi-nu-nuC-rotate}) and the optimal U(4) rotation angles $\varphi_s$ for different integer fillings $\nu\le 0$, where the valley-spin flavors $\{\eta_j,s_j\}$ ($1\le j\le 4$) in state (\ref{seq-Psi-nu-nuC-rotate}) are sorted in the order of $\{+,\uparrow\}$, $\{+,\downarrow\}$, $\{-,\downarrow\}$, $\{-,\uparrow\}$ (this flavor order maximizes the valley polarization of state (\ref{seq-Psi-nu-nuC-rotate}) and minimizes the perturbation energy in Eq. (\ref{seq:approx-E-H0v2})). When $\varphi_s$ is filled by ``-", it means there is no electron in the spin $s$ sector. (b) \emph{Right table}: the coefficients $\nu_s^{(j)}$ and the optimal rotation angles $\varphi_s$ if the valley-spin flavors of state (\ref{seq-Psi-nu-nuC-rotate}) are sorted in the order $\{+,\uparrow\}$, $\{+,\downarrow\}$, $\{-,\uparrow\}$, $\{-,\downarrow\}$. This flavor order yields an equal or higher perturbation energy according to Eq. (\ref{seq:approx-E-H0v2}) compared to the flavor order in the left table, as one can easily verify.}\label{Tab-nu123}
\end{table}

From the definition of the state $|\Psi_{\nu,\nu_C}\rangle$, and recall that $\nu_++\nu_-=4+\nu$ and $\nu_+-\nu_-=\nu_C$, one can calculate $\nu_s^{(j)}$ in Eq. (\ref{seq:approx-E-H0}), which depend on the order of valley-spin flavors $\{\eta_j,s_j\}$ chosen. For the valley-spin flavor orders defined in Eqs.~(\ref{seq-flavor-order1}) and (\ref{seq-flavor-order2}), the values of $\nu_s^{(j)}$ for all integer fillings $\nu\le 0$ and different Chern numbers $\nu_C$ are summarized in Tab. \ref{Tab-nu123}. 
In both cases, the perturbation energy expression in Eq. (\ref{seq:approx-E-H0}) holds. 

One can then rewrite the 2nd order perturbation energy as
\begin{equation}\label{seq:approx-E-H0v2}
E^{(2)'}_{\nu,\nu_C}(\varphi_s)\approx-N_M\sum_{s=\uparrow,\downarrow} \left[  \nu_{s}^{(1)}J_1(w_0) + \Big(\nu_{s}^{(2)}J_2(w_0)-\nu_{s}^{(1)}J_1(w_0)+\nu_{s}^{(3)}J_3(w_0)\Big)\sin^2\varphi_s \right].
\end{equation}
Since the 1st order perturbation energy is zero (Eq. (\ref{seq:nc-f-1st-pert-energy})), the lowest state is determined by the 2nd order perturbation energy in Eq. (\ref{seq:approx-E-H0v2}). Therefore, from Eq. (\ref{seq:approx-E-H0v2}), given a particular valley-spin flavor order (which determines the coefficients $\nu_s^{(j)}$), we find the lowest state is achieved when
\begin{equation}\label{seq:valley-polar-1}
\varphi_s=\pi/2 \ \  (\text{or}\ 0)\ \qquad \text{if}\ \nu_{s}^{(2)}J_2(w_0)-\nu_{s}^{(1)}J_1(w_0)+\nu_{s}^{(3)}J_3(w_0)>0 \ \  (\text{or} \ <0)\ .
\end{equation}

\subsubsection{Perturbation to the generic nonchiral-flat Chern states (part III): numerical calculation}\label{app:kineticpert-nonchiralflat3}

We numerically evaluate the values of $J_i(w_0)$ by dividing the MBZ into a $12\times 12$ momentum lattice (sufficiently large to simulate the thermodynamic limit), and summing over all the one electron-hole pair neutral excitations $|\ell,\Psi_{\nu,\nu_C}\rangle$ we analytically derived in Ref.~\cite{ourpaper5} (which are obtainable by diagonalizing a one-body Hamiltonian). The numerical values of $J_i(w_0)$ with the FMC (Eq.~(\ref{eq-Mq=0})) for twist angle $\theta=1.05^\circ$ are given in Tabs.~\ref{Tab-J123}, which are independent of the filling $\nu$. We also calculated their values without the FMC in Tab. \ref{Tab-J123-noFMC}, which are almost the same as that in the presence of the FMC, although they gain a very small $\nu$ dependence. This implies that the FMC is good approximation. Generically, we find $J_2(w_0)\gg J_1(w_0)-J_2(w_0)> J_3(w_0)>0$ for $w_0>0$. 

We then examine the $\nu_s^{(j)}$ coefficients of all possible valley-spin flavor orders, and identify the lowest states by Eq.~(\ref{seq:valley-polar-1}). We find that among all possible valley-spin flavor orders, the flavor order of Eq. (\ref{seq-flavor-order1}) (i.e., $\{\eta_j,s_j\}$ ($1\le j\le 4$) sorted in the order of $\{+,\uparrow\}$, $\{+,\downarrow\}$, $\{-,\downarrow\}$, $\{-,\uparrow\}$) yields the lowest states for all $\nu$ and $\nu_C$. The coefficients $\nu_s^{(j)}$ for this flavor order are summarized in the left table of Tab. \ref{Tab-nu123}, which satisfy
\begin{equation}\label{seq:nusj}
\sum_{s}\nu_s^{(1)}=|\nu_+-\nu_-|=|\nu_C|,\qquad  \sum_{s}\nu_s^{(2)}=4-|\nu|, \qquad \sum_{s}\nu_s^{(3)}=4-|\nu_+-2|-|\nu_--2|.
\end{equation}
In particular, one has $\nu_s^{(1)}\le \nu_s^{(2)}$ for each $s$. For any $\nu$ and $\nu_C$, we have $\sum_s\nu_s^{(2)}\ge \sum_s\nu_s^{(1)}$ and $\sum_s\nu_s^{(2)}\ge \sum_s\nu_s^{(3)}$. 
Besides, when the Chern number $|\nu_C|=4-|\nu|$, we have $\nu_s^{(1)}=\nu_s^{(2)}\ge\nu_s^{(3)}$. The conditions for the lowest states (which are achieved with the flavor order in Eq.~(\ref{seq-flavor-order1})) can be summarized as follows:

(i) if $|\nu_C|=0$, the lowest state favors $\varphi_\uparrow=\varphi_\downarrow=\pi/2$, namely, \emph{full intervalley coherence} with the valley polarization of all electrons in the $x$-$y$ plane of valley Bloch sphere. This agrees with our rigorous conclusion in Eq.~(\ref{seq:valley-pol1}) for the exact ground states at even fillings $\nu$ and Chern number $\nu_C=0$.

(ii) if $0<|\nu_C|<4-|\nu|$, the lowest state prefers $\varphi_\uparrow=\pi/2$ and $\varphi_\downarrow=0$. Therefore, the ground state is \emph{partially intervalley coherent}: the spin $\uparrow$ electrons are intervalley coherent, while the spin $\downarrow$ electrons are valley polarized. 

(iii) if $\nu_C=4-|\nu|$, the lowest state has $\varphi_\uparrow=\varphi_\downarrow=0$, namely, \emph{fully valley polarized} (polarized in the $z$ direction of valley Bloch sphere).

\begin{table}[tbp]
\centering
\begin{tabular}{c|c|c|c|c}
\hline
 $w_0/w_1$ & filling $\nu$ & $J_1(w_0)$ (meV) & $J_2(w_0)$ (meV) & $J_3(w_0)$ (meV)  \\
\hline
0 & -3 & 0.3018 & 0.3018 & 0\\
0.2 & -3 & 0.2650 & 0.2625 & 0.00008 \\
0.4 & -3 & 0.1735 & 0.1674 & 0.0003 \\
0.6 & -3 & 0.0751 & 0.0701 & 0.0004 \\
0.8 & -3 & 0.0174 & 0.0158 & 0.0003 \\
\hline
0 & -1 & 0.3018 & 0.3018 & 0\\
0.2 & -1 & 0.2650 & 0.2625 & 0.0003 \\
0.4 & -1 & 0.1735 & 0.1673 & 0.0010 \\
0.6 & -1 & 0.0751 & 0.0699 & 0.0014 \\
0.8 & -1 & 0.0174 & 0.0156 & 0.0009 \\
\hline
\end{tabular}
\caption{The numerically calculated perturbation energies $J_i$ ($i=1,2,3$) at twist angle $\theta=1.05^\circ$ without the FMC (Eq.~(\ref{eq-Mq=0})), and Coulomb screening length $\xi=10$nm. The calculation is done by dividing the MBZ into a $12\times12$ momentum lattice. Their values depend on $w_0$ (we assume $w_1=110$meV is fixed), and slightly depend on filling $\nu$. These values are approximately equal to that calculated with the FMC imposed in Tab. \ref{Tab-J123}.}\label{Tab-J123-noFMC}
\end{table}

From Eq.~(\ref{seq-Psi-nu-nuC-rotate}), we can explicitly write down the lowest many-body wavefunctions for any $\nu=\nu_++\nu_--4$ and $\nu_C=\nu_+-\nu_-$ as follows:
\begin{equation}\label{seq-Psi-nc-nf0}
\boxed{ |\Psi_{\nu,\nu_C}^{\text{nc-nf}}\rangle=\prod_{\mathbf{k}} \left[\prod_{j=1}^{\nu_L} \left( \prod_{e_Y=\pm} \frac{d^\dag_{\mathbf{k},e_{Y},\eta_j,s_j}+\eta_j e_{Y}d^\dag_{\mathbf{k},e_{Y},-\eta_j,s_j}}{\sqrt{2}}\right)\prod_{j=\nu_L+1}^{\nu_L+|\nu_C|} d^\dag_{\mathbf{k},\text{sgn}(\nu_C),\eta_j,s_j}\right]|0\rangle }\ ,
\end{equation}
where we have defined $\nu_L=\min(\nu_+,\nu_-)$, and  $\{\eta_j,s_j\}$ ($1\le j\le4$) are sorted with $j$ in the order of $\{+,\uparrow\}$, $\{+,\downarrow\}$, $\{-,\downarrow\}$, $\{-,\uparrow\}$. Note that this expression of wavefunction covers all the three cases, the $\nu_C=0$, $0<|\nu_C|<4-|\nu|$ and $|\nu_C|=4-|\nu|$. One can always further rotate the spin and charge within each valley without changing the total energy. As a result, one finds the generic ground state is given by
\begin{equation}\label{eq:H0perturbed-GS-rot}
|\Psi_{\nu,\nu_C}^{\text{nc-nf}},(\gamma_{\pm},\hat{\mathbf{s}}_+,\hat{\mathbf{s}}_-)\rangle =\prod_{\mathbf{k}} \left[ \prod_{j=1}^{\nu_{L}} \left(\prod_{e_Y=\pm} \frac{d^\dag_{\mathbf{k},e_{Y},\eta_j,\chi_j\hat{\mathbf{s}}_{\eta_j}}+e^{i\eta_j\gamma_{\chi_j}}\eta_j e_{Y}d^\dag_{\mathbf{k},e_{Y},-\eta_j,\chi_j\hat{\mathbf{s}}_{-\eta_j}}}{\sqrt{2}}\right) \prod_{j=\nu_L+1}^{\nu_L+|\nu_C|} d^\dag_{\mathbf{k},\text{sgn}(\nu_C),\eta_j,\chi_j\hat{\mathbf{s}}_{\eta_j}} \right]|0\rangle\ ,
\end{equation}
where $d^\dag_{k,e_Y,\eta,\hat{\mathbf{s}}}$ is the electron basis of band $m$ and valley $\eta$ with spin polarization along the direction of unit vector $\hat{\mathbf{s}}$, and $\{\eta_j,\chi_j\}$ ($1\le j\le4$) are in the order of $\{+,+\}$, $\{+,-\}$, $\{-,-\}$, $\{-,+\}$. The unit vector parameters $\gamma_\pm,\hat{\mathbf{s}}_+,\hat{\mathbf{s}}_-$ can be chosen arbitrarily, which are all degenerate.

In particular, for the Chern insulators with the highest possible Chern number $\nu_C=4-|\nu|$, the ground states are valley $z$ polarized, meaning that their wavefunctions are simply
\begin{equation}\label{seq-Psi-nc-nf2}
|\Psi_{\nu,\nu_C}^{\text{nc-nf}}\rangle=|\Psi_{\nu,\nu_C}(\varphi_s=0)\rangle=|\Psi_{\nu,\nu_C}\rangle\ , \qquad (|\nu_C|=4-|\nu|)
\end{equation}
with $|\Psi_{\nu,\nu_C}\rangle$ defined in Eq.~(\ref{seq-Psi-nu-nuC-rotate}).  
Any further U(2)$\times$U(2) rotations of these states are also degenerate.

\begin{table}[tbp]
\centering
\begin{tabular}{c|c|c|c|c}
\hline
 filling $\nu$ & Chern number $|\nu_C|$ & Little group & Little group generators & valley U(1)$_V$ symmetry \\
\hline
0 & 0 & U(1)$_C\times$SU(2) & $\tau^0s^{0,x,y,z}$ & no \\
0 & 2 & U(1)$_C\times$U(1)$\times$U(1) & $\tau^0s^0, \tau^0s^z, \tau^z(s^0-s^z)/2$ & no  \\
0 & 4 & U(2)$\times$U(2) & $\tau^0 s^{0,x,y,z}, \tau^z s^{0,x,y,z}$ & yes  \\
-1 & 1 & U(1)$_C\times$U(1)$\times$U(1) & $\tau^0s^0, \tau^0s^z, \tau^z(s^0-s^z)/2$ & no  \\
-1 & 3 & SU(2)$_K\times$U(1)$_C\times$U(1)$_V\times$U(1) & $\tau^0s^0, \tau^0s^z, \tau^zs^0, \tau^zs^z, (\tau^0+\tau_z)s^{x,y}/2$ & yes  \\
-2 & 0 & U(1)$_C\times$U(1)$\times$U(1) & $\tau^0s^0, \tau^0s^z, \tau^z(s^0-s^z)/2$ & no  \\
-2 & 2 & U(2)$\times$U(2) & $\tau^0 s^{0,x,y,z}, \tau^z s^{0,x,y,z}$ & yes  \\
-3 & 1 & SU(2)$_{K'}\times$U(1)$_C\times$U(1)$_V\times$U(1) & $\tau^0s^0, \tau^0s^z, \tau^zs^0, \tau^zs^z, (\tau^0-\tau_z)s^{x,y}/2$ & yes  \\
\hline
\end{tabular}
\caption{Little group (remaining symmetry subgroup of the nonchiral-nonflat U(2)$\times$U(2) group) of the nonchiral-nonflat state in Eq. (\ref{seq-Psi-nc-nf0}). We decompose the U(2)$\times$U(2) symmetry into SU(2)$_K\times$SU(2)$_{K'}\times$U(1)$_C\times$U(1)$_V$, with SU(2)$_\eta$ being the spin rotation symmetry of valley $\eta$ (generators $(\tau^0+\eta\tau_z)s^a/2$), U(1)$_C$ being the global charge U(1) symmetry (generator $\tau^0s^0$), and U(1)$_V$ being the valley U(1) symmetry (generator $\tau^zs^0$).}\label{Tab-little}
\end{table}

Generically, the nonchiral-nonflat state in Eq. (\ref{seq-Psi-nc-nf0}) break the nonchiral-nonflat U(2)$\times$U(2) symmetry, except for the $|\nu_C|=4-|\nu|$ states at even fillings $\nu$. Tab. \ref{Tab-little} shows the remaining symmetry little group of the states in Eq. (\ref{seq-Psi-nc-nf0}) at $\nu\le 0$. Note that if $|\nu_C|< 4-|\nu|$, one has $\nu_L>0$ in Eqs. (\ref{seq-Psi-nc-nf0}) and (\ref{eq:H0perturbed-GS-rot}), and the state $|\Psi_{\nu,\nu_C}^{\text{nc-nf}},(\gamma_\pm,\hat{\mathbf{s}}_+,\hat{\mathbf{s}}_-)\rangle$ is spontaneously breaking the valley U(1)$_V$ symmetry generated by $S^{z0}=\tau^zs^0$. This can be seen by noting that 
\begin{equation}
e^{i\beta S^{z0}/2}|\Psi_{\nu,\nu_C}^{\text{nc-nf}},(\gamma_\pm,\hat{\mathbf{s}}_+,\hat{\mathbf{s}}_-)\rangle= e^{i\beta/2} |\Psi_{\nu,\nu_C}^{\text{nc-nf}},(\gamma_\pm-\beta,\hat{\mathbf{s}}_+,\hat{\mathbf{s}}_-)\rangle\ ,
\end{equation}
namely, $e^{i\beta S^{z0}/2}$ maps the state with angles $\gamma_\pm$ to a different (but degenerate) state with angles $\gamma_\pm-\beta$. In fact, as shown in Tab. \ref{Tab-little}, such states with $|\nu_C|< 4-|\nu|$ still have multiple U(1) remaining symmetry groups, but these U(1) are combination of valley and spin rotations, instead of pure valley U(1)$_V$ rotations.

Due to the nonchiral interaction energy (see Eq.~\ref{seq-E1-nc}), the state with the lowest $|\nu_C|$ at a fixed filling $\nu$ is the lowest, and will be the ground state of filling $\nu$. We note that the $\nu=0,\pm2$, $\nu_C=0$ states we find here agree with the K-IVC state proposed in Ref.~\cite{bultinck_ground_2020} at $\nu=0,\pm2$.

\subsection{Another viewpoint: nonchiral perturbation to the (first) chiral-nonflat limit}\label{app:ncpert-chiralnonflat}

As an alternative to the study of the nonchiral-nonflat ground states, we can treat the nonchiral interaction terms as perturbation to the chiral-nonflat Chern insulator states we found in Eq.~(\ref{eq-Psi-nu-nuC2}). Here we assume $w_0<t$, so $w_0$ can be viewed as a perturbation to the approximate eigenstates in the chiral-nonflat limit. As we will show in this section, this will lead to the same lowest state as that given by Eq.~(\ref{eq-Psi-nc-nf}) (see also Eq.~(\ref{seq-Psi-nc-nf0})). 

To see this, we consider the following valley rotated state by 
\begin{equation}\label{seq-Psi-nu-nuC2-rotate}
|\widetilde{\Psi}_{\nu,\nu_C}(\varphi_s)\rangle =U'(\varphi_s)|\widetilde{\Psi}_{\nu,\nu_C}\rangle = U'(\varphi_s) \prod_{\mathbf{k}} \prod_{j=1}^{\nu_+}d^\dag_{\mathbf{k},+1,\eta_{j},s_{j}} \prod_{j=5-\nu_-}^{4}d^\dag_{\mathbf{k},-1,\eta_{j},s_{j}}|0\rangle\ ,
\end{equation}
where $|\widetilde{\Psi}_{\nu,\nu_C}\rangle$ is defined in Eq.~(\ref{eq-Psi-nu-nuC2}), and $\varphi_s$ ($s=\uparrow,\downarrow$) are two angles parameterizing the rotation.  The rotation 
\begin{equation}
U'(\varphi_s)=e^{i(\varphi_\uparrow+\varphi_\downarrow) \widetilde{S}^{y0}/4}e^{i(\varphi_\uparrow-\varphi_\downarrow) \widetilde{S}^{yz}/4}
\end{equation}
is given by the chiral-nonflat U(4) generators $\widetilde{S}^{ab}$ (the representation matrices in the Chern basis are $\tau^as^b$), so $\varphi_s$ is the valley rotation angle in the spin $s$ sector. 

Generically, we need to compare the energy of the states in Eq. (\ref{seq-Psi-nu-nuC2-rotate}) with all possible valley-spin flavors $\{\eta_j,s_j\}$ ($1\le j\le4$). As an example, in this section we will only present the results for the flavor order of Eq. (\ref{seq-flavor-order1}), namely, $\{\eta_j,s_j\}$ ($1\le j\le4$) sorted as $\{+,\uparrow\}$, $\{+,\downarrow\}$, $\{-,\downarrow\}$, $\{-,\uparrow\}$. Similar to that in App. \ref{app:kineticpert-nonchiralflat}, we checked all the flavor orders, and find this flavor order in Eq. (\ref{seq-flavor-order1}) yields the lowest energy state for all $\nu$ and $\nu_C$.

\subsubsection{1st order perturbation}

Before we add nonchiral perturbations to the chiral-nonflat limit, we want to get a more precise expression for the eigenstate $|\widetilde{\Psi}_{\nu,\nu_C}(\varphi_s)\rangle$ in the chiral-nonflat limit to start with. We include the perturbation of $H_0^+$ (the kinetic term in the chiral-nonflat limit) to the wavefunction $|\widetilde{\Psi}_{\nu,\nu_C}(\varphi_s)\rangle$ in Eq. (\ref{seq-Psi-nu-nuC2-rotate}) (which is exact in the chiral-flat limit) up to order $t^2$ ($t$ being the bandwidth of term $H_0^+$). Namely, the wavefunction in the chiral-nonflat limit which we start with is given by (see Ref.~\cite{winklerbook} for the generic perturbation theory):
\begin{equation}\label{seq-Psi-nu-nuC2-rotate-2nd}
|\widetilde{\Psi}^{(acc)}_{\nu,\nu_C}(\varphi_s)\rangle=\mathcal{N}^{-1/2}[1+(E_{0,\nu,\nu_C}-H_I^c)^{-1}(H_0^++\overline{H_0^+(E_{0,\nu,\nu_C}-H_I^c)^{-1}H_0^+}]|\widetilde{\Psi}_{\nu,\nu_C}(\varphi_s)\rangle\ ,
\end{equation}
where $\overline{Q}=Q-\langle \widetilde{\Psi}_{\nu,\nu_C}(\varphi_s)|Q|\widetilde{\Psi}_{\nu,\nu_C}(\varphi_s)\rangle$ is the part of operator $Q$ not mapping back to the state $|\widetilde{\Psi}_{\nu,\nu_C}(\varphi_s)\rangle$ itself, $\mathcal{N}=1+\langle \widetilde{\Psi}_{\nu,\nu_C}(\varphi_s)| H_0^+(E_{0,\nu,\nu_C}-H_I^c)^{-2}H_0^+ |\widetilde{\Psi}_{\nu,\nu_C}(\varphi_s)\rangle $ is the normalization factor up to order $t^2$, and $E_{0,\nu,\nu_C}$ is the energy of state $|\widetilde{\Psi}_{\nu,\nu_C}\rangle$ in the chiral-flat limit.

We then treat the state $|\widetilde{\Psi}^{(acc)}_{\nu,\nu_C}(\varphi_s)\rangle$ in Eq.~(\ref{seq-Psi-nu-nuC2-rotate-2nd}) as the accurate eigenstate in the chiral-nonflat limit, and add the nonchiral interaction $H_I^{nc}=H_I^{nc(1)}+H_I^{nc(2)}$ (defined in Eq.~(\ref{seq:HI-c-nc12})) and the second kinetic term $H_0^-$ as perturbation to the state $|\widetilde{\Psi}^{(acc)}_{\nu,\nu_C}(\varphi_s)\rangle$.  This ensures all the energies in this section are calculated up to order $w_0^2t^2$.

We now examine all the first order perturbations:

(i) The first order perturbation energy of the nonchiral term $H_I^{nc(2)}=\frac{1}{2\Omega_{\text{tot}}}\sum_{\mathbf{q,G}}(O_{\mathbf{-q,-G}}^0O_{\mathbf{q,G}}^1+O_{\mathbf{-q,-G}}^1O_{\mathbf{q,G}}^0)$ is
\begin{equation}\label{seq:c-nf-1st-pert-Hnc2}
\begin{split}
E^{(1)''}_{\nu,\nu_C}(H_I^{nc(2)},\varphi_s)&=\langle \widetilde{\Psi}^{(acc)}_{\nu,\nu_C}(\varphi_s) |H_I^{nc(2)} | \widetilde{\Psi}^{(acc)}_{\nu,\nu_C}(\varphi_s) \rangle \\
&= \mathcal{N}^{-1}\langle \widetilde{\Psi}_{\nu,\nu_C}(\varphi_s) |[1+(\overline{H_0^+(E_{0,\nu,\nu_C}-H_I^c)^{-1}H_0^+}+H_0^+)(E_{0,\nu,\nu_C}-H_I^c)^{-1}] H_I^{nc(2)} \\ &\qquad\qquad\qquad \times[1+(E_{0,\nu,\nu_C}-H_I^c)^{-1}(H_0^++\overline{H_0^+(E_{0,\nu,\nu_C}-H_I^c)^{-1}H_0^+})] | \widetilde{\Psi}_{\nu,\nu_C}(\varphi_s) \rangle\ .\\
\end{split}
\end{equation}
Using the properties we identified in Eqs.~(\ref{seq:nc-f-1st-pert-energy1})-(\ref{seq:nc-f-1st-pert-energy3}), we find both $H_I^{nc(2)}$ and $H_0^+$ flips the index $e_Y$, while $H_I^{nc(2)}$ ($H_0^+$) is odd (even) under electron momentum $\kk$ flipping operation $I$ (defined in Eq. (\ref{seq-operator-I})). Therefore, we find every term in Eq.~(\ref{seq:c-nf-1st-pert-Hnc2}) either flips $e_Y$ of at least one electron, or is odd $I$. Since state $| \widetilde{\Psi}_{\nu,\nu_C}(\varphi_s) \rangle$ has a conserved electron number in each $e_Y$ sector and is invariant under $I$, we find
\begin{equation}
E^{(1)''}_{\nu,\nu_C}(H_I^{nc(2)},\varphi_s)=0\ .
\end{equation}

(ii) The first order perturbation energy of the nonchiral term $H_I^{nc(1)}=\frac{1}{2\Omega_{\text{tot}}}\sum_{\mathbf{q,G}}O_{\mathbf{-q,-G}}^1O_{\mathbf{q,G}}^1$ is given by
\begin{equation}\label{seq-E1-nc0-rot}
E^{(1)''}_{\nu,\nu_C}(H_I^{nc(1)},\varphi_s)=\langle \widetilde{\Psi}^{(acc)}_{\nu,\nu_C}(\varphi_s) | H_I^{nc(1)} | \widetilde{\Psi}^{(acc)}_{\nu,\nu_C}(\varphi_s) \rangle= E^{(1,1)''}_{\nu,\nu_C}(H_I^{nc(1)},\varphi_s)+E^{(1,2)''}_{\nu,\nu_C}(H_I^{nc(1)},\varphi_s) + E^{(1,3)''}_{\nu,\nu_C}(H_I^{nc(1)},\varphi_s)\ ,
\end{equation}
where $E^{(1,1)''}_{\nu,\nu_C}(H_I^{nc(1)},\varphi_s)$ is proportional to $w_0^2$, while $E^{(1,2)''}_{\nu,\nu_C}(H_I^{nc(1)},\varphi_s)$ and $E^{(1,3)''}_{\nu,\nu_C}(H_I^{nc(1)},\varphi_s)$ are proportional to $w_0^2t^2$ (see definition below). The other terms are either zero because of properties in Eqs.~(\ref{seq:nc-f-1st-pert-energy1})-(\ref{seq:nc-f-1st-pert-energy3}), or are of order higher than $w_0^2t^2$. For instance, the term $\langle \widetilde{\Psi}_{\nu,\nu_C}(\varphi_s) | H_I^{nc(1)} (E_{0,\nu,\nu_C}-H_I^c)^{-1}H_0^+ | \widetilde{\Psi}_{\nu,\nu_C}(\varphi_s) \rangle$ (proportional to $w_0^2t$) are zero, because $H_I^{nc(1)}$ does not flip index $e_Y$ of an electron, while $H_0^+$ flips $e_Y$. An example of higher order term is $\langle L^\dag H_I^{nc(1)} L  \rangle$ where $L=(E_{0,\nu,\nu_C}-H_I^c)^{-1}\overline{H_0^+(E_{0,\nu,\nu_C}-H_I^c)^{-1}H_0^+}$, which is of order $w_0^2t^4$ and is thus ignored.

The first term in Eq.~(\ref{seq-E1-nc0-rot}) is given by
\begin{equation}\label{seq-E1-nc0-rot1}
\begin{split}
&E^{(1,1)''}_{\nu,\nu_C}(H_I^{nc(1)},\varphi_s)=\frac{1}{2\Omega_{\text{tot}}}\sum_{\mathbf{q,G}}\langle \widetilde{\Psi}_{\nu,\nu_C}(\varphi_s) |O_{\mathbf{-q,-G}}^1O_{\mathbf{q,G}}^1 | \widetilde{\Psi}_{\nu,\nu_C}(\varphi_s) \rangle =\frac{1}{2\Omega_{\text{tot}}}\sum_{\mathbf{q,G}}\langle \widetilde{\Psi}_{\nu,\nu_C} |U'^\dag (\varphi_s) O_{\mathbf{-q,-G}}^1O_{\mathbf{q,G}}^1 U'(\varphi_s)| \widetilde{\Psi}_{\nu,\nu_C} \rangle \\
&=\frac{1}{2\Omega_{\text{tot}}}\sum_{\mathbf{q,G}}V(\qq+\GG)\langle \widetilde{\Psi}_{\nu,\nu_C} | \left(\sum_{\kk,s', e_Y'}F_{-e_Y'}(\kk,\qq+\GG)^*d^\dag_{\mathbf{k},-e_Y',s'}(\tau_z\cos\varphi_{s'}+\tau_x\sin\varphi_{s'})d_{\mathbf{k+q},e_Y',s'}\right) \\
&\times\left(\sum_{\kk,s,e_Y}F_{e_Y}(\kk,\qq+\GG)d^\dag_{\mathbf{k+q},-e_Y,s}(\tau_z\cos\varphi_s+\tau_x\sin\varphi_s)d_{\mathbf{k},e_Y,s}\right) | \widetilde{\Psi}_{\nu,\nu_C} \rangle \\
&= \frac{1}{2\Omega_{\text{tot}}} \sum_{s=\uparrow,\downarrow} \sum_{\mathbf{k,q,G}} V(\qq+\GG)\left|F_{+1}(\kk,\qq+\GG)\right|^2\Big[\nu_s^{(2)}\cos^2\varphi_s +\nu_s^{(1)}\sin^2\varphi_s\Big] \\
&=N_MU_1\sum_{s=\uparrow,\downarrow}\Big[\nu_s^{(2)}\cos^2\varphi_s +\nu_s^{(1)}\sin^2\varphi_s\Big]\ ,
\end{split}
\end{equation}
where $U_1$ is defined in Eq.~(\ref{eq-U1}), and the coefficients $\nu_s^{(j)}$ are the same as those given in Tab. \ref{Tab-nu123} (the left table). The second term in Eq.~(\ref{seq-E1-nc0-rot}) is
\begin{equation}\label{seq-E1-nc0-rot2}
\begin{split}
&E^{(1,2)''}_{\nu,\nu_C}(H_I^{nc(1)},\varphi_s) =\frac{1}{2\Omega_{\text{tot}}}\sum_{\mathbf{q,G}}\langle \widetilde{\Psi}_{\nu,\nu_C} |H_0^+(E_{0,\nu,\nu_C}-H_I^c)^{-1} U'^\dag (\varphi_s) O_{\mathbf{-q,-G}}^1O_{\mathbf{q,G}}^1 U'(\varphi_s) (E_{0,\nu,\nu_C}-H_I^c)^{-1} H_0^+| \widetilde{\Psi}_{\nu,\nu_C} \rangle \\
&\qquad +(\mathcal{N}^{-1}-1)E^{(1,1)''}_{\nu,\nu_C}(H_I^{nc(1)}) \\
&=\frac{1}{2\Omega_{\text{tot}}}\sum_{\mathbf{q,G}}V(\qq+\GG)\langle \widetilde{\Psi}_{\nu,\nu_C} | H_0^+(E_{0,\nu,\nu_C}-H_I^c)^{-1} \left(\sum_{\kk,s' e_Y'}F_{-e_Y'}(\kk,\qq+\GG)^*d^\dag_{\mathbf{k},-e_Y',s'}(\tau_z\cos\varphi_{s'}+\tau_x\sin\varphi_{s'})d_{\mathbf{k+q},e_Y',s'}\right) \\
&\qquad \times\left(\sum_{\kk,s,e_Y}F_{e_Y}(\kk,\qq+\GG)d^\dag_{\mathbf{k+q},-e_Y,s}(\tau_z\cos\varphi_s+\tau_x\sin\varphi_s)d_{\mathbf{k},e_Y,s}\right) (E_{0,\nu,\nu_C}-H_I^c)^{-1}H_0^+ | \widetilde{\Psi}_{\nu,\nu_C} \rangle  \\
&\qquad +(\mathcal{N}^{-1}-1)E^{(1,1)''}_{\nu,\nu_C}(H_I^{nc(1)},\varphi_s)\ .
\end{split}
\end{equation}
Since the energy scale of $H_0^+$ is smaller than that of the nonchiral interaction $H_I^c$ (which is needed for the perturbative wave function $|\widetilde{\Psi}_{\nu,\nu_C}\rangle$ to be the ground state in chiral-nonflat limit), we can approximately regard $\epsilon_\mathcal{N}=\mathcal{N}-1=\langle \widetilde{\Psi}_{\nu,\nu_C}(\varphi_s)| H_0^+(E_{0,\nu,\nu_C}-H_I^c)^{-2}H_0^+ |\widetilde{\Psi}_{\nu,\nu_C}(\varphi_s)\rangle$ as a small quantity, and thus to the leading order of $\epsilon_\mathcal{N}$ we have $\mathcal{N}^{-1}-1\approx -\epsilon_\mathcal{N}$. With this approximation, and restricting the action of all the operators in Eq.~(\ref{seq-E1-nc0-rot2}) within the sub-Hilbert space of the one electron-hole pair exact excitations of the state $|\widetilde{\Psi}_{\nu,\nu_C}\rangle$ we derived in Ref.~\cite{ourpaper5}, we arrive at an energy
\begin{equation}
E^{(1,2)''}_{\nu,\nu_C}(H_I^{nc(1)},\varphi_s) \approx N_M[J_1(w_0)-J_2(w_0)]\sum_{s=\uparrow,\downarrow}\nu_s^{(2)}\sin^2\varphi_s\ ,
\end{equation}
where $J_1(w_0)$ and $J_2(w_0)$ are as defined in Eq.~(\ref{seq:approx-E-H0-J}). The fact that the energy coefficient is approximately $[J_1(w_0)-J_2(w_0)]$ can be seen by expanding the definition of $J_1(w_0)$ and $J_2(w_0)$ in Eq.~(\ref{seq:approx-E-H0-J}) with respect to the nonchiral term $H_I^{nc}$ up to the $w_0^2t^2$ order.

The third term in Eq.~(\ref{seq-E1-nc0-rot}) is given by
\begin{equation}\label{seq-E1-nc0-rot3}
\begin{split}
&E^{(1,3)''}_{\nu,\nu_C}(H_I^{nc(1)},\varphi_s) =\frac{1}{2\Omega_{\text{tot}}}\sum_{\mathbf{q,G}}\langle \widetilde{\Psi}_{\nu,\nu_C} |U'^\dag (\varphi_s) O_{\mathbf{-q,-G}}^1O_{\mathbf{q,G}}^1 U'(\varphi_s) (E_{0,\nu,\nu_C}-H_I^c)^{-1} \overline{H_0^+ (E_{0,\nu,\nu_C}-H_I^c)^{-1} H_0^+}| \widetilde{\Psi}_{\nu,\nu_C} \rangle +h.c. \\
\end{split}
\end{equation}
which is of order $w_0^2t^2$. 
Note that the operator $\overline{H_0^+ (E_{0,\nu,\nu_C}-H_I^c)^{-1} H_0^+}$ is the part of $H_0^+ (E_{0,\nu,\nu_C}-H_I^c)^{-1} H_0^+$ not mapping back to the state $| \widetilde{\Psi}_{\nu,\nu_C} \rangle$. Since $H_0^+$ creates an electron-hole pair by flipping $e_Y$, and $H_I^c$ conserves $e_Y$, one finds that $\overline{H_0^+ (E_{0,\nu,\nu_C}-H_I^c)^{-1} H_0^+}|\widetilde{\Psi}_{\nu,\nu_C} \rangle$ is necessarily a state with two electron-hole pairs away from the state $|\widetilde{\Psi}_{\nu,\nu_C} \rangle$ (i.e., a state of the form $c^\dag_1c^\dag_2c_3c_4|\widetilde{\Psi}_{\nu,\nu_C}\rangle$ with electron operators $c_{1,2,3,4}$ all different). However, in this paper, we are numerically calculating all the perturbation energies by restricting ourselves in the sub-Hilbert space of the one electron-hole pair exact excitations we found in Ref.~\cite{ourpaper5} (i.e., of the form $c^\dag_1c_2|\widetilde{\Psi}_{\nu,\nu_C}\rangle$ with different $c_{1,2}$). Therefore, under this restriction, we have
\begin{equation}\label{seq-E1-nc0-rot3-approximate}
E^{(1,3)''}_{\nu,\nu_C}(H_I^{nc(1)},\varphi_s)=0\ ,
\end{equation}
which we will adopt in this section. This completes the calculation of the three terms in Eq. (\ref{seq-E1-nc0-rot}).

(iii) The first order perturbation energy of $H_0^-$ is zero. To see this, we first note that
\begin{equation}\label{seq:H0-nogoldstone}
\begin{split}
&H_0^-| \widetilde{\Psi}_{\nu,\nu_C}(\varphi_s) \rangle=H_0^-U'(\varphi_s)| \widetilde{\Psi}_{\nu,\nu_C} \rangle = U'(\varphi_s)\sum_{\kk,s,e_Y} \epsilon_-(\kk)d^\dag_{\mathbf{k},e_Y,s}(\tau_z\cos\varphi+\tau_x\sin\varphi)d_{\mathbf{k},e_Y,s} |\widetilde{\Psi}_{\nu,\nu_C} \rangle \\
&= U'(\varphi_s)\sum_{\kk,s,e_Y} \sin\varphi_s \epsilon_-(\kk)d^\dag_{\mathbf{k},e_Y,s}\tau_xd_{\mathbf{k},e_Y,s} |\widetilde{\Psi}_{\nu,\nu_C} \rangle\ .
\end{split}
\end{equation}
We stress that the resulting state of Eq.~(\ref{seq:H0-nogoldstone}) is orthogonal to the gapless Goldstone modes of the chiral-nonflat U(4) \cite{ourpaper5}, since $\sum_{\kk} \epsilon_-(\kk)=0$ (similar to the argument in Eq.~(\ref{seq:kinetic-goldstone}) for the nonchiral-flat limit). Therefore, the perturbation by $H_0^-$ is nondegenerate. 
Then, using the properties in Eqs.~(\ref{seq:nc-f-1st-pert-energy1}) and~(\ref{seq:nc-f-1st-pert-energy2}), we find
\begin{equation}
\begin{split}
&E^{(1)''}_{\nu,\nu_C}(H_0^-,\varphi_s)=\langle \widetilde{\Psi}^{(acc)}_{\nu,\nu_C}(\varphi_s) |H_0^- | \widetilde{\Psi}^{(acc)}_{\nu,\nu_C}(\varphi_s) \rangle \\
&= \mathcal{N}^{-1}\langle \widetilde{\Psi}_{\nu,\nu_C}(\varphi_s) |[1+(\overline{H_0^+ (E_{0,\nu,\nu_C}-H_I^c)^{-1} H_0^+}+H_0^+)(E_{0,\nu,\nu_C}-H_I^c)^{-1}] H_0^-\\
&\qquad \qquad\qquad  \times[1+(E_{0,\nu,\nu_C}-H_I^c)^{-1}(H_0^++\overline{H_0^+ (E_{0,\nu,\nu_C}-H_I^c)^{-1} H_0^+})] | \widetilde{\Psi}_{\nu,\nu_C}(\varphi_s) \rangle\\
&=0\ ,
\end{split}
\end{equation}
which can be seen from the fact that the terms $H_0^+$, $E_{0,\nu,\nu_C}-H_I^c$ and the state $| \widetilde{\Psi}_{\nu,\nu_C}(\varphi_s) \rangle$ are invariant under the $\kk$ flipping operation $I$ (defined in Eq. (\ref{seq-operator-I})), while $H_0^-$ changes sign under $I$ (see Eqs.~(\ref{seq:nc-f-1st-pert-energy1}) and~(\ref{seq:nc-f-1st-pert-energy2})).

In summary, the total 1st order perturbation energy is given by
\begin{equation}\label{seq-1st-E-chiralnonflat}
\begin{split}
E^{(1)''}_{\nu,\nu_C}(\varphi_s)&=E^{(1)''}_{\nu,\nu_C}(H_I^{nc(2)},\varphi_s)+E^{(1)''}_{\nu,\nu_C}(H_I^{nc(1)},\varphi_s)+E^{(1)''}_{\nu,\nu_C}(H_0^-,\varphi_s) \\
&= N_M\sum_{s=\uparrow,\downarrow}\Big\{U_1\Big[\nu_s^{(2)}\cos^2\varphi_s +\nu_s^{(1)}\sin^2\varphi_s\Big]+[J_1(w_0)-J_2(w_0)]\nu_s^{(2)}\sin^2\varphi_s\Big\}\ .
\end{split}
\end{equation}

\subsubsection{2nd order perturbation}

Now we proceed to examine the 2nd order perturbations away from the chiral-nonflat limit. 
Since this is of higher order, the energy scale is smaller, and we shall approximate the chiral-nonflat wavefunction $| \widetilde{\Psi}_{\nu,\nu_C}^{(acc)}(\varphi_s) \rangle$ in Eq.~(\ref{seq-Psi-nu-nuC2-rotate-2nd}) into the less accurate expression $| \widetilde{\Psi}_{\nu,\nu_C}(\varphi_s) \rangle$ in Eq.~(\ref{seq-Psi-nu-nuC2-rotate}), which is sufficient for the purpose here.  
Since $H_I^{nc(1)}\propto w_0^2$, $H_I^{nc(2)}\propto w_0$, and $H_0^-\propto w_0t$, if we assume the FMC and keep only the perturbation energy up to order $w_0^2t^2$, we find the total 2nd order perturbation energy
\begin{equation}
\begin{split}
&E^{(2)''}_{\nu,\nu_C}(\varphi_s)=\langle\widetilde{\Psi}_{\nu,\nu_C}(\varphi_s) |(\overline{H}_I^{nc(1)}+H_I^{nc(2)}+H_0^-) (E_{0,\nu,\nu_C}-H_I^c)^{-1} (\overline{H}_I^{nc(1)}+H_I^{nc(2)}+H_0^-)|\widetilde{\Psi}_{\nu,\nu_C}(\varphi_s)  \rangle\\
&\qquad \approx \langle\widetilde{\Psi}_{\nu,\nu_C}(\varphi_s) | (H_I^{nc(2)}+H_0^-) (E_{0,\nu,\nu_C}-H_I^c)^{-1} (H_0^-+H_I^{nc(2)}) |\widetilde{\Psi}_{\nu,\nu_C}(\varphi_s)\rangle\ ,
\end{split}
\end{equation}
where $\overline{H}_I^{nc(1)}=H_I^{nc(1)}-\langle \widetilde{\Psi}_{\nu,\nu_C}(\varphi_s)|H_I^{nc(1)}|\widetilde{\Psi}_{\nu,\nu_C}(\varphi_s)\rangle$. Here we have used the fact that 
\begin{equation}
\begin{split}
&\langle \widetilde{\Psi}_{\nu,\nu_C}(\varphi_s)| H_I^{nc(2)}(E_{0,\nu,\nu_C}-H_I^c)^{-1}H_0^-|\widetilde{\Psi}_{\nu,\nu_C}(\varphi_s)\rangle=0\ ,\\
&\langle\widetilde{\Psi}_{\nu,\nu_C}(\varphi_s) | H_0^- (E_{0,\nu,\nu_C}-H_I^c)^{-1} \overline{H}_I^{nc(1)}|\widetilde{\Psi}_{\nu,\nu_C}(\varphi_s)\rangle=0\ ,
\end{split}
\end{equation} 
which can be seen by noting that the former term flips $e_Y$, and the latter term is odd under the $\kk$ flipping operation $I$, as can be seen from Eqs.~(\ref{seq:nc-f-1st-pert-energy1}-\ref{seq:nc-f-1st-pert-energy3}). 
Besides, we have ignored the term $\langle \widetilde{\Psi}_{\nu,\nu_C}(\varphi_s)| \overline{H}_I^{nc(1)}(E_{0,\nu,\nu_C}-H_I^c)^{-1}\overline{H}_I^{nc(1)}|\widetilde{\Psi}_{\nu,\nu_C}(\varphi_s)\rangle$, which is of higher order $w_0^4$. 
Therefore, from Eq.~(\ref{seq:H0-nogoldstone}) and the identities
\begin{equation}
U'^\dag(\varphi_s)O^1_{\kk,\GG}U(\varphi_s)=\left(\sum_{\kk,s,e_Y}F_{e_Y}(\kk,\qq+\GG)d^\dag_{\mathbf{k+q},-e_Y,s}(\tau_z\cos\varphi_s+\tau_x\sin\varphi_s)d_{\mathbf{k},e_Y,s}\right)\ ,\  U'^\dag(\varphi_s)O^0_{\kk,\GG}U(\varphi_s)=O^0_{\kk,\GG}\ ,
\end{equation}
we find the 2nd order perturbation energy taking the form
\begin{equation}\label{seq-2nd-E-chiralnonflat}
E^{(2)''}_{\nu,\nu_C}(\varphi_s)=-N_M\sum_{s=\uparrow,\downarrow}\left\{\nu^2U_2\Big[\nu_s^{(2)}\cos^2\varphi_s +\nu_s^{(1)}\sin^2\varphi_s\Big]+\nu_s^{(3)}J_3(w_0) \sin^2\varphi_s\right\}\ ,
\end{equation}
where $U_2$ is defined in Eq.~(\ref{seq:U2-def}), and the coefficient $J_3(w_0)$ is approximately equal to the expression for $J_3(w_0)$ defined in Eq.~(\ref{seq:approx-E-H0-J}) to the $w_0^2t^2$ order. Note that the $U_2$ term has the same form as the $U_1$ term of the first order perturbation energy of $H_I^{nc(1)}$ in Eq.~(\ref{seq-E1-nc0-rot1}). 

\subsubsection{The total perturbation energy}

We have calculated the 1st order perturbation energy $E^{(1)''}_{\nu,\nu_C}(\varphi_s)$ in Eq. (\ref{seq-1st-E-chiralnonflat}), and the 2nd order perturbation energy $E^{(2)''}_{\nu,\nu_C}(\varphi_s)$ in Eq. (\ref{seq-2nd-E-chiralnonflat}). The total perturbation energy is then $E^{''}_{\nu,\nu_C}(\varphi_s)=E^{(1)''}_{\nu,\nu_C}(\varphi_s)+E^{(2)''}_{\nu,\nu_C}(\varphi_s)$, which can be written as
\begin{equation}\label{seq-pert-energy-to-chiral-nonflat}
E^{''}_{\nu,\nu_C}(\varphi_s)=N_M\sum_{s=\uparrow,\downarrow}\Big\{(U_1-\nu^2U_2)\nu_s^{(2)}-\Big[(\nu_s^{(2)}-\nu_s^{(1)})(U_1-\nu^2U_2)-\nu_s^{(2)}(J_1(w_0)-J_2(w_0))+\nu_s^{(3)}J_3(w_0)\Big]\sin^2\varphi_s\Big\}\ .
\end{equation}
For the flavor order in Eq. (\ref{seq-flavor-order1}), $\nu_s^{j}$ ($j=1,2,3$) are the same as those given in Tab. \ref{Tab-nu123} (the left table). 
By the numerical values of the interaction constants, we generically have $U_1-\nu^2U_2\gg J_1-J_2 > J_3>0$ for any $0<w_0/w_1<1$ and $|\nu|\le 3$. Thus, we find the lowest energy is achieved at exactly the same $\varphi_s$ we found below Eq. (\ref{seq:valley-polar-1}) and in Tab. \ref{Tab-nu123} (the left table), namely, 

(i) $\varphi_\uparrow=\varphi_\downarrow=\pi/2$ if $\nu_C=0$, 

(ii) $\varphi_\uparrow=\pi/2$ and $\varphi_\downarrow=0$ if $0<|\nu_C|<4-|\nu|$, and 

(iii) $\varphi_\uparrow=\varphi_\downarrow=0$ if $|\nu_C|=4-|\nu|$. 

Therefore, again we find the same conclusion as that we found below Eq.~(\ref{seq:valley-polar-1}). In particular, one can check that the wave functions of these states here are also exactly the same as that in Eq. (\ref{seq-Psi-nc-nf0}) (see App. \ref{app:ncpert-chiralnonflat3} below for an example).

\subsubsection{Coincidence of ground states obtained from perturbations to the nonchiral-flat and chiral-nonflat limits}\label{app:ncpert-chiralnonflat3}

Generically, when $\varphi_\uparrow=\varphi_\downarrow=\pi/2$, using the fact that $\eta_{4-j}=-\eta_j$ and $s_{4-j}=s_{j}$, we find the rotated state is given by
\begin{equation}\label{seq-Psi-nc-nf1}
\begin{split}
&|\widetilde{\Psi}_{\nu,\nu_C}(\varphi_s=\pi/2)\rangle =U'(\pi/2) \prod_{\mathbf{k}} \prod_{j=1}^{\nu_+}d^\dag_{\mathbf{k},+1,\eta_{j},s_{j}} \prod_{j=1}^{\nu_-}d^\dag_{\mathbf{k},-1,-\eta_{j},s_{j}}|0\rangle \\
&= \prod_{\mathbf{k}} \prod_{j=1}^{\nu_+}\frac{d^\dag_{\mathbf{k},+1,\eta_j,s_j}+\eta_j d^\dag_{\mathbf{k},+1,-\eta_j,s_j}}{\sqrt{2}}  \prod_{j=1}^{\nu_-} \frac{d^\dag_{\mathbf{k},-1,-\eta_j,s_j}-\eta_j d^\dag_{\mathbf{k},-1,\eta_j,s_j}}{\sqrt{2}}|0\rangle\ ,\\
&= \prod_{\mathbf{k}} \prod_{e_Y=\pm} \prod_{j=1}^{\nu_{e_Y}} (-\eta_j)^{(1-e_Y)/2}\frac{d^\dag_{\mathbf{k},e_{Y},\eta_j,s_j}+\eta_j e_{Y}d^\dag_{\mathbf{k},e_{Y},-\eta_j,s_j}}{\sqrt{2}}  |0\rangle\ .
\end{split}
\end{equation}
When $\nu_C=0$ (i.e., $\nu_+=\nu_-$), this chiral-nonflat U(4) valley rotated state $|\widetilde{\Psi}_{\nu,\nu_C}(\varphi_s=\pi/2)\rangle$ is the lowest state we found below Eq. (\ref{seq-pert-energy-to-chiral-nonflat}), and it is exactly the same as the nonchiral-flat U(4) valley rotated state (with rotation angle $\varphi_s=\pi/2$) for $\nu_C=0$ in Eq.~(\ref{seq-Psi-nc-nf0}). Namely, the lowest nonchiral-nonflat state we found by perturbing the nonchiral-flat limit is the same as that we found by perturbing the chiral-nonflat limit. This is because the state of Eq.~(\ref{seq-Psi-nc-nf1}) with rotation angles $\varphi_s=\pi/2$ is not only a state in the chiral-nonflat U(4) multiplet, but also belongs to the nonchiral-flat U(4) multiplet of $|\Psi_{\nu,\nu_C}\rangle$. For $\nu_C=0$, this is true only if $\varphi_s=\pi/2$. Thus, for $\nu_C=0$, we find the intervalley coherent state Eq.~(\ref{seq-Psi-nc-nf0}), which is equal to the state in Eq.  (\ref{seq-Psi-nc-nf1}), simultaneously minimizes the kinetic energy and the nonchiral interaction energy, therefore should be the ground state in the entire perturbative nonchiral-nonflat regime.

For the states with $\nu_C=4-|\nu|$, however, the chiral-nonflat U(4) rotated wavefunction $|\widetilde{\Psi}_{\nu,\nu_C}(\varphi_s)\rangle$ becomes no different from the nonchiral-flat U(4) rotated wavefunction $|\Psi_{\nu,\nu_C}(\varphi_s)\rangle$ in Eq.~(\ref{seq-Psi-nc-nf0}) for any $\varphi_s$. The same is true in the spin $\downarrow$ sector of states with $0<|\nu_C|<4-|\nu|$, where the state $|\widetilde{\Psi}_{\nu,\nu_C}(\varphi_s)\rangle$ with $\varphi_\uparrow=\pi/2$ and any $\varphi_\downarrow$ is a state in both the chiral-nonflat U(4) multiplet and the nonchiral-flat U(4) multiplet. Therefore, in these cases, we cannot easily identify the lowest state valley polarization angle $\varphi_s$ (in the spin $\downarrow$ sector in the $0<|\nu_C|<4-|\nu|$ case). Instead, we need perform calculations at finer energy scales ($\propto w_0^2t^2$), from which we have showed that a $z$-direction valley Bloch sphere polarization is preferred.

\section{Perturbation of the out-of-plane Magnetic Field}\label{app:magnetic}

We now consider the effect of an out-of-plane magnetic field $B$. Since the Zeeman energy ($\sim 0.1meV$ per Tesla) is much smaller than the interaction energy, we only consider the orbital effects of $B$. The most important orbital effect of magnetic field $B$ for a gapped insulator of Chern number $\nu_C$ is given by the Streda formula \cite{streda1982}, which gives the change of number of occupied electrons $N=\nu N_M$ as
\begin{equation}
\frac{dN}{d(\Phi/\Phi_0)} =N_M\nu_C\ ,
\end{equation}
where $\Phi=B\Omega_M$ is the magnetic flux per unit cell, $\Omega_M$ is the moir\'e unit cell area, $N_M$ is the total number of moir\'e unit cells, and $\Phi_0=h/e$ is the flux quanta. Electrons are adiabatically pumped between the conduction and valence states via the edge states of the Chern insulator \cite{asboth2017,lian2020}.

We consider the Chern insulator states $|\Psi_{\nu,\nu_C}\rangle$ defined in Eq.~(\ref{seq-Psi-nu-nuC}) in the nonchiral-flat limit. We want to estimate the free energy change of the state. The increased number of electrons of the insulator state due to the Streda formula is thus
\begin{equation}
\Delta N(B)=N(B)-N(0)=\nu_C N_M\frac{\Phi}{\Phi_0}=N_M \nu_C\frac{eB}{h \Omega_M}\ .
\end{equation}
However, since the vector potential of the magnetic field $B$ breaks the translation symmetry of moir\'e unit cells, it makes the calculation of interaction energy in magnetic field $B$ generically difficult. As an estimation, we can use the orbital magnetic moment $\hat{M}$ of the projected Hamiltonian $H=H_0+H_I$ to estimate the change in energy $\langle H\rangle$ as $-\hat{M}B$. 
However, due to the PH symmetry $P$, as explained in the next paragraph, under the exact flat band assumption $H_0=0$, one can show that the orbital magnetic moment \cite{shi_orbmag_2007} of the flat bands is zero.
For many-body states within the flat bands, the total orbital moment is thus zero,  implying the interaction energy to be nearly unchanged at small $B$.

Now we show that the energy splitting of the two Chern bands due to the effective orbital moment is zero to first order of $B$. 
For simplicity, we first focus on the $\eta=+$ valley. 
In the Chern band basis, the sewing matrices have the form $B^{C_{2z}T}(\kk)=\zeta^x$, $B^{P}(\kk)=-i\zeta^z$. 
(Recall that in the energy band basis there is $B^{C_{2z}T}(\kk)=\zeta^0$, $B^{P}(\kk)=-i\zeta^y$ in the valley $\eta=+$ (\cref{eq:gauge-0}) and the Chern band basis diagonalizes the $\zeta^y$ matrix.)
We assume the coupling term of orbital moment to the magnetic field has the form $M(\kk) B$ for an electron at quasimomentum $\kk$, where $M(\kk)$ is a matrix in the Chern band basis. 
Since $C_{2z}T$ flips the magnetic field, $M(\kk)$ must anti-commute with $\zeta^x K$. 
Thus $M(\kk)$ must be proportional to $\zeta^z$. 
On the other hand, the action of $P$ is the same as inversion and hence it keeps the direction of magnetic field unchanged. 
(Magnetic field is a pseudo-vector that transforms as a vector under proper rotations and is invariant under inversion.)
Thus, the coupling $M(\kk) B$ still respects the $P$ symmetry: It must satisfy $M(-\kk) = - \zeta^z M(\kk) \zeta^z$.  
Therefore, $M(\kk)$ must have the form
\begin{equation}
    M(\kk) = m(\kk) \zeta^z,\qquad m(\kk) = - m(-\kk)\ .
\end{equation}
Since $m(\kk)$ is an odd function in $\kk$, the total magnetic moment of each Chern band $M=\int_{\text{MBZ}} d^2\kk M(\kk)$ vanishes.
This conclusion can also be verified by computing directly the orbital magnetic moment \cite{shi_orbmag_2007} using the Bistritzer-Macdonald TBG model (with exact $P$ symmetry).

Meanwhile, the chemical potential for Chern insulator state $|\Psi_{\nu,\nu_C}\rangle$ can be estimated by Eqs.~(\ref{eq:chemical-potential-shift}) and~(\ref{eq:AG-general}) as 
\begin{equation}
\mu_\nu=\frac{\nu}{N_M^2\Omega_M}\sum_{\GG}V(\GG)\left(\sum_\kk \alpha_0(\mathbf{k,G})\right)^2 =\nu U_0\ ,
\end{equation}
where we have defined
\begin{equation}
U_0=\frac{1}{N_M^2\Omega_M}\sum_{\GG}V(\GG)\left(\sum_\kk \alpha_0(\mathbf{k,G})\right)^2\ .
\end{equation}
Therefore, the change of the free energy $F=\langle H_I-\mu_\nu N\rangle$ of the Chern insulator state $|\Psi_{\nu,\nu_C}\rangle$ for small $B$ is approximately
\begin{equation}\label{seq:free-energy-B}
\Delta F(B)=-\mu_\nu \Delta N(B)=- \nu\nu_C U_0N_M\frac{\Phi }{\Phi_0}=- \nu\nu_C U_0\frac{e N_M }{h\Omega_M} B \ .
\end{equation}
Therefore, for $B>0$, we find that the state with larger $\nu\nu_C>0$ gains more free energy.

\begin{figure}[htbp]
\begin{centering}
\includegraphics[width=0.9\linewidth]{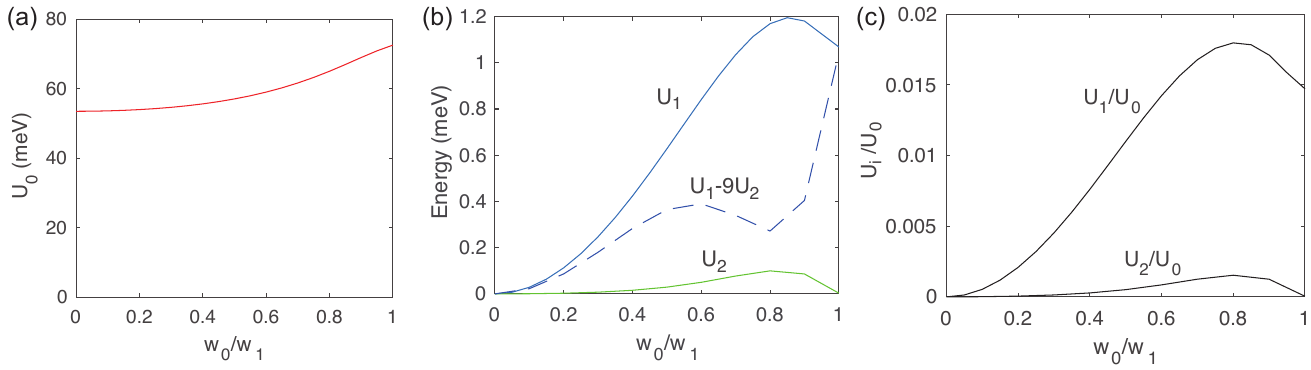}
\end{centering}
\caption{Energies $U_0$ (a), $U_1$ and $U_2$ (b) calculated with respect to $w_0/w_1$ for twist angle $\theta=1.05^\circ$ and screening length $\xi=10$nm, where the FMC is not imposed. The calculation is done by discretizing the MBZ into a $12\times12$ momentum lattice (which is large enough to simulate the thermodynamic limit). Generically, we find $U_1-\nu^2U_2>0$ for any $0< w_0/w_1\le 1$ and any $|\nu|\le 3$. The values of $U_0$ and $U_1$ are independent of whether the FMC holds, while $U_2=0$ if FMC holds. The ratio $U_1/U_0$ and $U_2/U_0$ are also shown in (c).}
\label{fig:U1U0}
\end{figure}

In Eq.~(\ref{seq-E1-nc}) we have estimated the interaction energy $E^{(nc)}_{\nu,\nu_C}$ of the Chern number $\nu_C$ state to be linear in $|\nu_C|$ due to the nonchiral interaction terms. Besides, when the kinetic term is taken into account, the valley polarization ($z$ direction or in-plane in the valley space) of the ground state also contributes an energy $E^{(2)'}_{\nu,\nu_C}$ (see Eq.~(\ref{seq:approx-E-H0v2})). In a magnetic field $B$, we therefore estimate the free energy (the $|\nu_C|$ dependent part) of the Chern insulator state $|\Psi_{\nu,\nu_C}\rangle$ as
\begin{equation}
\begin{split}
&F_{\nu,\nu_C}(B)\approx E^{(nc)}_{\nu,\nu_C}+E^{(2)'}_{\nu,\nu_C}+\Delta F(B)\\
&=N_M|\nu_C|(U_1-\nu^2U_2)-N_M(4-|\nu|)[J_1(w_0)-J_2(w_0)-J_{3}(w_0)]\delta_{\nu_C,4-|\nu|} -\nu\nu_C U_0\frac{e N_M }{h\Omega_M} B \\
&=N_M\left(|\nu_C|(U_1-\nu^2U_2)-(4-|\nu|)[J_1(w_0)-J_2(w_0)-J_{3}(w_0)]\delta_{\nu_C,4-|\nu|}-\nu\nu_CU_0\frac{\Phi}{\Phi_0}\right)\ ,
\end{split}
\end{equation}
where $U_1$ and $U_2$ are defined in Eqs.~(\ref{seq:U1-def}) and~(\ref{seq:U2-def}), 
and $J_i(w_0)$ is defined in Eq.~(\ref{seq:approx-E-H0-J}). Since $|J_1(w_0)-J_2(w_0)-J_3(w_0)|\ll U_1, U_2$ (see Tabs. \ref{Tab-J123} and \ref{Tab-J123-noFMC}) and Fig. \ref{fig:U1U0}), we can ignore the $|J_1(w_0)-J_2(w_0)-J_3(w_0)|$ term, thus the free energy is approximately given by Eq.~(\ref{eq:free-energy}) in the main text. Besides, for $|\nu|\le 3$, we always have $U_1-\nu^2U_2>0$ (Fig. \ref{fig:U1U0}). 
Therefore, for $B>0$ and $\nu=\pm1,\pm2$, we expect the ground state to transit from the lowest Chern number state with $\nu_C=\text{sgn}(\nu)\text{mod}(\nu,2)$ to the largest Chern number state with $\nu_C=\text{sgn}(\nu)(4-|\nu|)$ to become the ground state when
\begin{equation}
B>B_{\nu}^*=\frac{U_1-\nu^2U_2}{|\nu|U_0}\frac{h}{e\Omega_M}\ ,
\end{equation}
and this transition is a first-order transition. In particular, the critical magnetic field for transition is lower if $|\nu|$ is larger. For $\nu=\pm3$, there are just the Chern number $\nu_C=\pm1$ states, so the transition with $B$ does not exist.

As an estimation, near the magic angle $\theta=1.05^\circ$ and relaxation $w_0\approx 0.8 w_1$, if we take the top/bottom gate screening length $\xi\approx 10$nm (see definition in Eq.~(\ref{seq-Vq})), we numerically find (see Fig. \ref{fig:U1U0})
\begin{equation}
U_1/U_0\approx 0.02\ ,\qquad U_2/U_0\approx 0.0015\ .
\end{equation}
This gives a critical magnetic field for $\nu=\pm1,\pm2$ as
\begin{equation}
B_{1}^*\approx 0.5\ \text{Tesla}\ ,\qquad B_{2}^*\approx 0.2\ \text{Tesla}\ .
\end{equation}

\section{Hartree-Fock Hamiltonian for the (Chern) insulator states}\label{app:HF-Hamiltonian}

We here derive the effective Hartree-Fock Hamiltonian for the exact Chern insulator ground states $|\Psi_{\nu}^{\nu_+,\nu_-}\rangle$ in Eq.~(\ref{eq:U(4)U(4)-GS}) in the (first) chiral-flat limit: 
\begin{equation}
|\Psi_{\nu}^{\nu_+,\nu_-}\rangle =\prod_{\mathbf{k}\in\text{MBZ}} \left(\prod_{j_1=1}^{\nu_+}d^\dag_{\mathbf{k},+1,\eta_{j_1},s_{j_1}} \prod_{j_2=1}^{\nu_-}d^\dag_{\mathbf{k},-1,\eta_{j_2}',s_{j_2}'}\right)|0\rangle\ ,
\end{equation}
and for $\nu_C=0$ at even fillings in the nonchiral-flat limit (where the state is still exact). We note that the FMC is not needed for the derivation of Hartree-Fock Hamiltonian here. 
The fact that all the exact ground states we found (in the chiral-flat limit and nonchiral-flat limit) are Fock states indicates that Hartree-Fock may provide a good approximation in these cases. 

\subsection{The chiral-flat limit}
In the chiral-flat limit, the interaction Hamiltonian is $H=H_I=\frac{1}{2\Omega_{\text{tot}}}\sum_{\mathbf{q}\in\text{MBZ}}\sum_{\mathbf{G}\in\mathcal{Q}_0} O_{\mathbf{-q,-G}} O_{\mathbf{q,G}}$, with the operator $O_{\mathbf{q,G}}=O_{\mathbf{q,G}}^0=\sum_{\mathbf{k},e_Y,\eta,s} F^0_{\mathbf{k,q,G},e_Y}\left(d^\dagger_{\kk+\qq,e_Y,\eta,s} d_{\kk,e_Y,\eta,s}-\frac{1}{2}\delta_{\mathbf{q,0}}\right)$ given by Eq.~(\ref{eq:chiral-OqG}). We can calculate the Hartree-Fock mean fields
using the ground state $|\Psi_{\nu}^{\nu_+,\nu_-}\rangle$, thus obtaining a quadratic Hartree-Fock Hamiltonian as
\begin{equation}
H_{HF}(|\Psi_{\nu}^{\nu_+,\nu_-}\rangle)=\sum_{\kk}\sum_{\eta,s,e_Y} h_{HF,\nu,\nu_C}^{e_Y,\eta,s} (\kk) d^\dagger_{\kk,e_Y,\eta,s} d_{\kk,e_Y,\eta,s}\ ,
\end{equation}
where the single-particle Hartree-Fock Hamiltonian is defined by
\begin{equation}\label{seq:HF-chiral-flat}
\begin{split}
&h_{HF,\nu,\nu_C}^{e_Y,\eta,s} (\kk) =\frac{1}{\Omega_{\text{tot}}}\sum_\GG \Big[V(\GG)\Big(\sum_{\kk',e_Y',\eta',s'}M_{e_Y}(\kk,\GG) M_{e_Y'}(\kk',-\GG) \langle d^\dagger_{\kk',e_Y',\eta',s'} d_{\kk',e_Y',\eta',s'} -\frac{1}{2} \rangle\Big) \\
&\qquad\qquad\qquad -\frac{1}{2}\Big(V(\qq+\GG)\sum_{\qq}|M_{e_Y}(\kk,\qq+\GG)|^2 \langle d^\dagger_{\kk+\qq,e_Y,\eta,s} d_{\kk+\qq,e_Y,\eta,s} -d_{\kk+\qq,e_Y,\eta,s} d^\dagger_{\kk+\qq,e_Y,\eta,s} \rangle\Big) \Big] \\
=& \frac{1}{\Omega_{\text{tot}}}\sum_\GG \Big( \nu V(\GG)[\alpha_0(\kk,\GG)+ie_Y\alpha_2(\kk,\GG)]\sum_{\kk'}\alpha_0(\kk',-\GG) \\
&\qquad \qquad \qquad -(\nu_{e_Y,\eta,s}-\frac{1}{2})\sum_{\qq}V(\qq+\GG) [\alpha_0(\kk,\qq+\GG)^2+\alpha_2(\kk,\qq+\GG)^2] \Big)\\
=& \frac{1}{\Omega_{\text{tot}}}\sum_\GG \Big( \nu V(\GG)\alpha_0(\kk,\GG)\sum_{\kk'}\alpha_0(\kk',-\GG)  -(\nu_{e_Y,\eta,s}-\frac{1}{2})\sum_{\qq}V(\qq+\GG) [\alpha_0(\kk,\qq+\GG)^2+\alpha_2(\kk,\qq+\GG)^2] \Big)\\
=&\nu R_+(\kk)-(2\nu_{e_Y,\eta,s}-1)R_-(\kk)\ ,
\end{split}
\end{equation}
where $\nu_{e_Y,\eta,s}=0$ or $1$ is the filling of the Chern band $e_Y,\eta,s$ in state $|\Psi_{\nu}^{\nu_+,\nu_-}\rangle$, and we have defined 
\begin{equation}\label{seq:HF-chiral-flat-R}
\begin{split}
&R_+(\kk)=\frac{1}{\Omega_{\text{tot}}}\sum_\GG V(\GG)\alpha_0(\kk,\GG)\sum_{\kk'}\alpha_0(\kk',-\GG)\ , \\
&R_-(\kk)=\frac{1}{2\Omega_{\text{tot}}}\sum_{\qq,\GG}V(\qq+\GG) [\alpha_0(\kk,\qq+\GG)^2+\alpha_2(\kk,\qq+\GG)^2] \ .
\end{split}
\end{equation}
In particular, this Hartree-Fock Hamiltonian is exactly equal to the electron Hamiltonian for the exactly solvable electron (hole) excitations in an empty (occupied) Chern band $e_Y,\eta,s$ we derived in Ref.~\cite{ourpaper5}, where the electron (hole) excitations correspond to the conduction (valence) Hartree-Fock bands here. Note that $h_{HF,\nu,\nu_C}^{e_Y,\eta,s} (\kk)$ is simply a $1\times1$ matrix, thus the Hartree-Fock band energy is simply
\begin{equation}
\omega_{e_Y,\eta,s}(\kk) =h_{HF,\nu,\nu_C}^{e_Y,\eta,s} (\kk)\ .
\end{equation}

\subsection{The nonchiral-flat limit}
In the nonchiral-flat limit, if $\nu$ is even and $\nu_C=0$, the state $|\Psi_\nu^{\nu_+,\nu_-}\rangle=|\Psi_\nu\rangle$ is still an exact ground state. Note that the state $|\Psi_\nu\rangle$ has each valley-spin flavor either fully occupied or fully empty. In this case, the effective Hartree-Fock Hamiltonian under the energy band basis $n=\pm1$ can be written as
\begin{equation}
H_{HF}(|\Psi_{\nu}\rangle)=\sum_{\kk}\sum_{\eta,s,m,n} \left[h_{HF,\nu}^{\eta,s} (\kk)\right]_{mn} c^\dagger_{\kk,m,\eta,s} c_{\kk,n,\eta,s}\ ,
\end{equation}
with a $2\times 2$ Hartree-Fock Hamiltonian matrix
\begin{equation}\label{seq:HF-nonchiral-flat}
\begin{split}
&\left[h_{HF,\nu}^{\eta,s} (\kk)\right]_{mn}=\frac{1}{\Omega_{\text{tot}}}\sum_\GG \Big[V(\GG)\Big(\sum_{\kk',m',n',\eta',s'}M_{mn}^{(\eta)}(\kk,\GG) M_{m'n'}^{(\eta')}(\kk',-\GG) \langle c^\dagger_{\kk',m',\eta',s'} c_{\kk',n',\eta',s'} -\frac{1}{2}\delta_{m'n'} \rangle\Big) \\
&\quad -\frac{1}{2}\Big(V(\qq+\GG)\sum_{\qq,m',n'}[M^{(\eta)}(\kk,\qq+\GG)]^\dag_{mm'}M_{n'n}^{(\eta)}(\kk,\qq+\GG) \langle c^\dagger_{\kk+\qq,n',\eta,s} c_{\kk+\qq,m',\eta,s} -c_{\kk+\qq,m',\eta,s} c^\dagger_{\kk+\qq,n',\eta,s} \rangle\Big) \Big] \\
&=\frac{1}{\Omega_{\text{tot}}}\sum_\GG \Big[\nu V(\GG)M_{mn}^{(\eta)}(\kk,\GG)\sum_{\kk'} \alpha_0(\kk',-\GG) -(\nu_{\eta,s}-\frac{1}{2})\Big(V(\qq+\GG)\sum_{\qq,m'}[M^{(\eta)}(\kk,\qq+\GG)]^\dag_{mm'}M_{m'n}^{(\eta)}(\kk,\qq+\GG)\Big) \Big]\\
&=[\nu R_+(\kk)-(2\nu_{\eta,s}-1)R_-(\kk)]_{mn}\ ,
\end{split}
\end{equation}
where we have defined the $2\times2$ matrices
\begin{equation}
\begin{split}
&[R_+(\kk)]_{mn}=\frac{1}{\Omega_{\text{tot}}}\sum_\GG V(\GG)M_{mn}^{(\eta)}(\kk,\GG)\sum_{\kk'} \alpha_0(\kk',-\GG)\ , \\
&[R_-(\kk)]_{mn}=\frac{1}{2\Omega_{\text{tot}}}\sum_{\qq,\GG}V(\qq+\GG)\sum_{\qq,m'}[M^{(\eta)}(\kk,\qq+\GG)]^\dag_{mm'}M_{m'n}^{(\eta)}(\kk,\qq+\GG) \ ,
\end{split}
\end{equation}
and $\nu_{\eta,s}=0$ or $1$ if the valley spin flavor $\eta,s$ is fully empty or fully occupied.

This Hartree-Fock Hamiltonian is also equal to the Hamiltonian of the exactly solvable electron (hole) excitations in an empty (occupied) valley-spin flavor $\eta,s$ in the nonchiral-flat limit derived in Ref.~\cite{ourpaper5}, with the electron (hole) excitations therein corresponding to the Hartree-Fock conduction (valence) bands here.

\section{The stabilizer Code Limit}\label{app:stabilizer}

As shown in Ref.~\cite{ourpaper3}, in the chiral limit $w_0=0$, if $M_{m,n}^{\left(\eta\right)}\left(\mathbf{k},\mathbf{q+G}\right)$ is independent of $\mathbf{k}$, the Hamiltonian $H=H_I$ becomes similar to a stabilizer code with all of its terms $O_{\mathbf{-q,-G}}O_{\mathbf{q,G}}$ mutually commuting. In this section, we show that all the many-body eigenstates are exactly solvable. 

In this stabilizer code limit, by Eq.~(\ref{eq:chiral-OqG}) we have
\begin{equation}
O_{\mathbf{q,G}}=\sum_{\mathbf{k},e_Y,\eta,s} \beta_{e_Y}(\mathbf{q+G})  (d_{\mathbf{k+q},e_Y,\eta,s}^\dag d_{\mathbf{k},e_Y,\eta,s} -\frac{1}{2})\ ,
\end{equation}
where $ d_{\mathbf{k},e_Y,\eta,s}^\dag$ ($e_Y=\pm1$) is the Chern basis defined in Eq.~(\ref{eq:Chern-band}), and
\begin{equation}
\beta_{e_Y}(\mathbf{q+G})=\sqrt{V(\mathbf{G}+\mathbf{q})} M_{e_Y}(\kk,\qq+\GG)=\sqrt{V(\mathbf{G}+\mathbf{q})} [\alpha_0(\mathbf{k,q+G})+ie_Y\alpha_2(\mathbf{k,q+G})]\ ,
\end{equation}

and we have assumed $\alpha_0(\mathbf{k,q+G})$ and $\alpha_2(\mathbf{k,q+G})$ are independent of $\mathbf{k}$. One then has $[O_{\mathbf{q,G}},O_{\mathbf{q',G'}}]=0$ and thus the Hamiltonian $H_I$ in Eq.~(\ref{seq-pHI}) is a sum of commuting terms, and thus is similar to \emph{stabilizer Code} (see proof and discussion in Ref.~\cite{ourpaper3}).

We now show that the Hamiltonian in this stabilizer code limit can be transformed into an extended Hubbard model with zero hopping, i.e., a purely classical electrostatic problem. 
It is not difficult to find that
\begin{equation}
d_{e_Y,\eta,s,\mathbf{R}_M}^\dag=\frac{1}{\sqrt{N_M}}\sum_{\mathbf{k}}e^{i\mathbf{k\cdot R}_M}d^\dag_{\mathbf{k},e_Y,\eta,s}
\end{equation}
form a complete orthonormal basis, and satisfies
\begin{equation}
[O_{\mathbf{q,G}},d^\dag_{e_Y,\eta,s,\mathbf{R}_M}]=\beta_\theta(\mathbf{q+G})e^{-i\mathbf{q\cdot R}_M} d^\dag_{e_Y,\eta,s,\mathbf{R}_M}\ ,
\end{equation}
where $\mathbf{R}_M$ are the moir\'e unit cell sites defined at AA stacking centers of TBG. 

Using this new basis, we can define a Fourier transformation of $O_{\qq,\GG}$ in continuous space as
\begin{equation}
\begin{split}
&O(\mathbf{r})=\frac{1}{\Omega_{\text{tot}}}\sum_\mathbf{q,G}e^{i\mathbf{(q+G)\cdot r}}O_{\mathbf{q,G}}= 
\frac{1}{\Omega_{\text{tot}}}\sum_{\mathbf{k,q,G},e_Y,\eta,s}\beta_{e_Y}(\mathbf{q+G})e^{i\mathbf{(q+G)\cdot r}} (d_{\mathbf{k+q},e_Y,\eta,s}^\dag d_{\mathbf{k},e_Y,\eta,s}-\frac{1}{2}) \\
&=\frac{1}{\Omega_{\text{tot}}N_M}\sum_{\mathbf{k,q,G},\RR_M,\RR_M',e_Y,\eta,s}\beta_{e_Y}(\mathbf{q+G})e^{i\mathbf{(q+G)\cdot r}-i\mathbf{(k+q)\cdot R}_M+i\mathbf{k\cdot R'}_M} (d_{e_Y,\eta,s,\RR_M}^\dag d_{e_Y,\eta,s,\RR_M}-\frac{1}{2})\\
&=\frac{1}{\Omega_{\text{tot}}}\sum_{\mathbf{q,G,R}_M,e_Y,\eta,s}\beta_{e_Y}(\mathbf{q+G})e^{i\mathbf{(q+G)\cdot r}-i\mathbf{q\cdot R}_M} (d_{e_Y,\eta,s,\RR_M}^\dag d_{e_Y,\eta,s,\RR_M}-\frac{1}{2})\\
&=\sum_{\mathbf{R}_M,e_Y,\eta,s}\widetilde{\beta}_{e_Y} (\mathbf{r-R}_M) (d_{e_Y,\eta,s,\RR_M}^\dag d_{e_Y,\eta,s,\RR_M}-\frac{1}{2})\ ,
\end{split}
\end{equation}
where $\widetilde{\beta}_{e_Y} (\mathbf{r})$ is the Fourier transform of $\beta_{e_Y}(\mathbf{q})$. The interaction Hamiltonian is then given by
\begin{equation}
H_{I}=\frac{1}{2}\int d^2\mathbf{r} O(\mathbf{r})^2\ .
\end{equation}
From the form of $O(\mathbf{r})$, we know the Hamiltonian consists of density-density interaction between different moir\'e sites. Equivalently, we can rewrite the interaction Hamiltonian as
\begin{equation}
H_I=\frac{1}{2}\sum_{e_Y,s,\eta, e_Y',s',\eta'}\sum_{\RR_M,\RR_M'} U^{e_Y,e_Y'}_{\RR_M-\RR_M'} n_{e_Y,\eta,s,\RR_M} n_{e_Y',\eta',s',\RR_M'}\ ,
\end{equation}
where
\begin{equation}
\begin{split}
&U^{e_Y,e_Y'}_{\RR_M-\RR_M'}=\int d^2\mathbf{r} \widetilde{\beta}_{e_Y} (\mathbf{r-R}_M) \widetilde{\beta}_{e_Y'} (\mathbf{r-R}_M')=\frac{1}{\Omega_{\text{tot}}}\sum_{\qq,\GG} \beta_{e_Y}(\mathbf{q+G}) \beta_{e_Y'}(\mathbf{-q-G})e^{i(\qq+\GG)\cdot(\RR_M-\RR_M')}\ ,\\
&n_{e_Y,\eta,s,\RR_M}=d^\dag_{e_Y,\eta,s,\RR_M} d_{e_Y,\eta,s,\RR_M}-\frac{1}{2}\ .
\end{split}
\end{equation}
$U^{e_Y,e_Y'}_{\RR_M-\RR_M'}$ is the density-density interaction between site $\mathbf{R}_M$ and $\mathbf{R}_M'$, which only depends on $\RR_M-\RR_M'$. The number operator $n_{e_Y,\eta,s,\RR_M}$ at each $\mathbf{R}_M$ can be $-1/2$ or $1/2$. The many-body eigenstates and energies are then given by assigning each electron occupation operator $n_{e_Y,\eta,s,\RR_M}$ a number $\pm1/2$.

We note that the function $\widetilde{\beta}_{e_Y} (\mathbf{r})$ is not necessarily local. In fact, the fermion operator $d^\dag_{e_Y,\eta,s,\RR_M}$ cannot be local, because $d^\dag_{\mathbf{k},e_Y,\eta,s}$ form a Chern band basis and do not have global gauge in the moir\'e BZ. However, $d^\dag_{e_Y,\eta,s,\RR_M}$ at different $\mathbf{R}_M$ are indeed orthogonal.

\end{document}